%% file: ms_arXiv.tex
\documentclass[12pt,preprint]{aastex}

\shorttitle{PSF Correction}
\shortauthors{Levinson}
\usepackage{graphicx}    
\usepackage{amsmath}   
\usepackage{lscape}
\usepackage{multirow}
\usepackage{natbib}
\usepackage{afterpage}
\usepackage[bottom]{footmisc}

\begin{document}

\title{Analytic PSF Correction for Gravitational Flexion Studies}
\author{Rebecca Levinson}
\affil{MIT Kavli Institute, Cambridge, MA 02139}
\email{rsobel@mit.edu}

\begin{abstract}
Given a galaxy image, one cannot simply measure its flexion.  An image's spin one and three shape properties, typically associated with F\-- and G\--flexion, are actually complicated functions of the galaxy's intrinsic shape and the telescope's PSF, in addition to the lensing properties.  The same is true for shear.  In this work we create a completely analytic mapping from apparent measured galaxy flexions to gravitational flexions by (1) creating simple models for a lensed galaxy and for a PSF whose distortions are dominated by atmospheric smearing and optical aberrations, (2) convolving the two models, and (3) comparing the pre\-- and post\--convolved flexion-like shape variations of the final image.  For completeness, we do the same for shear.  As expected, telescope astigmatism, coma, and trefoil can corrupt measurements of shear, F\--flexion, and G\--flexion, especially for small galaxies.  We additionally find that PSF size dilutes the flexion signal more rapidly than the shear signal.  Moreover, mixing between shears, flexions, and asymmetric aberrations can create additive offsets in lensing measurements that vary with both galaxy size and galaxy ellipticity and flexion values.  But all is not lost; by measuring the patterns, we can correct for them.
\end{abstract}

\input{section_1_introduction}
\input{section_2_aberrations}
\input{section_3_models}

\input{section_4_extracting_gravitational_lensing_parameters}

\input{section_5_conclusion}

\input{section_6_biblio}
\appendix
\input{section_7_appendix}

\end{document}

%% file: section_1_introduction.tex
\section{Introduction}
		
	Weak gravitational lensing is one of the key tools with which we can probe the dark matter content of our universe and by extension explore its structure and formation history.  \citet{TysonWenk1990} were the first to measure the statistical apparent alignment of many galaxies caused by gravitational shear.  Since their initial measurement, weak lensing has been used to measure mass structure and substructure in regimes outside of strong lensing's reach.  Flexion, the next order gravitational effect, has the potential to bridge the gap between shear and strong lensing by probing scales between the two and providing orthogonal constraints on the mass estimates of each.  \citet{GoldbergBacon2005} made the first measurement of flexion, and \citet{VelanderKuijken2011} and \citet{CainSchechter2011} made notable progress in this area.
		
	But lensing measurements must still be honed if they are to live up to their theoretical potential.  Observationally, weak lensing is the science and art of measuring small shape distortions imparted to galaxy images by foreground masses-- a measurement which faces many practical encumbrances.  Weak lensing signals are by definition small; shear measurements are swamped by scatter in intrinsic galaxy ellipticities; flexion signals mix with galaxy ellipticity and gravitational shear and are then quickly degraded by low signal to noise \citep{ViolaMelchior2012, VelanderKuijken2011, RoweBacon2012}. 

	To circumvent these issues, we might (and do) average shears and flexions over many objects.  But averages aren't a foolproof way to boost the signal.  Imperfect masking of stars and neighboring galaxies limit available lensing targets over which averages can be made \citep{CainSchechter2011, RoweBacon2012} and can impart net biases in the flexion measurements for those which we do consider \citep{VelanderKuijken2011}.  For ground based images, the atmosphere dilutes the lensing signal, biasing measured lensing signals low.  Telescope jitter can add asymmetries to the PSF and consequently impart directional bias to shear and flexion measurements.   Optical aberrations and detector effects add field variable asymmetric distortions to the images, such that the PSF varies in both magnitude and direction with field location \citep{Schroeder1987, SchechterLevinson2011}.  Accordingly, directional offsets in shear and flexion measurements are variable in the field.
			
	The 2008 and 2010 GRavitational lEnsing Accuracy Testing challenges, GREAT08 and GREAT10, \citep{BridleBalan2010, KitchingBalan2012} were controlled lensing simulations specifically deployed to test ``the accuracy of current shape measurement methods" used to detect the weak lensing shear signal.  The GREAT10 challenge was an improvement over its predecessor in that it simulated the field variability of the shear-like signal often present in a point spread function (PSF).  In evaluating the results of the GREAT10 challenge, \citet{KitchingBalan2012} note two biases relating to the PSF that we quote here:
\begin{itemize}
\item ``Despite the PSF being known exactly we find contributions to biases from PSF size, but less so from PSF ellipticity. The methods with the largest biases have a strong PSF\--size correlation."
\item ``For large galaxies well sampled by the PSF, with scale radii $\gtrsim 2$ times the mean PSF size we find that methods meet requirements on bias parameters for the most ambitious experiments. However if galaxies are [relatively] unresolved with radii $\lesssim 1$ time the PSF size, biases become significant."
\end{itemize}
\noindent Properly accounting for the PSF, whether symmetric or asymmetric, seems to demand future improvement for various lensing measurement techniques, especially when analyzing galaxies of small size relative to the PSF.  If lensing surveys are to be conducted from the ground, as is proposed with with Large Synoptic Survey Telescope \citep{NWNH2010}, accounting for large, atmospherically induced PSFs will be critical. The GREAT challenges do not address measurements of gravitational flexion and how they may be affected by symmetric and asymmetric PSFs.
	
	There are two primary methods for determining weak lensing flexion (and by default some information about shear) in galaxy images: by measuring moments of galaxy images and by fitting the images to models.   Moment measuring techniques notably include Higher Order Lensing Image Characteristics (HOLICS) of \citet{OkuraUmetsu2007}, based on the work of \citet{KaiserSquires1995} (KSB).  For KSB, weighted second image moments are related to shear.  HOLICs extends the principles of KSB, measuring moments through fourth to obtain the flexion signal as well.  Image fitting methods include  shapelets \citep{Refregier2003, RefregierBacon2003} and the Analytic Image Method (AIM) \citep{CainSchechter2011}.  For image fitting techniques, a model of an unlensed galaxy is artificially lensed by shear and flexion and then fit to the galaxy image.  For the shapelets method, image data are decomposed into cartesian or polar shapelets, a basis of two-dimensional Gauss-Hermite polynomials (cartesian) or modified Laguerre polynomials (polar), and the coefficients of the decomposition are fit against an artificially sheared and flexed model galaxy represented in that same basis. For AIM, a flexed galaxy model, comprised of an unlensed galaxy model lensed by (variable) shear and flexion, is fit directly to the image data, with no prior decomposition step.
		
	Different groups deal with the PSF in different ways.  \citet{KaiserSquires1995} proposed a semi-empirical method of PSF correction whereby biases in the shear are corrected ignoring the effects of seeing, and then the `seeing induced suppression [of shear] as a function of image size' is calibrated out using artificially smeared and degraded HST data.  The KSB method accounts for anisotropies in the PSF, but overall the form of the PSF required by this deconvolution method needs to be rather simple. \citet{OkuraFutamase2012} expanded on the traditional KSB approach using Elliptical-HOLICs, dividing the PSF into elliptical components orthogonal and parallel to the galaxy shear rather than into isotropic and anisotropic components.  
		
	\citet{MelchiorViola2011} introduced an alternate PSF deconvolution technique for moments methods that avoids some of the shortcomings of the traditional KSB approach.  DEIMOS, short for `deconvolution in moments space', is a method for separating the PSF's moments from a sheared galaxy's moments, without assuming an a priori form for the PSF, or splitting the PSF into isotropic and anisotropic (or orthogonal/parallel) components.  While the weighting functions required to accurately measure moments in noisy fields necessarily make this method mathematically approximate, the technique is still highly accurate, and only requires the same computational time as the traditional KSB approach.  This same principle of deconvolving the PSF in moments space which is applied to shear measurements in \citet{MelchiorViola2011} could be extended to higher order for flexion studies.  

	Groups using the shapelets method account for the PSF before even fitting the galaxy data to a model.  As detailed in \citet{Kuijken2006} and \citet{VelanderKuijken2011}, a model for the PSF is created by decomposing stellar images into shaplets.  Next, a model for each lensed galaxy is created by superposing circularly symmetric shapelets.  The galaxy model is then translated, sheared and flexed, and finally convolved with the model PSF.  This lensed, convolved galaxy model is represented in the polar shapelets basis.  Finally, the image data is decomposed in the polar shapelets basis, and this decomposed galaxy data is fit to the aforementioned galaxy model.  The resulting best fit lensing parameters are therefore pre-corrected for the PSF.  This method of PSF `deconvolution' allows for as complex a PSF as the image resolution and the CPU performing the convolution can handle.  
			
	While one can always treat each individual case numerically, analytic estimates permit the design of observing programs without resorting to massive simulation.  We therefore seek an analytic conversion from measured image properties to true lensing parameters.  In order to achieve this end we create simple, analytic models for the galaxy and the PSF.  For the galaxy model, we draw heavily on the work of \citet{CainSchechter2011} and AIM.  For the PSF model, we use the fact that, in the presence of atmospheric seeing, geometric optics suffice to describe the telescope aberrations, and the combination of seeing and the optically induced aberrations will dominate over other effects.  We therefore draw from \citet{JarvisSchechter2008} to construct an analytic PSF model at individual field positions and refer the reader to \citet{SchechterLevinson2011} for a model of PSF field variations.   We convolve the models for the lensed galaxy and aberrated PSF, create a mapping from gravitational to measured lensing values, and invert to find the mapping from measured to gravitatoinal.

	The outline of the paper is as follows.  In \S 2, we demonstrate that weak lensing deflections and telescope aberrations are functionally equivalent, and thus will manifest in images in very similar ways.  In \S 3, we derive a simple model for a lensed galaxy by assuming that the unlensed galaxy can be approximated by a Gaussian radial profile.  Next, we derive a simple model for an aberrated PSF by assuming that atmospheric smearing produces a stellar image with a Gaussian radial profile.  For transparency, we choose to work directly with the Gaussian models of the galaxies, treating lensing and aberration effects as perturbations to Gaussians, and only switching to the shapelets basis of \citet{Refregier2003} and \citet{RefregierBacon2003} to perform convolutions between the PSF and galaxy, \S 4.  In \S 4, we additionally present the form of the final convolved galaxy image, and infer the transformation from the measured to gravitational lensing terms.  In \S 5 we conclude.  A review of the complex vector notation that will be used throughout this work to express the aberration and lensing models and the effects of lensing shear and flexions on images can be found in appendix A.

%% file: section_2_aberrations.tex
\section{Aberrations: glass or mass, it's all the same \label{sec:aberrations}}

Optical aberrations and gravitational lensing distortions are formally identical because they are computed nearly identically-- from path length differences of rays, using the same small angle approximations and expansions in circular bases to extract leading order terms.  We here outline the parallel procedures used to develop the two theories and so demonstrate their equivalence.

\subsection{Telescope aberrations}

Telescope aberrations are computed from the optical path length difference between the ray striking the center of the pupil (which defines the center of the unaberrated object in the image plane) and a ray from the same source object striking an arbitrary location on the pupil.  These path length differences are non-linear functions of the mirror shape, but in practice they are linearized in order to quantify the most extreme distortions to the image.  Because the pupil is usually circular, the natural basis in which to represent the aberrations as they vary on the pupil are Zernike polynomials \citep{Zernike1934}, namely ($tilt_1$, $tilt_2$, $defocus$, $astigmatism_1$, $astigmatism_2$, $coma_1$, $coma_2$, $trefoil_1$, $trefoil_2$, $spherical$, and higher-order aberrations).  The higher-order aberrations will generally be negligibly small in magnitude \citep{SchechterLevinson2011}.\footnote{The low order aberrations may be small in magnitude by design, but misalignment of the telescope optics can quickly render them large again.}

The mathematical forms for these aberrations as they vary with pupil coordinates $\rho$ and $\phi$ are given in tables \ref{tab:opticalgravitational} and \ref{tab:vecopticalgravitational}.  The tables exclude terms that displace measured image locations rather than distort images.  In practice, the orthonormality of the Zernikes is often dropped when discussing the aberrations, and we have done so here.\footnote{It is safe to drop the orthonormality in this discussion as we're not trying to compute the magnitude of each aberration type from first principles, but rather trying to group the functional dependences, i.e even if some $5th \ order \ coma$ sneaks into what we're calling $coma$ it doesn't matter.}  The derivation of the telescope aberrations through 3rd order can be found in chapter 5 in \citet{Schroeder1987}. For field variations and extension through 5th order,\footnote{Trefoil is in fact considered to be a 5th order aberration due to its 3rd order variation in field coordinate as well as pupil coordinate in an aligned system.  It is thus smaller than the other listed aberrations.  However, no other 3rd or 5th order aberrations vary with a similar spin symmetry.  We include it because of its relevance to flexion measurements} see \citet{SchechterLevinson2011}.

\subsection{Weak lensing aberrations}

A gravitational lens is entirely analogous to a telescope, with the lens plane of a weak lensing system akin to the pupil plane of an aberrated lens system.  Distortions are introduced to the weak lens system in the same way as they were in the telescope system, namely through the optical path length differences between the ray defining the center of an object in the lens/pupil plane (which defines the center of the undistorted object in the image plane) and a ray from the same source object striking an arbitrary location in the lens plane.  These path length differences are non-linear functions of the lensing potential, but in practice they are linearized in order to quantify the most extreme distortions to the image.  Due to convention and a desire to express lensing effects in terms of same-order derivatives of the lensing potential, the lensing community has adopted the following basis for representing aberrations as they vary with distance from the center of the source on the lensing plane ($convergence$, $shear_1$, $shear_2$, $F\text{\--}flexion_1$, $F\text{\--}flexion_2$, $G\text{\--}flexion_1$, $G\text{\--}flexion_2$). 

The mathematical forms for these aberrations as they vary with lens plane coordinates centered on the object, $\theta$ and $\omega$, are given in tables \ref{tab:opticalgravitational} and \ref{tab:vecopticalgravitational}.  {\it They are identical to the telescope aberrations but with pupil coordinates replaced by lensing plane coordinates.}  Note that the net deflection imparted to the image by the lens would have the same form as $tilt_1$ and $tilt_2$, but this effect is not measurable in most weak lensing studies as the undeflected galaxy position is unknowable, so we have excluded those terms here.

\begin{deluxetable}{l l l l}
     \tablecaption{Optical and gravitational aberration patterns.  Telescope pupil radial and angular coordinates are $\rho$ and $\phi$.  Source object radial and angular coordinates at the lens plane, centered on the object, are $\theta$ and $\omega$.  The aberrations are identical. \label{tab:opticalgravitational}}
     \tablewidth{17cm}
     \tablehead{
          \colhead{telescope aberration} &
          \colhead{pupil plane variation} &
          \colhead{gravitational aberration} & 
          \colhead{lens plane variation}
      }
     \startdata
	$defocus$		& $\rho^2$				& $convergence$	& $\theta^2$			\\
	$astigmatism_1$	& $\rho^2\cos 2\phi$		& $shear_1$			& $\theta^2\cos 2\omega$\\
	$astigmatism_2$	& $\rho^2\sin 2\phi$		& $shear_2$			& $\theta^2\sin 2\omega$	\\     
	$coma_1$		& $\rho^3\cos \phi$		& $F\text{\--}flexion_1$	& $\theta^3\cos \omega$	\\
	$coma_2$		& $\rho^3\sin \phi$		& $F\text{\--}flexion_2$	& $\theta^3\sin \omega$	\\     
	$trefoil_1$		& $\rho^3\cos 3\phi$		& $G\text{\--}flexion_1$	& $\theta^3\cos 3\omega$ \\
	$trefoil_2$		& $\rho^3\sin 3\phi$		& $G\text{\--}flexion_2$	& $\theta^3\sin 3\omega$  \\
	$spherical$		& $\rho^4$			& 					&					
     \enddata
\end{deluxetable}

\begin{deluxetable}{l l l l}
     \tablecaption{Vector notation for optical and gravitational aberration patterns.  Telescope pupil vector is $\vec{\rho}$.  Source object vector at the lens plane is $\vec{\theta}$.  The aberrations are identical. \label{tab:vecopticalgravitational}}
     \tablewidth{17cm}
     \tablehead{
          \colhead{telescope aberration} &
          \colhead{pupil plane variation} &
          \colhead{gravitational aberration} & 
          \colhead{lens plane variation}
      }
     \startdata
	$defocus$	& $\rho^2$			& $convergence$	& $\theta^2$			\\
	$astigmatism$	& $\vec{\rho}^2$		& $shear$			& $\vec{\theta}^2$		\\
	$coma$		& $\rho^2\vec{\rho}$		& $F\text{\--}flexion$	& $\theta^2\vec{\theta}$	\\
	$trefoil$		& $\vec{\rho}^3$		& $G\text{\--}flexion$	& $\vec{\theta}^3$		\\
	$spherical$	& $\rho^4$			& 				&						
     \enddata
\end{deluxetable}

Some may be unfamiliar with the lensing aberrations as computed from optical path length differences as opposed to deflections, but one can compute distortions equally well from either.  Deflections of light from a source are simply gradients in that light's wavefront.  Consider that, in the thin lens approximation, the deflection of light from a source galaxy at the lens plane is given by the gradient of the lensing potential, $\Phi$.  It follows that $\Phi$ is analogous to the wavefront of the source at the lens plane.\footnote{A traditional, unaberrated wavefront at a telescope pupil consists of parallel rays emanating from a single point in space.  The telescope's pupil is thus uniformly illuminated.  For a gravitational lensing system, the image of the galaxy at the lens plane is composed of rays emanating from multiple locations on the galaxy, not from a single point.  Therefore the lens plane `pupil'  does not contain a wavefront in the classical sense, but rather is filled with an unevenly illuminated image of the source galaxy.  As is discussed in section \S \ref{sec:models_gobal}, the differing illuminations of the telescope pupil and gravitational lens `pupil' can change the {\it effects} of their (identical) wavefront aberrations.}

In the weak lensing approximation, one is assuming that the wavefront can be linearized locally.  Using the notation discussed in the appendix, the optical path length difference between the ray from the center of the source and a ray emanating from elsewhere in the source, $\Phi(\theta) - \Phi(\theta_0)$, is given by

\begin{align}
\label{eq:linearized lensing potential}
\Phi(\theta) - \Phi(\theta_0) = 
	&\left( \frac{ \vec{\partial} }{\partial \theta} \Phi |_{\theta_0} \cdot \vec{\theta} \right)&
	\quad &(\text{deflection})&
	\\ \nonumber
	+  &\frac{1}{2} 
		\left[      \frac{1}{2}  \left(\frac{           \partial^2 }{ \partial \theta^2 }  \Phi |_{\theta_0}
			\theta^2 \right)
			+  \frac{1}{2} \left( \frac{ \vec{\partial}^2 }{ \partial \theta^2 } \Phi |_{\theta_0} 
			\cdot \vec{\theta}^2 \right) \right]&
	\quad &(\text{convergence \& shear})&
	\\ \nonumber
	+ &\frac{1}{4}\frac{1}{3} 
		\left[   \frac{3}{2} \left( \frac{           \partial^2 }{ \partial \theta^2 } 
			     \frac{ \vec{\partial}} {\partial \theta} \Phi |_{\theta_0} 
			    	\cdot \vec{\theta} \right) \theta^2 
			+  \frac{1}{2} \left( \frac{ \vec{\partial}^3 }{ \partial \theta^3 } \Phi |_{\theta_0} 
				\cdot \vec{\theta}^3 \right)
		\right].&
	\quad &(\text{F\-- \& G\--flexion})&
\end{align}

\noindent  The net deflection of the object from the source to the lens plane is the vector first derivative of the lensing potential, $\frac{ \vec{\partial} }{\partial \theta} \Phi |_{\theta_0}$.  $Convergence$ and $shear$ are the spin zero and spin two second derivatives, $\frac{1}{2}\frac{ \partial^2 }{ \partial \theta^2 }  \Phi |_{\theta_0}$ and $\frac{1}{2}\frac{ \vec{\partial}^2 }{ \partial \theta^2 } \Phi |_{\theta_0}$.  $F\text{\--}flexion$ and $G\text{\--}flexion$ are the spin one and spin three derivatives, $\frac{1}{2} \frac{ \partial^2 }{ \partial \theta^2 } \frac{ \vec{\partial}} {\partial \theta} \Phi |_{\theta_0}$ and $ \frac{1}{2}\frac{ \vec{\partial}^3 }{ \partial \theta^3 } \Phi |_{\theta_0}$.  Each term has the radial variation given in table \ref{tab:vecopticalgravitational}.  

To compute the resultant distortion imparted to the image one takes the gradient of this optical path length difference and recovers that the deflection term will be the same for all rays, simply displacing an image between source and image planes.  Only second and third derivatives will vary with ray position and thus contribute to shape distortion.  The above derivation could be inferred from many lensing works, though the fundamentals of lensing, flexion, and the imaginary number notation are set in \citet{BartelmannSchneider2001}, \citet{GoldbergNatarajan2002} and \citet{GoldbergBacon2005}, and \citet{BaconGoldberg2006} respectively.

%% file: section_3_models.tex
\section{Models for a lensed galaxy and an aberrated PSF}
\label{sec:models_gobal}

We aim here to capture the effects of shear, flexion, and corruptions to their measurable signals by an aberrated PSF.  Desiring only leading order influences on the asymmetries of the true and convolved galaxy, we use the simplest physically plausible models for both the galaxy and the PSF: 
\begin{itemize}
	\item An elliptical Gaussian model for the unlensed galaxy.  
	\item A circular Gaussian model for the effects of atmospheric seeing, in absence of telescope aberrations.
\end{itemize}
\noindent While adding more complexity to the radial profiles of the unlensed galaxy and atmospherically aberrated PSF models might yield results which can be better fine-tuned to a particular telescope's PSF (or a particular galaxy's morphology given adequate sampling to determine it), adding additional complexity to models can often obscure the physically motivated trends in the results.  We therefore use simple Gaussians for the radial profiles of unlensed galaxies and symmetrically aberrated PSFs.

To create the galaxy model, we use shear and flexion to lens a model for an initially unlensed galaxy.  The lensing wavefront variations are given by the rightmost components of table \ref{tab:vecopticalgravitational}.  In the weak lensing approximation, these delays are imparted onto the image of the galaxy at the plane of the lens.  The final galaxy model only accounts for shape distortions imparted by the lensing terms onto the initially unlensed galaxy, and no further aberrations.

To create the PSF model, we assume the PSF obtains an initial broad shape from the atmosphere and is then further aberrated by defocus, astigmatism, coma, and trefoil arising in the telescope.\footnote{Spherical aberration is very small in telescopes by design and cannot be reintroduced by misalignment errors.  While it can be reintroduced by despace errors between optical elements, these are usually well-controlled.} These leading order aberrations impart onto the image wavefront delays at the uniformly illuminated pupil of the telescope, see table \ref{tab:vecopticalgravitational}.  Analogous to the galaxy model, the final PSF model is the total image created by the symmetric atmospheric broadening and the possibly asymmetric effects of the telescope aberrations. 

For the galaxy model, we consider all terms through second order in shear.  These second order terms are cumbersome, but we keep them out of foresight, not malice.  They will in some cases have a larger effect on the final PSF convolved image shape than terms varying linearly in telescope aberration asymmetries.

In order to provide physical intuition for the galaxy and PSF models, we will express them first using the complex number notation shorthand.  Only when necessary will we switch to shapelets coefficients for convolution.

\subsection{Generic model for a lensed galaxy, a review}

A lensed galaxy in the image plane, $I(\vec{\theta})$, is exactly represented by

\begin{equation}
\label{eq:IeqI} 
I(\vec{\theta}) = I\left( \vec{\beta}(\vec{\theta}) \right).
\end{equation}

\noindent  for source plane coordinate $\vec{\beta}$ and image plane coordinate $\vec{\theta}$.  However, as detailed in the previous section, the deflection from source to image plane is approximated by a truncated Taylor series of the lensing potential, and thus the mapping of coordinates from source to image plane $\vec{\beta}(\vec{\theta})$ is imperfect.  Likewise, if we express $\vec{\beta} = \vec{\theta} +  \vec{\delta\theta}$, $\vec{\delta\theta}$ is imperfect.

As most galaxy models have some form of exponential radial profile (e.g. Gaussian, Sersic), it is often best to move the approximation for $\vec{\delta\theta}$, the difference in angular distance between ray offset from galaxy center in the source and image planes, out of the exponent.  Fortunately, $\vec{\delta\theta}$ is small, and this can be achieved via another Taylor expansion as is done in \citet{GoldbergBacon2005}, and expanded here in vector notation as

\begin{equation}
\label{eq:IeqItaylor} 
I(\vec{\theta}) \approx I(\vec{\beta})|_{\vec{\beta}=\vec{\theta}} 
	+  \frac{ \vec{\partial} }{\partial\beta} I(\vec{\beta}) |_{\vec{\beta}=\vec{\theta}} \cdot \vec{\delta\theta} 
	+ \frac{1}{2} \left( \frac{1}{2} \frac{ \vec{\partial}^2 }{ \partial \beta^2 } I(\vec{\beta}) |_{\vec{\beta}= \vec{\theta}} \cdot (\vec{\delta\theta})^2
	+ \frac{1}{2} \frac{          \partial^2 }{ \partial \beta^2 } I(\vec{\beta}) |_{\vec{\beta}= \vec{\theta}} (\delta\theta)^2 \right).
\end{equation}

In practice, the second derivative in this second, galaxy Taylor expansion is often dropped, but we have included it here for completeness, as all terms through second order in shear (galaxy ellipticity) will be addressed fully as will terms that vary as the product of ellipticity and flexion. Those varying as the square of flexion will be dropped as they are yet smaller.  For the remainder of this work, the first derivative in equation \eqref{eq:IeqItaylor} will be referred to as the linear galaxy expansion term and the second as the quadratic galaxy expansion term.  These expansions of the galaxy model are not to be confused with flexion and shear, which are expansions of the lensing potential.

Given that the conversion from source to image plane coordinates is the gradient of the linearized lensing potential from \S \ref{sec:aberrations}, equation \eqref{eq:linearized lensing potential}, one can compute $\vec{\beta}(\vec{\theta})$ in vector form to be

\begin{align}
\label{eq:source_2_image} 
 \vec{\beta}(\vec{\theta}) = \vec{\theta}  &- \vec{g}\vec{\theta}^{*} 
		- \frac{1}{4} \vec{\psi}_1^{*}\vec{\theta}^2 
		- \frac{1}{2} \vec{\psi}_1 \theta^2 
		- \frac{1}{4} \vec{\psi}_3 (\vec{\theta^{*}})^2 \\ \nonumber
\text{where  }	
	\kappa		&= \frac{1}{2}\frac{ \partial^2 }{ \partial \theta^2 }  \Phi |_{\theta_0} \\ \nonumber
	\vec{g}		&= \frac{1}{2}\frac{ \vec{\partial}^2 }{ \partial \theta^2 } \Phi |_{\theta_0} \quad 
				/ (1- \kappa) \\ \nonumber
	\vec{\psi}_1 	&= \frac{1}{2} \frac{ \partial^2 }{ \partial \theta^2 } 
					\frac{ \vec{\partial}} {\partial \theta} \Phi |_{\theta_0} \quad 
				/ (1- \kappa) \\ \nonumber
	\vec{\psi}_3 	&= \frac{1}{2}\frac{ \vec{\partial}^3 }{ \partial \theta^3 } \Phi |_{\theta_0} \quad
				/ (1- \kappa). \nonumber
\end{align}

\noindent In the expression for $\beta$, we have dropped the convergence in favor of using reduced shear, $g$, and reduced unitful flexions, $\psi_1$ and $\psi_3$ \citep{SchneiderEr2008}.

Unitless flexions are more appropriate for the following discussion as these quantities measure galaxy shape and thus are true properties of a measured image rather than properties of a deduced lensing potential.  We therefore switch to unitless flexions $\vec{F}$ and $\vec{G}$, where $\vec{\psi}_1$ and $\vec{\psi}_3$ are rendered unitless by multiplying by the galaxy half-light radius $r_{hl}$.  The equation for the source plane coordinate $\vec{\beta}(\vec{\theta})$ becomes

\begin{equation}
\label{eq:source_2_image2} 
\vec{\beta}(\vec{\theta}) = \vec{\theta} - \vec{g}\vec{\theta}^{*}
	- \frac{1}{4 r_{hl}} \left(
		\vec{F}^{*}\vec{\theta}^2 
		+ 2 \vec{F} \theta^2 
		+ \vec{G} (\vec{\theta^{*}})^2
		\right).
\end{equation}

\noindent Subtracting $\vec{\theta}$ from the above expression thus yields $\vec{\delta\theta}$ as
\begin{equation}
\label{eq:deltheta}
\vec{\delta\theta} = - \vec{g}\vec{\theta}^{*}
	- \frac{1}{4 r_{hl}} \left(
		\vec{F}^{*}\vec{\theta}^2 
		+ 2 \vec{F} \theta^2 
		+ \vec{G} (\vec{\theta^{*}})^2
		\right),
\end{equation}

\noindent a third order approximation of the deflection of rays about the center of the galaxy image.

\subsection{Test Case: Circular Gaussian model for the unlensed galaxy}

First we will consider a circular model for the unlensed galaxy.  While this model may seem overly simplistic, as a galaxy's intrinsic ellipticity is generally too large to be ignored, this test case provides insight into the final model for the image which uses an elliptical Gaussian as the prior for the unlensed galaxy.

   To second order in shear, the model for a lensed (initially circular) galaxy with unlensed Gaussian width $\alpha$ is

\begin{align}
\label{eq:Lensed_circular_galaxy_model} 
I(\vec{\theta}) = \frac{I_0}{2 \pi \alpha^2} 
	& \text{exp}\left(- \frac{\theta^2}{2 \alpha^2} \right) \times 
	\Bigg[ 1
	\\ \nonumber
	+ \Bigg\{ &
		\frac{1}{ \alpha^2 } 
			\left( \vec{g} \cdot \vec{\theta}^2 \right)
		+ \frac{1}{4 \sqrt{2 \text{ln}(2)} \alpha^3 } \Big[
			3 \left( \vec{F} \cdot \vec{\theta} \right) \theta^2 
			+ \left( \vec{G} \cdot \vec{\theta}^3 \right)
		\Big]
	\Bigg\}
	\\ \nonumber
	+ \Bigg\{ &
		- \frac{1}{4 \alpha^2} g^2 \theta^2
		+  \frac{1}{ 4 \alpha^{4} } g^2 \theta^4
		+ \frac{1}{ 4 \alpha^{4} } \left(\vec{g}^2 \cdot \vec{\theta}^4 \right)
		\\ \nonumber
		& - \frac{1}{4 \sqrt{2 \text{ln}(2)} \alpha^3 } \Big[
			3 \left( \frac{1}{3} \left(2\vec{g}\vec{F}^{*} + \vec{g}^{*}\vec{G } \right)
				 \cdot \vec{\theta} \right) \theta^2 
			+ \left( \left( \vec{g}\vec{F} \right)
				 \cdot \vec{\theta}^3 \right)
		\Big]
		\\ \nonumber
		& + \frac{1}{8 \sqrt{2 \text{ln}(2)} \alpha^{5} } \Big[
			\left( \left( 3 \vec{g}\vec{F}^{*} + \vec{g}^{*}\vec{G} \right)
				\cdot \vec{\theta} \right) \theta^4
			+ \left( \left(3 \vec{g}\vec{F} \right) 
				\cdot \vec{\theta}^3 \right) \theta^2
			+ \left( \left( \vec{g}\vec{G} \right) 
				\cdot \vec{\theta}^5 \right)
		\Big]
	\Bigg\}
	\Bigg].
\end{align}

\noindent  In the above expression and all following discussion, we simplify all terms into `polynomial' form, where products of aberrations multiply powers of $\vec{\theta}$.

The first row of the expression is the circular, unlensed galaxy simply moved to the image plane.  The first braced term is the linear part of the galaxy expansion and varies linearly with shear and flexion.  The second braced term containing the last three rows are the quadratic part of the galaxy expansion.  These vary as the square of shear and the products of shear and flexion.  For the moment, we shall retain all terms in equation \eqref{eq:Lensed_circular_galaxy_model} without further reduction.  However, the observant reader will likely note that the flexion-like spin one and spin three terms with cubed radial dependence in row 4 can be incorporated into the linear expansion term through a transformation of $\vec{F}$ and $\vec{G}$ to new variables $\vec{\tilde{F}}$ and $\vec{\tilde{G}}$ by shear.

\subsection{Elliptical Gaussian model for the unlensed galaxy}

	Inserting a more physically realistic, intrinsically elliptical galaxy into equation \eqref{eq:IeqItaylor}, we find

\begin{align}
\label{eq:Lensed_elliptical_galaxy_model} 
I_{e}(\vec{\theta}) = \frac{I_0}{2 \pi \alpha^2} 
	& \text{exp}\left(- \frac{1}{2 \alpha^2} 
		\left( (1+e^2)\theta^2 - 2 \left( \vec{e} \cdot \vec{\theta}^2 \right) \right)
	\right) \times 
	\Bigg[ 1 
	\\ \nonumber
	+ \Bigg\{ &
		\frac{1}{ \alpha^2 } 
			\left( \vec{g} \cdot \vec{\theta}^2 \right)
		+ \frac{1}{4 \sqrt{2 \text{ln}(2)} \alpha^3 } \Big[
			3 \left( \vec{F} \cdot \vec{\theta} \right) \theta^2 
			+ \left( \vec{G} \cdot \vec{\theta}^3 \right)
		\Big]
	\Bigg\}
	\\ \nonumber
	+ \Bigg\{ &
		- \frac{1}{4 \alpha^2} g^2 \theta^2
		+  \frac{1}{ 4 \alpha^{4} } g^2 \theta^4
		+ \frac{1}{ 4 \alpha^{4} } \left( \vec{g}^2 \cdot \vec{\theta}^4 \right)
		\\ \nonumber
		& - \frac{1}{4 \sqrt{2 \text{ln}(2)} \alpha^3 } \Big[
			3 \left( \frac{1}{3} \left(2\vec{g}\vec{F}^{*} + \vec{g}^{*}\vec{G } \right)
				 \cdot \vec{\theta} \right) \theta^2 
			+ \left( \left( \vec{g}\vec{F} \right)
				 \cdot \vec{\theta}^3 \right)
		\Big]
		\\ \nonumber
		& + \frac{1}{8 \sqrt{2 \text{ln}(2)} \alpha^{5} } \Big[
			\left( \left( 3 \vec{g}\vec{F}^{*} + \vec{g}^{*}\vec{G} \right)
				\cdot \vec{\theta} \right) \theta^4
			+ \left( \left(3 \vec{g}\vec{F} \right) 
				\cdot \vec{\theta}^3 \right) \theta^2
			+ \left( \left( \vec{g}\vec{G} \right) 
				\cdot \vec{\theta}^5 \right)
		\Big]
	\Bigg\}
	\\ \nonumber
	+ \Bigg\{ &
		- \left(2 \vec{e} \vec{\theta^{*}} \right) 
			\cdot 
			\Big(  \frac{1}{ \alpha^2 } \left( \vec{g} \vec{\theta^{*}} \right)    
				+ \frac{1}{4 \sqrt{2 \text{ln}(2)} \alpha^3 } \Big[
					\vec{F}^{*}\vec{\theta}^2 + 2 \vec{F} \theta^2 
					+  \vec{G} (\vec{\theta^{*}})^2 \Big] 
			\Big)
	\Bigg\}
	\Bigg]
\end{align}

\noindent where $\vec{e}$ is the intrinsic galaxy ellipticity, apart from induced shear $\vec{g}$.  The Gaussian width is $\alpha$ in the limit of zero ellipticity. 

The first row of the model is now the elliptical unlensed galaxy moved to the image plane.  The next four rows are unchanged from the model created using a circular unlensed galaxy; the first braced term is the linear galaxy expansion and the second braced term is the quadratic lensing expansion.  However, these terms now multiply an elliptical Gaussian `base' galaxy rather than a circular Gaussian.  In addition to modifying the exponential term, two new terms, both second order in asymmetries, are added to the expansion.  These terms, contained in the last bracked row of the equation, are introduced by cross terms between the intrinsic galaxy ellipticity and the lensing terms which then carry down into the Taylor expansion.

It is useful to express the model for the galaxy image as a circular Gaussian plus perturbations for two reasons:  (1) to restrict the discussion to the most significant asymmetric terms only, i.e those varying to the lowest order in shear, intrinsic ellipticity, and flexion, and (2) to better facilitate convolutions in the shapelets basis.  Therefore we will once again Taylor expand the expression, reducing the model to a circular Gaussian plus perturbations.  We find

\begin{subequations}
\label{eq:Lensed_elliptical_galaxy_model2}
\begin{align}
\label{eq:base}
I(\vec{\theta}) = \frac{I_0}{2 \pi \alpha^2} 
	& \text{exp}\left(- \frac{\theta^2}{2 \alpha^2} \right) \times 
	\Bigg[ 1 
	\\ \label{eq:linear}
	+ \Bigg\{ &
		\frac{1}{ \alpha^2 } 
			\left( \vec{\tilde{g}} \cdot \vec{\theta}^2 \right)
		+ \frac{1}{4 \sqrt{2 \text{ln}(2)} \alpha^3 } \Big[
			3 \left( \vec{\tilde{F}} \cdot \vec{\theta} \right) \theta^2 
			+ \left( \vec{\tilde{G}} \cdot \vec{\theta}^3 \right)
		\Big]
	\Bigg\}
	\\ \label{eq:quad_shear}
	+ \Bigg\{ &
		- \frac{1}{4 \alpha^2} \left( \tilde{g}^2 + 2 (\vec{g} \cdot \vec{e}) \right) \theta^2
		+  \frac{1}{ 4 \alpha^{4} } \tilde{g}^2 \theta^4
		+ \frac{1}{ 4 \alpha^{4} } \left( \vec{\tilde{g}}^2 \cdot \vec{\theta}^4 \right)
		\\ \label{eq:quad_flex}
		& + \frac{1}{8 \sqrt{2 \text{ln}(2)} \alpha^{5} } \Big[
			\left( \left( 3 \vec{\tilde{g}}\vec{F}^{*} + \vec{\tilde{g}}^{*}\vec{G} \right)
				\cdot \vec{\theta} \right) \theta^4
		+ \left( \left(3 \vec{\tilde{g}}\vec{F} \right) 
				\cdot \vec{\theta}^3 \right) \theta^2
		+ \left( \left( \vec{\tilde{g}}\vec{G} \right) 
				\cdot \vec{\theta}^5 \right)
		\Big]
	\Bigg\}
	\Bigg].
\end{align}
\end{subequations}

The first row and braced term in the model, \eqref{eq:base} and \eqref{eq:linear}, are the same Gaussian model and linear expansion terms one would obtain by using a circular Gaussian as the unlensed galaxy model.  However, shear and flexion in \eqref{eq:linear} have been converted to new effective shears and flexions via the following transformations with galaxy ellipticity  

\begin{align}
\label{eq:mapping} 
& \vec{\tilde{g}} = \vec{g} + \vec{e} \\ \nonumber
& \vec{\tilde{F}} = \vec{F} - \frac{1}{3}(2\vec{\tilde{h}}\vec{F}^{*} + \vec{\tilde{h}}^{*}\vec{G}) \\ \nonumber 
& \vec{\tilde{G}} = \vec{G} - \vec{\tilde{h}}\vec{F} \\ \nonumber
& \text{where} \quad \vec{\tilde{h}} = \vec{g} + 2\vec{e}.
\end{align}

\noindent We could have made a similar transformation of $F$ and $G$ to tilde space for the model created using a circular Gaussian for the unlensed galaxy, equation \eqref{eq:Lensed_circular_galaxy_model}.  In so doing, we would have removed the redundant  $\vec{\theta}\theta^2$ and $\vec{\theta}^3$ terms there as we have done here.

The above transformations quantify the extent to which the nature of the observed object changes depending on how much intrinsic galaxy ellipticity, lensing shear, and each of the two types of flexions is present.  Lensing is measured as distortions to the galaxy image that are `shear-like' and `flexion-like' in the linear regime.  For the Gaussian model, these variations are $\vec{\theta}^2$, $\vec{\theta}\theta^2$, and $\vec{\theta}^3$.  As the ratios of intrinsic galaxy ellipticity, gravitational induced shear, and F\-- and G\--flexions change, so do the amounts of each term that will contribute to any particular shape distortion in the non-linear regime.  In the non-linear regime, shear and intrinsic galaxy ellipticity can combine with F\--flexion (spin one, third derivative of the lensing potential) to contribute to the G\--flexion-like galaxy distortion and vice versa.  The resulting quadratic `pseudoflexion' adds to the linear flexion to create the total effective flexion signal.

\subsection{Consequences of pseudoflexions}

The effective F\--flexion, $\vec{\tilde{F}}$, can be modified by a mixing between the linear F\--flexion and shear (or ellipticity) or between G\--flexion and shear.  G\--flexion is simpler; the effective signal, $\vec{\tilde{G}}$ can only be altered by a combination of F\--flexion and shear (or ellipticity).  As an extreme example of lensing signals mixing to masquerade as each other, consider a galaxy whose intrinsic spin three shape accidentally cancels out its linear G\--flexion lensing signal.  In this case, shear (or ellipticity) and F\--flexion could still produce a spin three pseudoflexion signal, which would impart a measurable $\tilde{G}$ lensing signal on the galaxy.

\begin{figure}[htb]
\epsscale{0.60}
\plotone{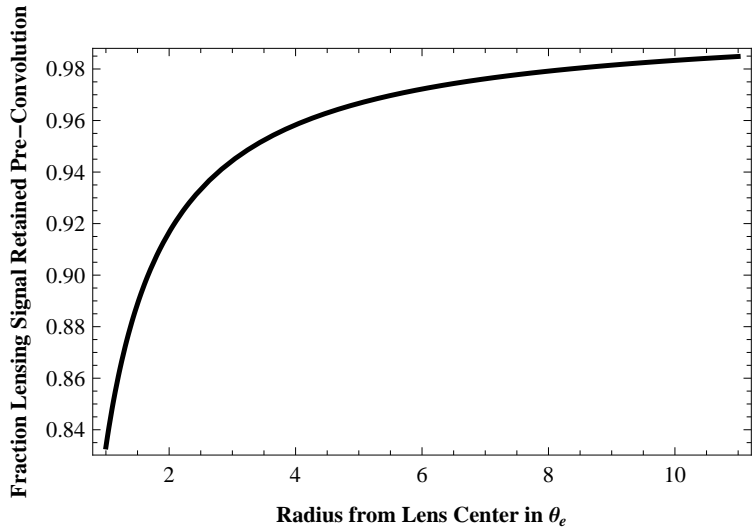}
\caption{Fraction of flexion signal observed for a circular galaxy lensed by a SIS.  Any ratio of $\frac{r_{hl}}{\theta_{e}}$ will produce these ratios of effective and true flexions.  All apparent signal loss occurs prior to convolution with a PSF, and is caused by linear flexion signals mixing with shear terms to create a `pseudoflexion' of opposite sign to the linear term. \label{fig:fracsigretained_preconv}}
\end{figure}

The above is an extreme example, but the quadratic pseudoflexions introduced by the non-linear expansion to the lensed galaxy model will generally contribute to the effective flexions by some amount.  Lensing systems consistent with a Singular Isothermal Sphere (SIS) profile will have aligned shear and flexions in the ratio $-g:-2\frac{r_{hl}}{\theta_e}g^2:6\frac{r_{hl}}{\theta_e}g^2$, where $r_{hl}$ and $\theta_e$ are the half light radii of the galaxy and the Einstein ring radius of the lens. Because shear and F\--flexion have the same sign, unless the source galaxy has significant intrinsic ellipticity opposing the shear, the effective $\tilde{G}$\--flexion will always be reduced by the combination of shear and F\--flexion.  Likewise, the effective $\tilde{F}$\--flexion, which is negative, will be rendered less negative by combinations of shear and itself and by shear and G\--flexion.  For a true SIS, the effective $\tilde{F}$\--flexion will be reduced by the same fraction as $\tilde{G}$\--flexion.  Figure \ref{fig:fracsigretained_preconv} shows the fractional reduction of apparent observable flexion, ($\tilde{F}$ or $\tilde{G}$ equivalently) for a SIS as a function of radial distance from the lens.  For a SIS lens at the Einstein ring, flexion and shear will reduce the apparent flexions by a full sixth, assuming no intrinsic galaxy ellipticity.  

If one were to consider only terms in the lensed galaxy model that vary like first order shears and flexions (i.e $\vec{\theta}^2$, $\vec{\theta} \theta^2$, and $\vec{\theta}^3$), one would not be able to distinguish between effective shears and flexions and the true lensing terms without some foreknowledge of either the shear or the intrinsic galaxy ellipticity.  This degeneracy has nothing to do with convolution, but rather the ability to resolve second order lensing effects.  Likewise, to first order in shear and flexion, one cannot distinguish between a model that uses a circular or an elliptical prior for the unlensed galaxy.

\subsection{An Aside: Shear-flexion `cross-talk' and pseudoflexion, similarities and differences}

	Using HOLICs, \citet{ViolaMelchior2012} were the first to quantify the corruption effects of shear-flexion mixing on flexion measurements.  They named the effect `cross-talk'.  
	
	The HOLICs method for computing flexion relies on measuring spin one and three distortion estimators.  These distortion estimators are the spin symmetric third moments of the galaxy intensity, divided by the spin zero fourth moment, (see eqs. (18) and (19) in \citet{ViolaMelchior2012}). The estimators have units of inverse galaxy scale length. 

	\citet{OkuraUmetsu2008} relate these estimators to flexions by making several simplifying assumptions, most notably that terms second order in shear and higher can be safely ignored.  The result of these simplifying assumptions are linear mappings between flexions and the distortion estimators.  \citet{ViolaMelchior2012} build on \citet{OkuraUmetsu2008}, retaining higher order terms, including products of shear and flexion.  In so doing they find, as we do here, that the distortion estimators are in fact functions of both flexion {\it and} the products of shear and flexion.  To test their results, \citet{ViolaMelchior2012} create synthetic flexed galaxies and measure their distortion estimators.  They then compute the expected distortion estimators using the known gravitational values used to create the synthetic galaxies.  They find that the expected distortion estimators computed using shear-flexion cross talk match the measured results to good agreement.  In contrast the distortion estimators computed only using terms linear in shear and flexion are much smaller than the measured distortions.  
	
	The non-linear mapping between distortion estimators and flexions proposed by \citet{ViolaMelchior2012} is now an implicit part of the HOLICs code for converting image moments into flexions.	
	
	Based on Viola's results one can conclude that, using the linear relationship between the distortion estimators and flexion, the HOLICs method would predict flexions which are too large.  Examining \eqref{eq:mapping} of this work, one can see that a linear fitting model will predict flexions which are smaller than the gravitational flexion values.  While these two statements would seem opposed, both are true.  
	
	The spin one and three distortion estimators in the HOLICs methods measure {\it all} spin one and three distortions, irrespective of the order of radial variation on the galaxy.\footnote{Albeit not equally- the moments of a cubic and quintic function with the same spin will not necessarily be the same, even if they have identical coefficients.} Therefore the distortion estimators are sensitive to the quintic spin one and three terms created by shear flexion mixing.   The spin three quintic cross term between shear and F\--flexion drives the net spin three moment up, resulting in an overestimated flexion.  

In contrast, modeling techniques, such as AIM, only see the corruption to the flexion-like, {\it cubic} spin one and three terms.  The pseudoflexion in these cubic terms forces the linear estimate of flexion lower than the true gravitational flexion, as discussed in the previous subsection.  Thus, the HOLICs and model fitting methods, while predicting opposite biases in the linear regime, are in complete agreement.

\subsection{A minimal galaxy model for analyzing the effects of PSF convolution}
\label{sec:minimal_model}

Up to and including terms which vary as the product of galaxy ellipticity (shear induced and intrinsic) and flexion, the model created by assuming a circular unlensed galaxy is no worse an approximation for analyzing the effects of convolution on shear and flexion than the model created using an elliptical unlensed galaxy.  In every place in the elliptical lensed galaxy model where shear and flexion appear, either the lensing terms can be mapped to effective shears and flexions in a way consistent with second order approximations, or the terms can be dropped from both the elliptical and circular lensed galaxy models without loss of accuracy to either.  The result is that the elliptical lensed galaxy model can be rendered formally equivalent to the circular model, but with newly named, effective lensing variables.

To elaborate, in equation \eqref{eq:quad_flex} every flexion term appears in product with shear $\vec{\tilde{g}}$, and so to second order may be replaced by its respective $\tilde{F}$ or $\tilde{G}$ equivalent.  In equation \eqref{eq:quad_shear}, where shear cannot simply be mapped to tilde space, the offending $\theta^2$ term can be modified or dropped as it is completely radially symmetric, and thus its inclusion in the model only modifies the steepness of the galaxy profile.  We are approximating the symmetric shape of the unlensed galaxy by a Gaussian, in order to better understand the effects and interplays of asymmetries caused by gravitational lensing and an asymmetric PSF caused by telescope aberrations.  Perturbations to the symmetric shape that affect the cuspiness of the final image are no more or less significant than the initial assumption that the unlensed  galaxy model is Gaussian. Therefore both $\theta^2$ terms and the $\theta^4$ symmetric term in equation \eqref{eq:quad_shear} may be folded into the Gaussian radial approximation without any additional losses.  With the conversion of shear and flexion to tilde space and the exclusion of symmetric terms, the model created using a circular Gaussian is identical to that created using an elliptical Gaussian unlensed galaxy.

We note here that \citet{ViolaMelchior2012} did explore the effects of higher order radial variations on shear and flexion measurements and found them to be non-negligible for flexion measurements obtained using moments.  For this work where we seek general trends imparted to the flexion signal by the PSF and fast, analytic remediation for the same, we are ignoring these non-Gaussian radial variations of the galaxy.  In the limit of low signal to noise or large PSF, variability in radial profiles galaxy to galaxy will wash out and this assumption should simply be crude, not devastating.

We also argue that, for the discussion of the deconvolution of the shear and flexion terms, we may additionally drop the terms from the model varying as $\vec{\tilde{g}}^2 \cdot \vec{\theta}^4$ and $\left( \vec{\tilde{g}}\vec{G} \right) \cdot \vec{\theta}^5$ in equation parts \eqref{eq:quad_shear} and \eqref{eq:quad_flex}.  While the elliptical and circular models are formally equivalent even without this step, dropping these superfluous terms makes the following discussion simpler.  Each of these terms are second order in galaxy ellipticity and shear, and thus only their convolutions with symmetric parts of the PSF will be retained in the final expression for the convolved galaxy image\-- all convolutions of these second order terms with asymmetric PSF terms that vary as astigmatism, coma, or trefoil must necessarily be third order small.  The former term has exclusive spin four symmetry and the latter has exclusive spin five symmetry.  The convolution of a spin m symmetric function with a spin 0 (i.e. symmetric) one will itself have spin m, and so these terms cannot affect measurements of shear and flexions which have spins one, two, and three.  Thus, while $\vec{\tilde{g}}^2 \cdot \vec{\theta}^4$ and $\left( \vec{\tilde{g}}\vec{G} \right) \cdot \vec{\theta}^5$ in equation parts \eqref{eq:quad_shear} and \eqref{eq:quad_flex} will affect the appearance of the pre\-- and post\--convolved galaxy, they will not affect measurements of shear or flexions in either and so can be dropped for this analysis.

The galaxy model we will be using for the remainder of this paper is thus

\begin{subequations}
\label{eq:Lensed_galaxy_model_final}
\begin{align}
\nonumber 
I(\vec{\theta}) = \frac{I_0}{2 \pi \alpha^2} &
	\text{exp}\left(- \frac{\theta^2}{2 \alpha^2} \right) \times \Bigg[ 
	\\ \label{eq:Lensed_galaxy_model_simple}  
	1
	& + \Bigg\{
		\frac{1}{ \alpha^2 } 
			\left(\vec{\tilde{g}} \cdot \vec{\theta}^2 \right)
		+ \frac{1}{4 \sqrt{2 \text{ln}(2)} \alpha^3 } \Big[
			3 \left( \vec{\tilde{F}} \cdot \vec{\theta} \right) \theta^2 
			+ \left( \vec{\tilde{G}} \cdot \vec{\theta}^3 \right)
		\Big]
	\Bigg\}
	\\ \label{eq:Lensed_galaxy_model_perturbation} 
	& + \Bigg\{
		\frac{1}{4 \sqrt{2 \text{ln}(2)} \alpha^{5} }  \times  \frac{1}{2}\Big[ 
	    	\left( \left( 3 \vec{\tilde{g}}\vec{F}^{*} + \vec{\tilde{g}}^{*}\vec{G} \right)
				\cdot \vec{\theta} \right) \theta^4
		+ \left( \left(3 \vec{\tilde{g}}\vec{F} \right) 
				\cdot \vec{\theta}^3 \right) \theta^2
		\Big]
	\Bigg\}\footnotemark
	\Bigg] 
\end{align}
\end{subequations}

\begin{figure}[htb]
\vspace{1.625 truein}
\includegraphics{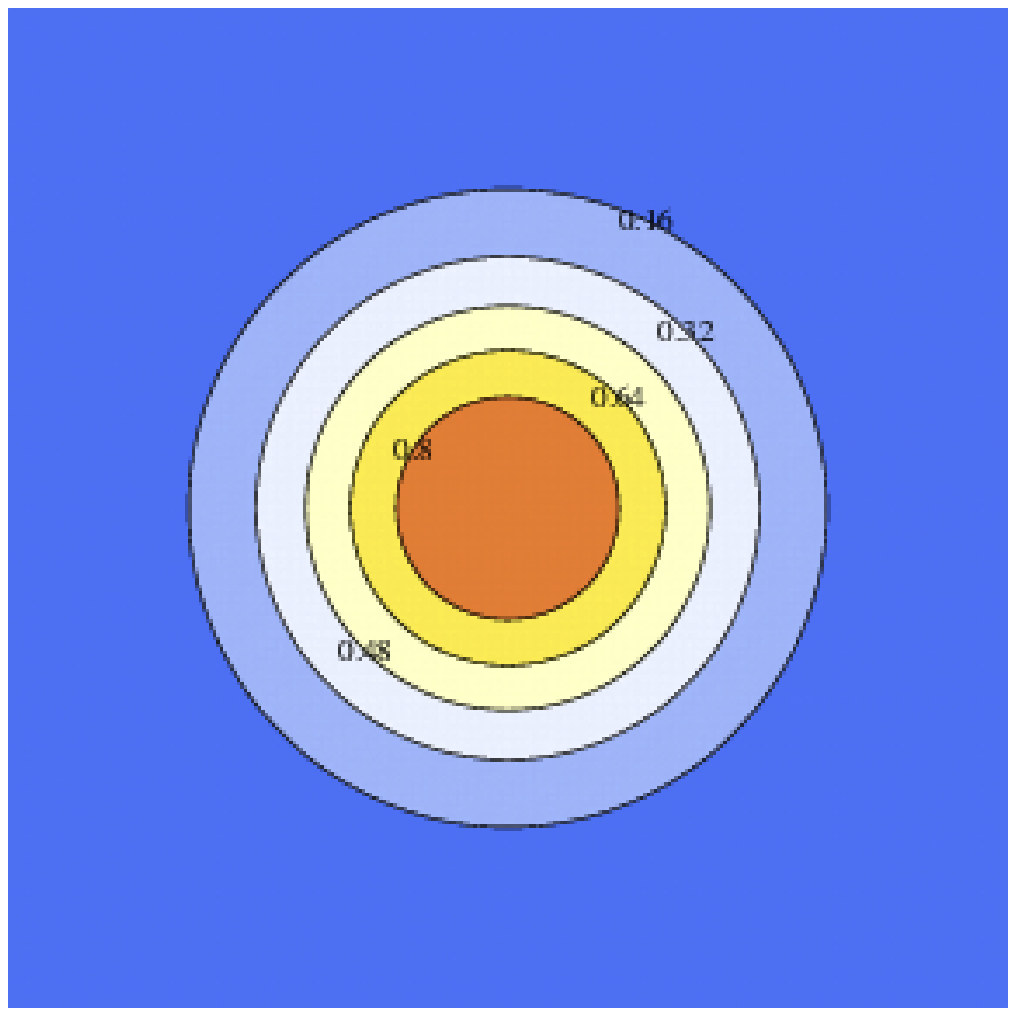}
\includegraphics{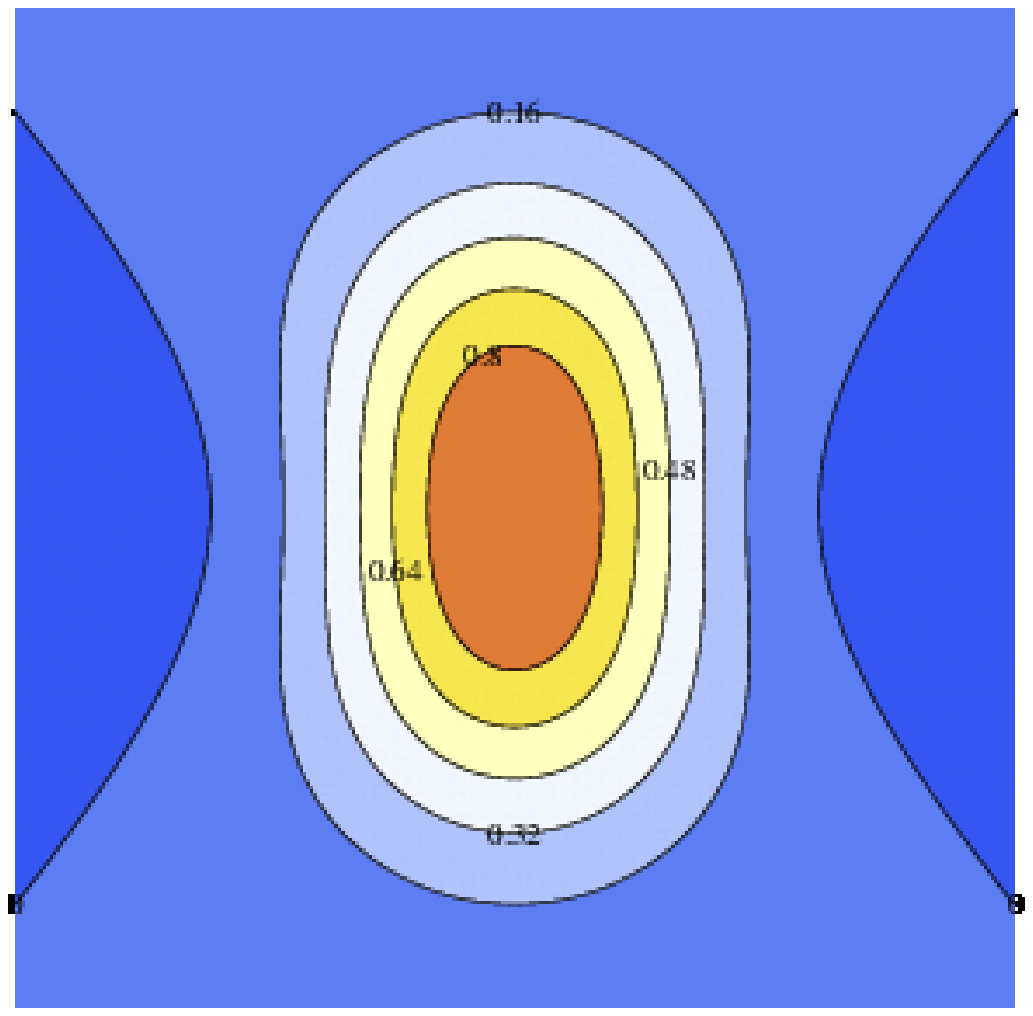}
\includegraphics{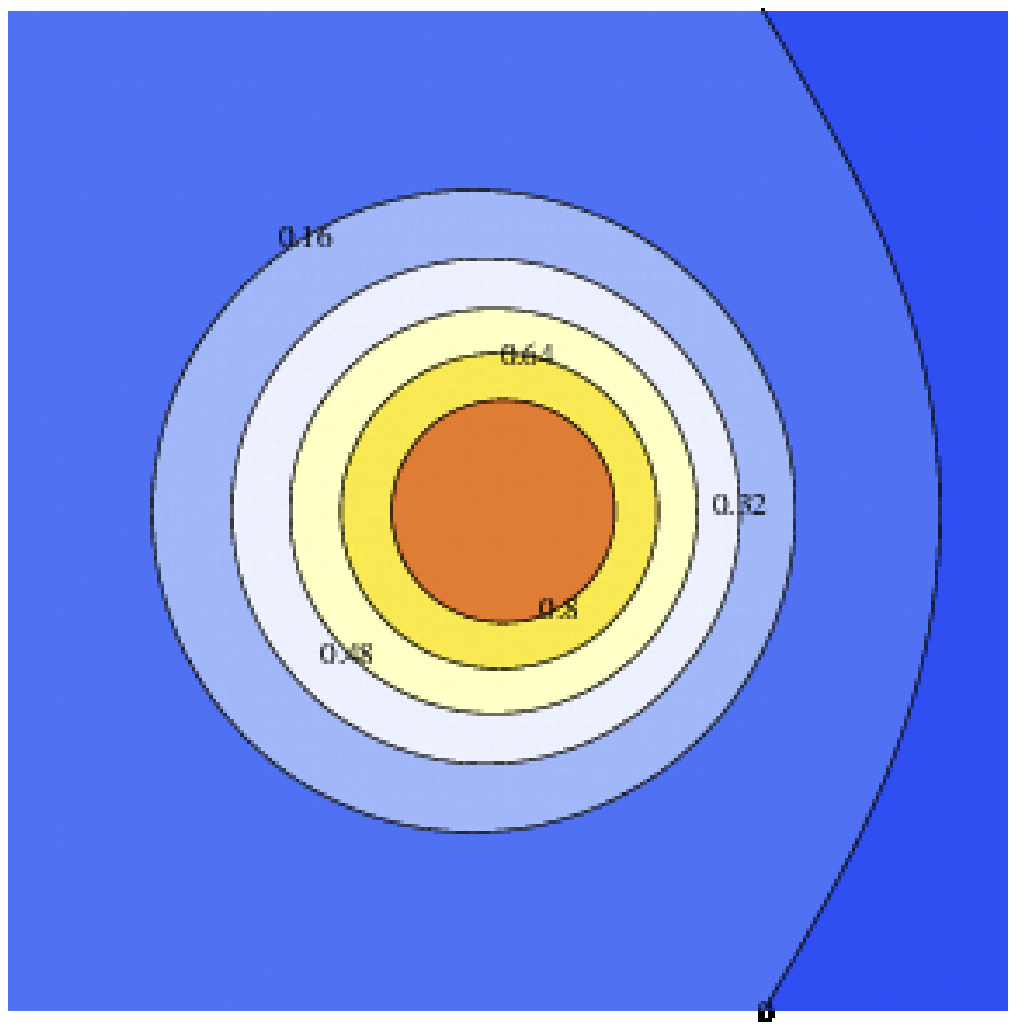}
\includegraphics{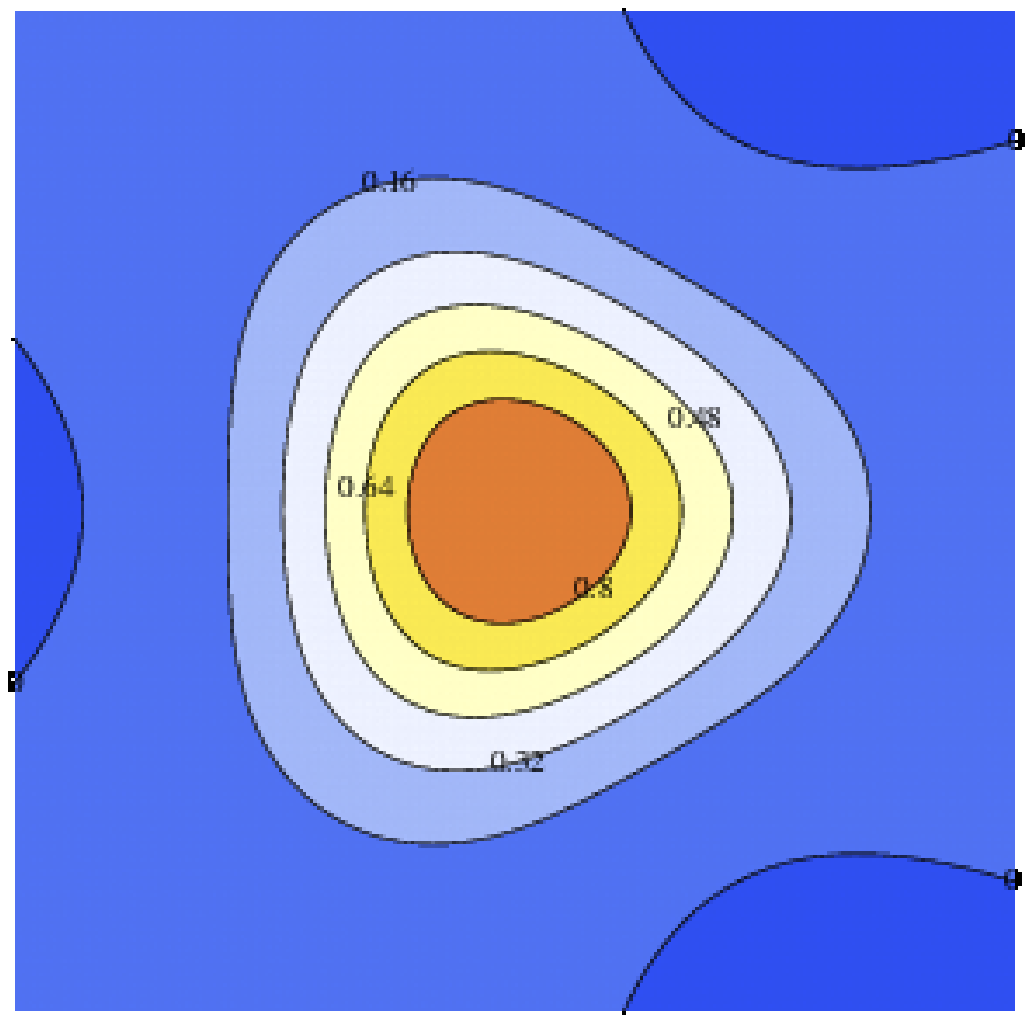}
\vspace{1.625 truein}
\includegraphics{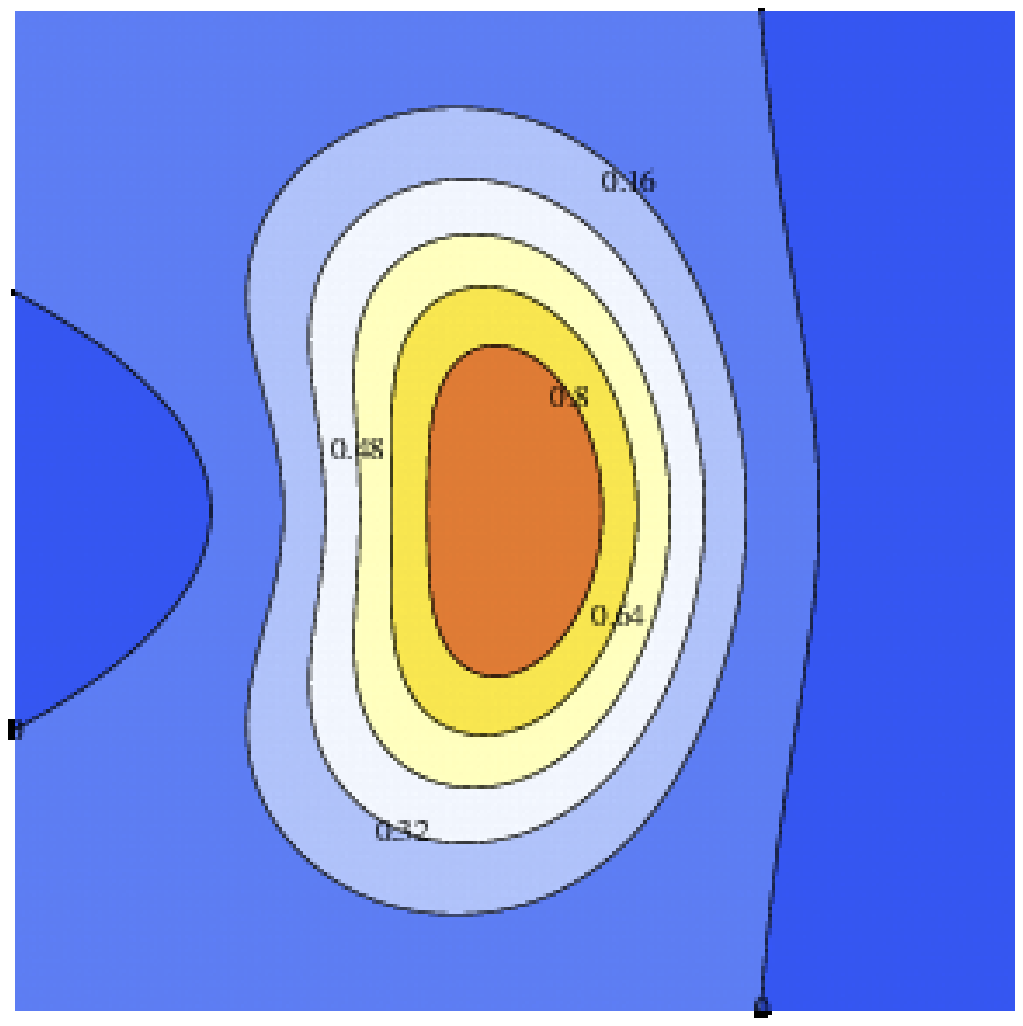}
\includegraphics{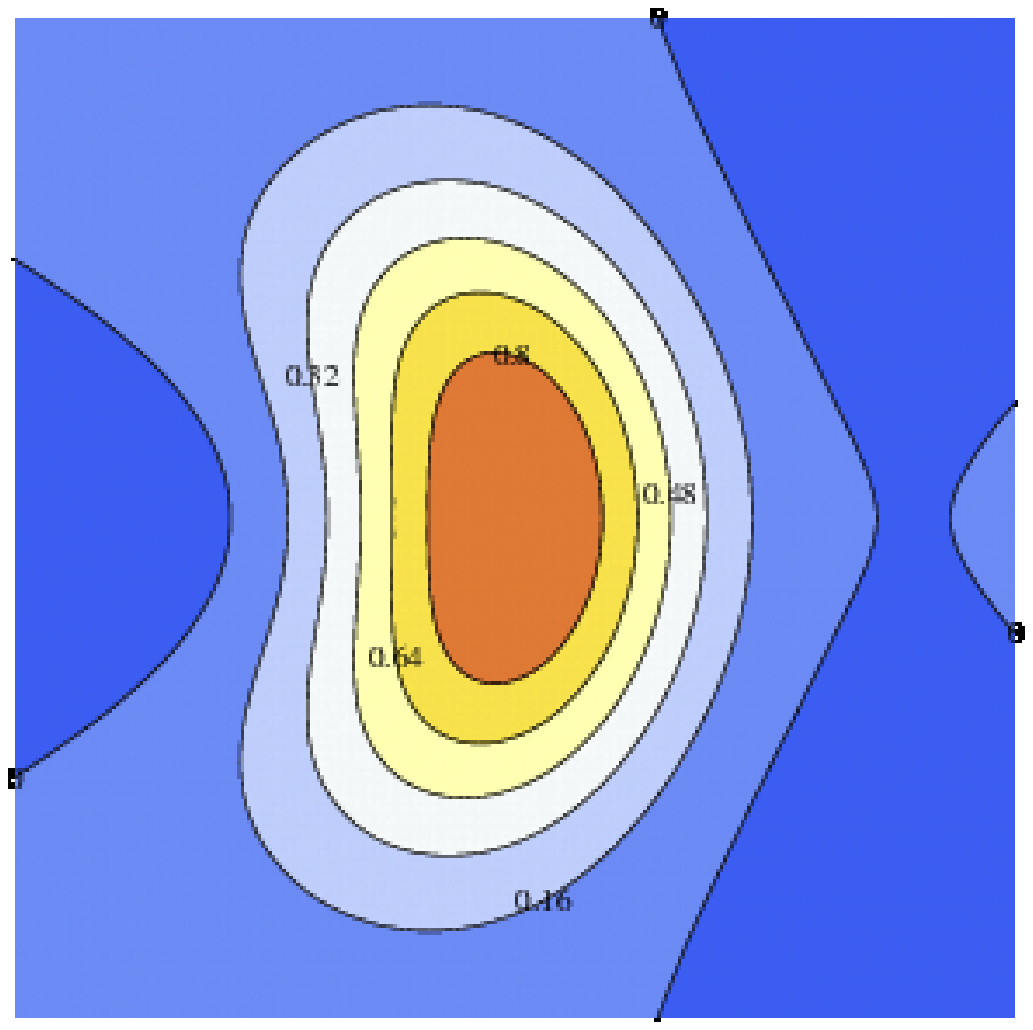}
\caption{From left to right, top to bottom: a circular galaxy (a) unlensed, (b) lensed by shear $= -0.3$, (c) lensed by F\--flexion $= -0.09$ only, (d) lensed by G\--flexion $= 0.27$ only, (e) lensed by all three in an aligned orientation, using a model linear in lensing terms, and (f) with quadratic spin one and three perturbation terms included.  Ratios of shears and flexions simulate a SIS lens with $\frac{r_{hl}}{\theta_{e}} = \frac{1}{2}$.\label{fig:lensingeffects}}
\end{figure}

\footnotetext{The flexions $\vec{F}$ and $\vec{G}$ multiplying $\vec{\tilde{g}}$ in this non-linear term are intentionally left as true flexions and not effective flexions $\vec{\tilde{F}}$ and $\vec{\tilde{G}}$.  While we earlier argued that, to second order, these terms {\it could} be replaced by their effective flexion counterparts, we do not {\it need} to replace them here.  As the model is more accurate if we do not, we leave in the true flexions.}   

\noindent where the first order effects of the spin one, two, and three terms, $\vec{\tilde{F}}$, $\vec{\tilde{g}}$, and $\vec{\tilde{G}}$ are given in equation part \eqref{eq:Lensed_galaxy_model_simple} and are shown separately and together for a circular galaxy lensed by a SIS in figure \ref{fig:lensingeffects}.  The second order spin one and three effects are given in equation part \eqref{eq:Lensed_galaxy_model_perturbation}.  The full model with these second order effects is also shown in figure \ref{fig:lensingeffects}.
\footnote{The models in figure \ref{fig:lensingeffects} have background artifacts, most notably a `pinching' in the sheared model galaxy and an `island' on the right of the fully lensed model galaxy.  The `pinching' in the sheared model is wholly accounted for by the approximation that we use for the purpose of convolving the galaxy model with seeing, but do not make in the AIM model for fitting data.  This same approximation partially accounts for the `island' in the fully lensed model.  Even with this approximation, these artifacts are in fact benign, but are accentuated by the contour levels and zoom of the plots.  

The first order Taylor expansion of an elliptical Gaussian with no flexion produces a nearly elliptical, positive intensity peak with broad, but shallow dimples along either side of the major axis.  The peak intensity of these dimples is a few percent of the intensity of the central peak.  The apparent `pinching' in the shear model is actually the cross over from positive to negative intensity caused by these low intensity dimples-- the outer edges of the dimples are out of the frame of the image.  When {\it fitting data}, we use the AIM model, in which we do NOT Taylor expand the elliptical Gaussian unlensed galaxy into a circular Gaussian plus elliptical perturbations as we have done here.  Rather, we retain the ellipticity and its degenerate shear counterpart in the Gaussian's exponent and only expand the flexion terms.  Therefore we do not get these negative intensity artifacts in an elliptical or sheared model galaxy.  

A model galaxy which has been lensed by F\--flexion or G\--flexion alone has one or three low intensity dimples in addition to the central peak.  The peak depth of these is only a few tenths of a percent of the central peak intensity.  The `island' in the plotted full model is actually where the sky level recovers to zero after having dropped to negative intensity.}

\subsection{Shapelet basis decomposition of the model}

The above model for the galaxy is physically intuitive, but for the convolution, it serves to move to the shapelets basis detailed in \citet{Refregier2003, RefregierBacon2003, MasseyRefregier2005, MasseyRowe2007}.  Polar shapelets are functions of the associated Laguerre Polynomials, and are convenient for representing and manipulating perturbations of Gaussian-like functions with various spin symmetries.  As F\--flexion, shear, and G\--flexion have spins one, two, and three respectively, these lensing terms are contained in a small number of coefficients in this basis when using simple models for the unflexed galaxies.  

\begin{landscape}

\begin{deluxetable}{ | c |  l  |  l  |} 
     \tiny
     \tablecaption{Conversion between lowest order shapelets and perturbations to a circular Gaussian with Gaussian width $\alpha$ and radial coordinate $\theta$.  Perturbations to the Gaussian have the form $\frac{1}{\sqrt{\pi}\alpha}\text{exp}(-\frac{\theta^2}{2\alpha^2})\big[(\vec{v} \cdot \vec{\theta}^m) \theta^{n-m}\big]$, where the magnitude and direction of the perturbation are given by $\vec{v}$, and the radial and spin variance of the perturbation are $n$ and $m$ respectively.  \label{tab:shapelets_conversion}}
     \tablewidth{0pt}
     \tablehead{
     	\colhead{Spin} &
          \colhead{Gaussian to Shapelets} &
          \colhead{Shapelets to Gaussian} 
          }
     \startdata
	& $\frac{1}{\sqrt{\pi}\alpha}\text{exp}(-\frac{\theta^2}{2\alpha^2})\big[\quad \big] = \big[\quad \big]$
	& $\big[\quad \big] = \frac{1}{\sqrt{\pi}\alpha}\text{exp}(-\frac{\theta^2}{2\alpha^2})\big[\quad \big]$ 
	\\[1.0ex] \tableline \tableline & &
	\\[-2.0ex] \multirow{3}{*}{0} 	
	& $1 = \chi_{00}$ 				& $\chi_{00} = 1$ \\
	& $\theta^2 = \chi_{20} + \chi_{00}$	& $\chi_{20} = \theta^2 -1$ \\
	& $\theta^4 = 2\chi_{40} + 4\chi_{20} + 2\chi_{00}$	
								& $\chi_{40} = \frac{1}{2}\theta^4 -2\theta^2 + 1$ 
	\\[1.0ex] \tableline & &
	\\[-2.0ex] \multirow{3}{*}{1}
	& $(\vec{v} \cdot \vec{\theta}) = 
			\frac{1}{2} \vec{v} \chi_{11} + \frac{1}{2} \vec{v}^{*} \chi_{1-1}$
	& $\vec{v} \chi_{11} + \vec{v}^{*} \chi_{1-1} = 
			2 (\vec{v} \cdot \vec{\theta}) $ 
	\\
	& $(\vec{v} \cdot \vec{\theta})\theta^2 = 
			\vec{v} \left(\frac{1}{\sqrt{2}}\chi_{31} + \chi_{11}\right) + \text{c.c.}$
	& $ \vec{v}\chi_{31} + \vec{v}^{*}\chi_{3-1} = 
			(\vec{v} \cdot \vec{\theta}) \left(\sqrt{2}\theta^2 - 2\sqrt{2}\right)$ 
	\\
	& $(\vec{v} \cdot \vec{\theta})\theta^4 = 
			\vec{v} \left(\sqrt{3}\chi_{51} + 3\sqrt{2} \chi_{31} + 3\chi_{22}\right) + \text{c.c.}$
	& $ \vec{v}\chi_{51} + \vec{v}^{*}\chi_{5-1} =
			(\vec{v} \cdot \vec{\theta}) \left( \frac{1}{\sqrt{3}} \theta^4 
			- 2\sqrt{3}\theta^2 + 2\sqrt{3} \right) $
	\\[1.0ex] \tableline & &
	\\[-2.0ex] \multirow{2}{*}{2}
	& $(\vec{v} \cdot \vec{\theta}^2) = 
			\frac{1}{\sqrt{2}} \vec{v} \chi_{22} + \text{c.c.}$
	& $\vec{v} \chi_{22} + \vec{v}^{*} \chi_{2-2} = 
			\sqrt{2} (\vec{v} \cdot \vec{\theta}^2) $ 
	\\
	& $(\vec{v} \cdot \vec{\theta}^2)\theta^2 = 
			\vec{v}\left(\frac{\sqrt{3}}{\sqrt{2}}\chi_{42} 
			+ \frac{\sqrt{3}}{\sqrt{2}}\chi_{22}\right) + \text{c.c.}$
	& $ \vec{v}\chi_{42} + \vec{v}^{*}\chi_{4-2} = 
			(\vec{v} \cdot \vec{\theta}^2) \left(\frac{\sqrt{2}}{\sqrt{3}}\theta^2 - \sqrt{6}\right)$ 
	\\[1.0ex] \tableline & &
	\\[-2.0ex] \multirow{2}{*}{3}
	& $(\vec{v} \cdot \vec{\theta}^3) = 
			\frac{\sqrt{3}}{\sqrt{2}} \vec{v} \chi_{33} + \text{c.c.}$
	& $\vec{v} \chi_{33} + \vec{v}^{*} \chi_{3-3} = 
			\frac{\sqrt{2}}{\sqrt{3}} (\vec{v} \cdot \vec{\theta}^3) $ 
	\\
	& $(\vec{v} \cdot \vec{\theta}^3)\theta^2 = 
			\vec{v}\left(\sqrt{6}\chi_{53} + 2\sqrt{6}\chi_{33}\right) + \text{c.c.}$
	& $ \vec{v}\chi_{53} + \vec{v}^{*}\chi_{5-3} = 
			(\vec{v} \cdot \vec{\theta}^3) \left(\frac{1}{\sqrt{6}}\theta^2 - 2\frac{\sqrt{2}}{\sqrt{3}}\right)$   
	\\[1.0ex] \tableline & &
	\\[-2.0ex] 4
	& $(\vec{v} \cdot \vec{\theta}^4) = \sqrt{6} \vec{v} \chi_{44} + \text{c.c.}$
	& $\vec{v} \chi_{44} + \vec{v}^{*} \chi_{4-4} = \frac{1}{\sqrt{6}} (\vec{v} \cdot \vec{\theta}^4) $ 
	\\[1.0ex] \tableline & &
	\\[-2.0ex] 5
	& $(\vec{v} \cdot \vec{\theta}^5) = \sqrt{30} \vec{v} \chi_{55} + \text{c.c.}$
	& $\vec{v} \chi_{55} + \vec{v}^{*} \chi_{5-5} = \frac{1}{\sqrt{30}} (\vec{v} \cdot \vec{\theta}^5) $ 	
     \enddata
\end{deluxetable}
\end{landscape}

\begin{landscape}
\begin{figure}[H]
\caption{Gaussian and spin zero through three perturbations with lowest radial dependences, in association with the left hand column of table \ref{tab:shapelets_conversion}.  These perturbations are analogous to the shapelets basis functions, however they are not orthonormal.  The left hand figures are for perturbations along the cartesian x axis (i.e $\vec{v}$ in table \ref{tab:shapelets_conversion} is real).  The right hand figures are for perturbations maximally orthogonal to the cartesian x axis ($\vec{v}$ in table \ref{tab:shapelets_conversion} is strictly non-real).  The Gaussian and spin zero perturbations have only a real component.  \label{fig:gaussian_perturbations}}
\vspace{1.41 truein}
	\includegraphics{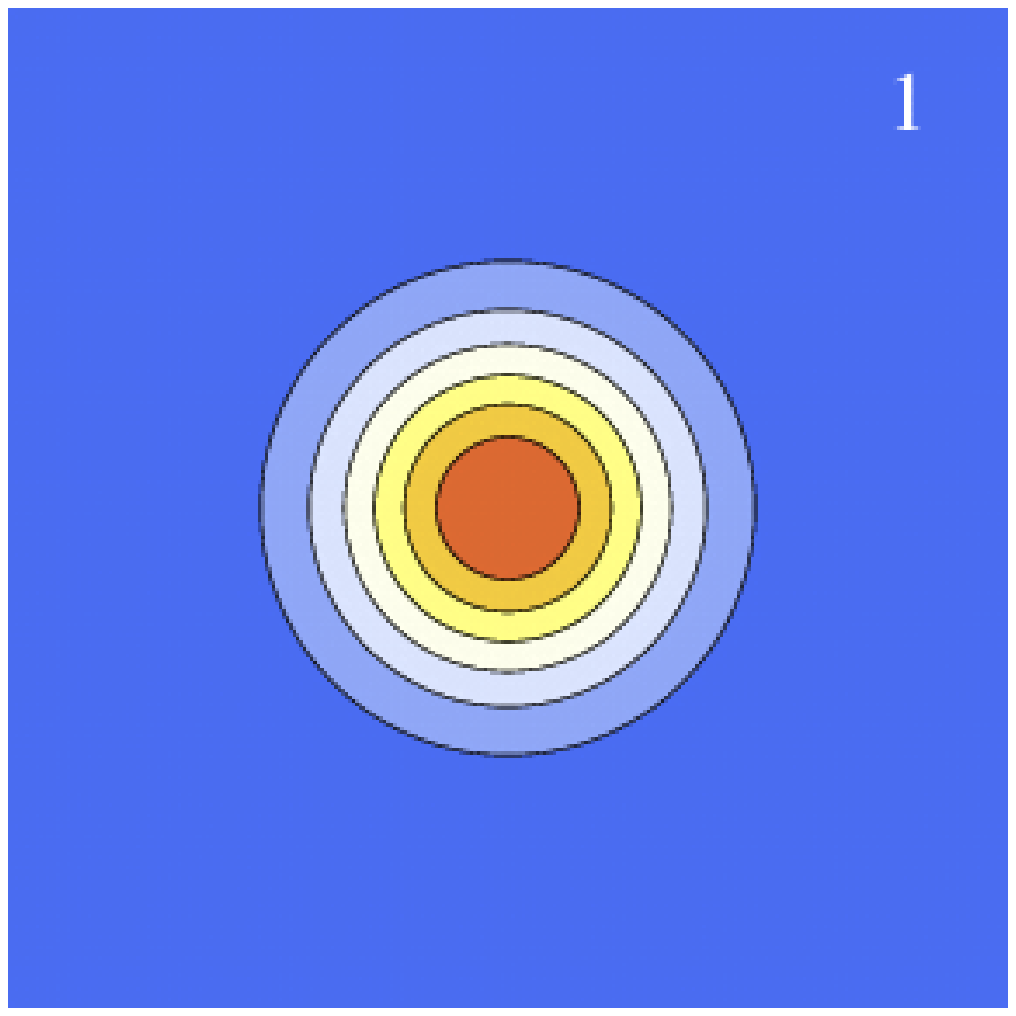}
	\includegraphics{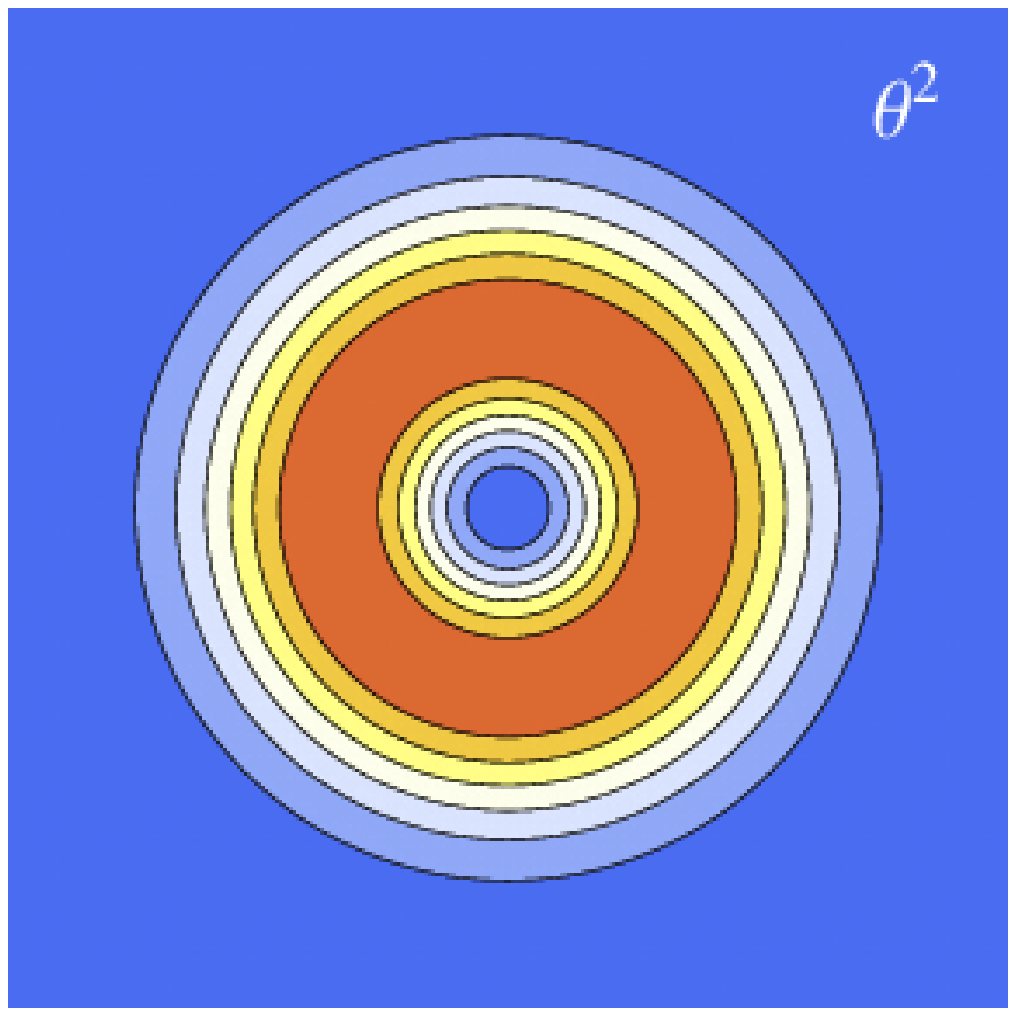}
	\includegraphics{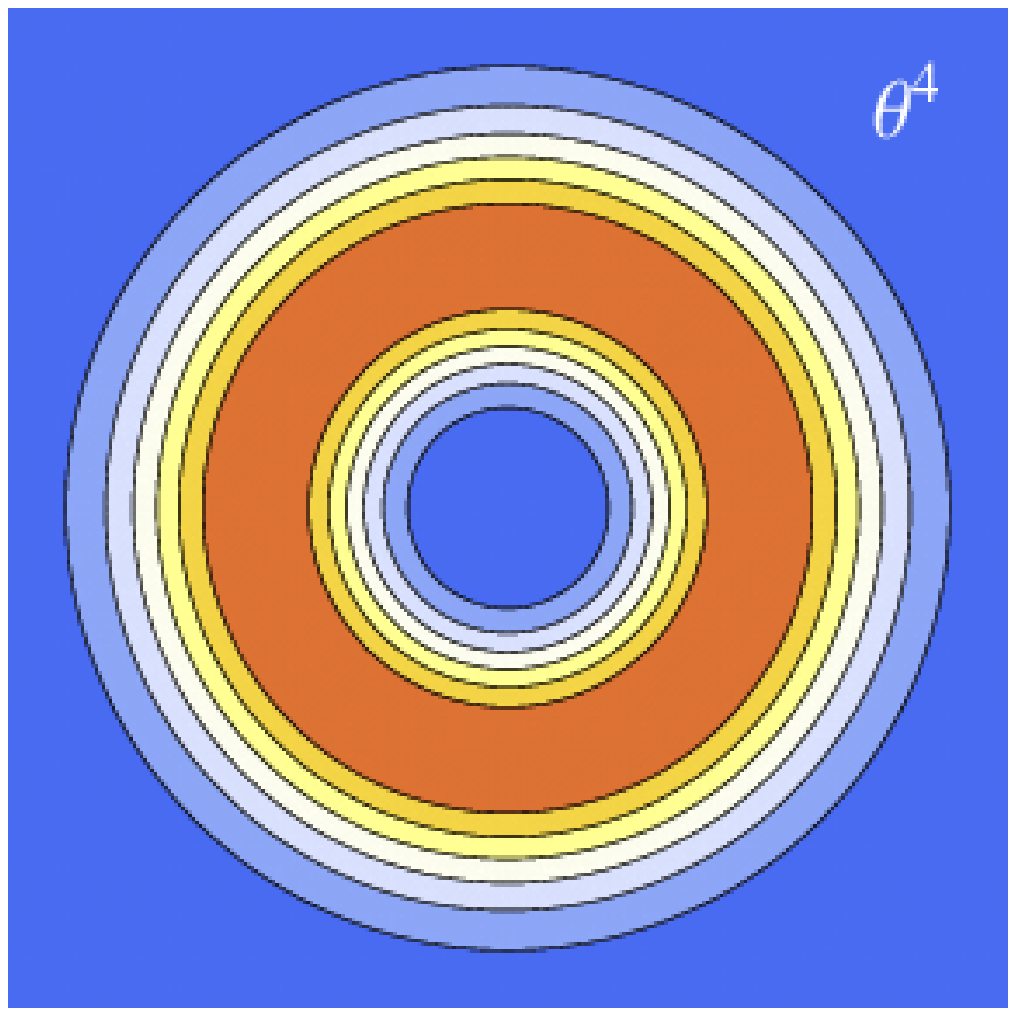}
\vspace{1.41 truein}
	\includegraphics{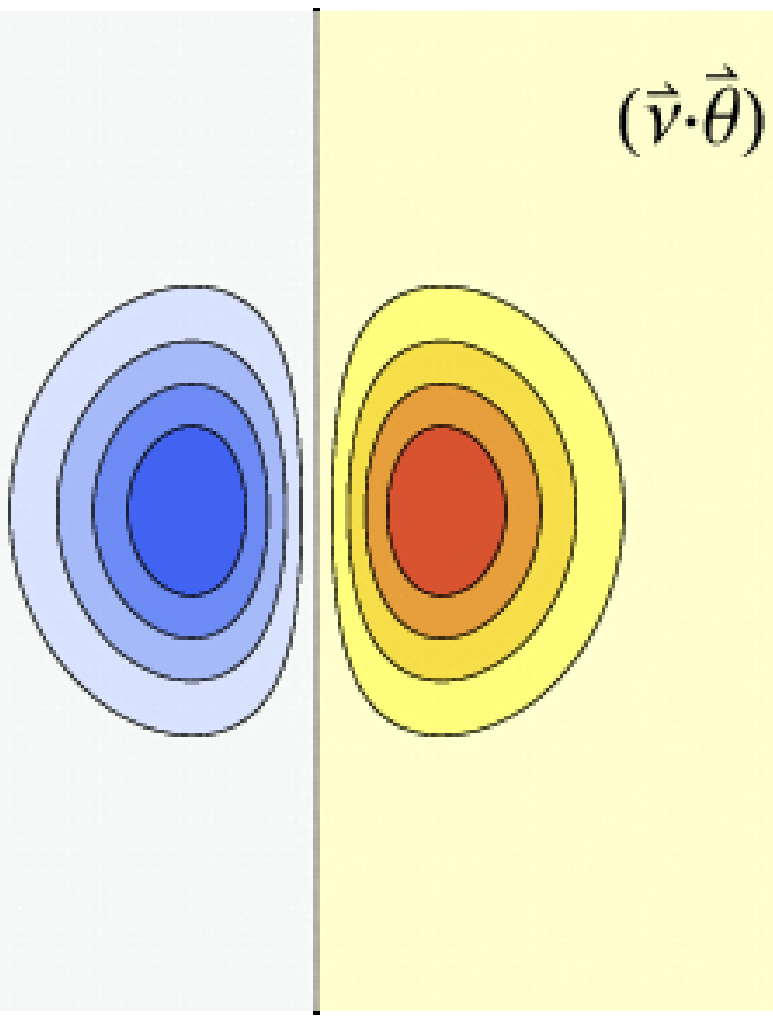}
	\includegraphics{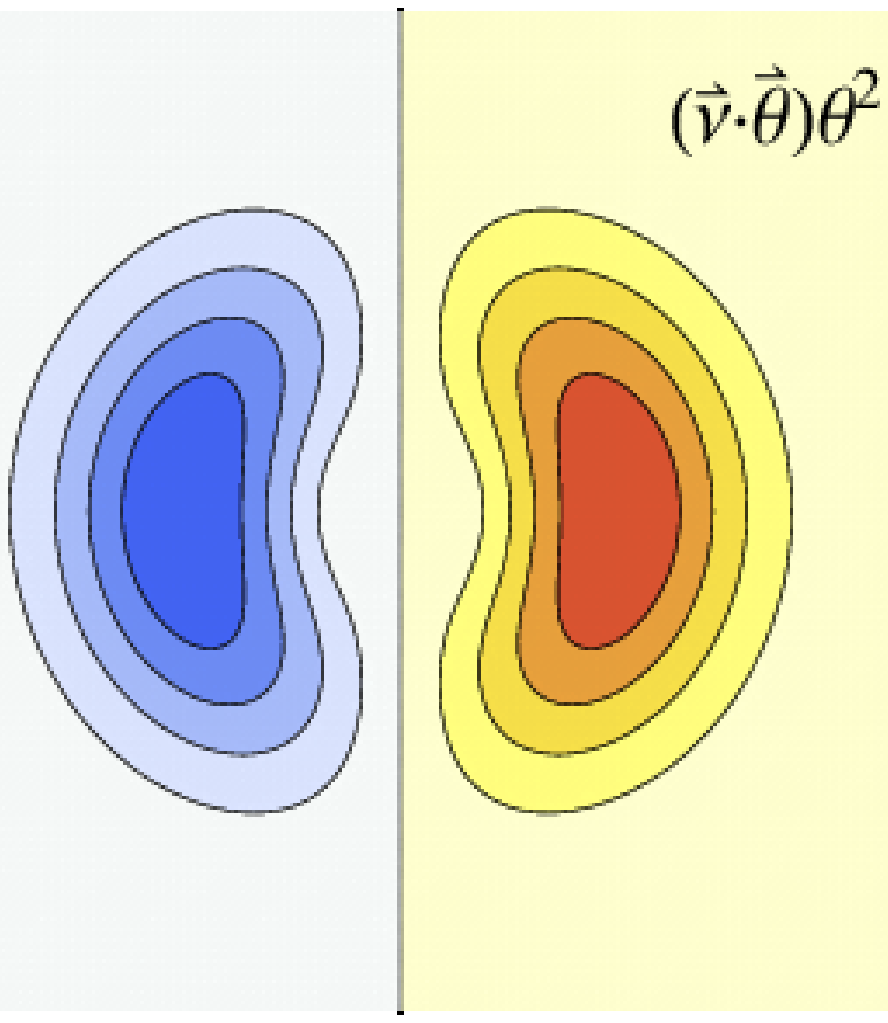}
	\includegraphics{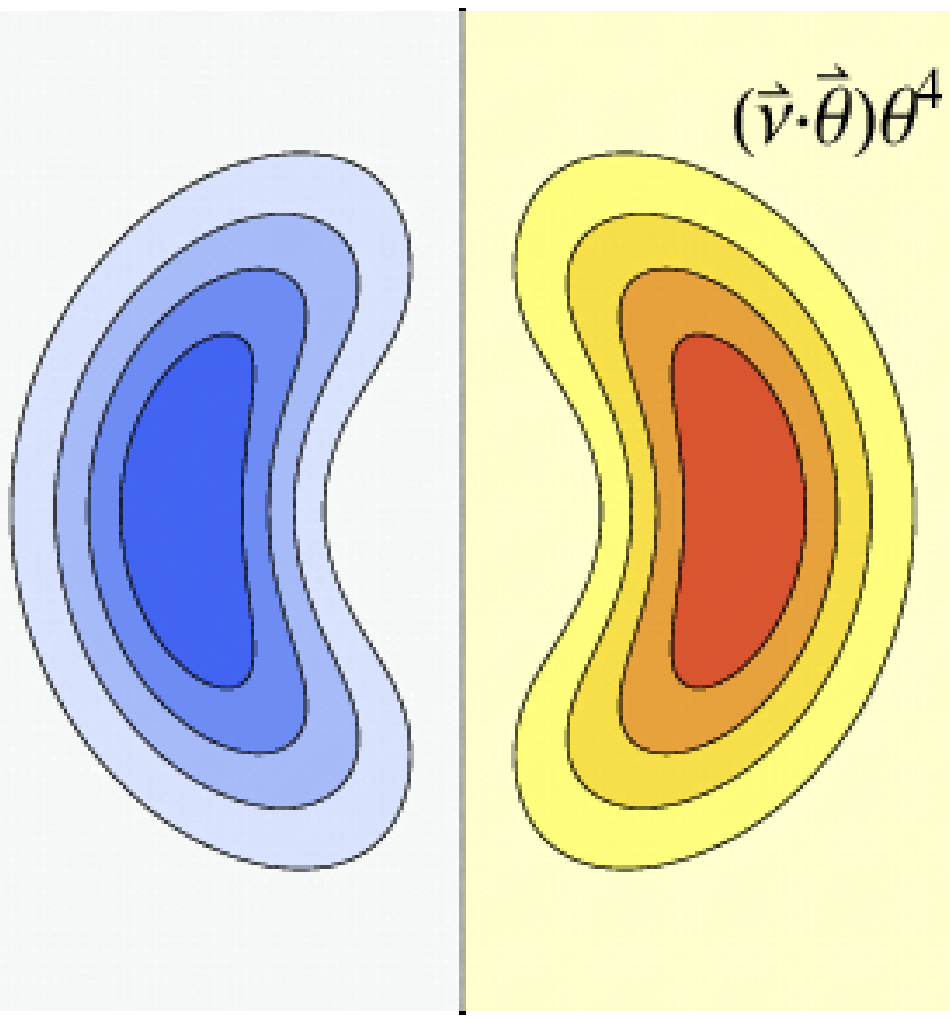}
	\includegraphics{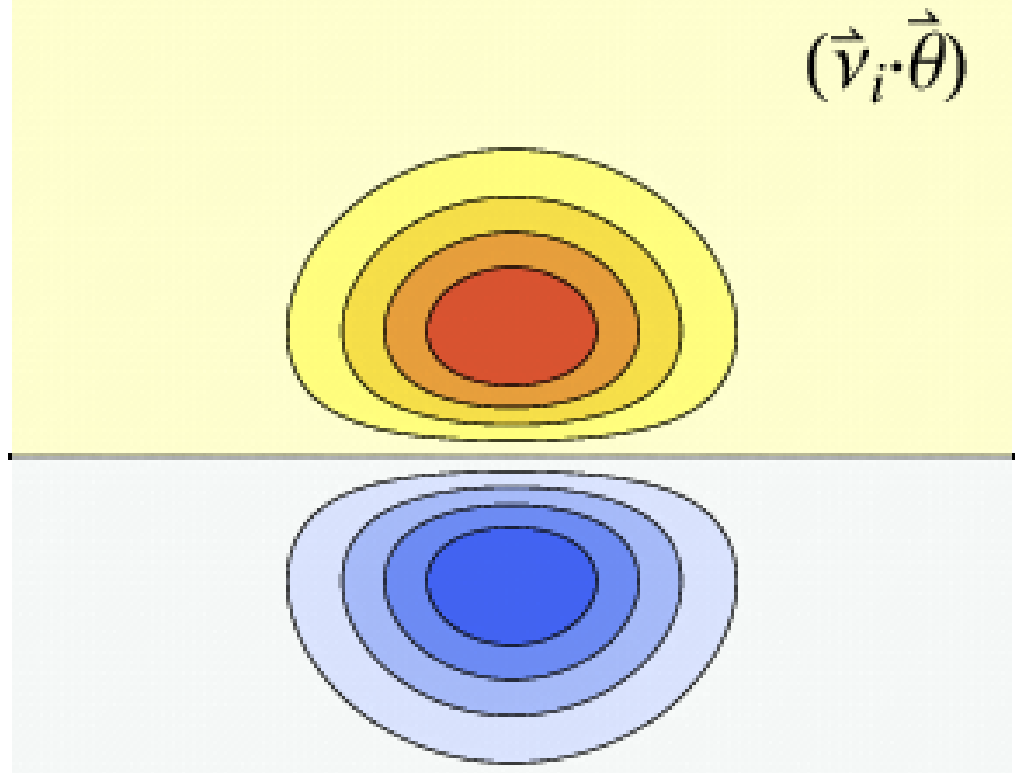}
	\includegraphics{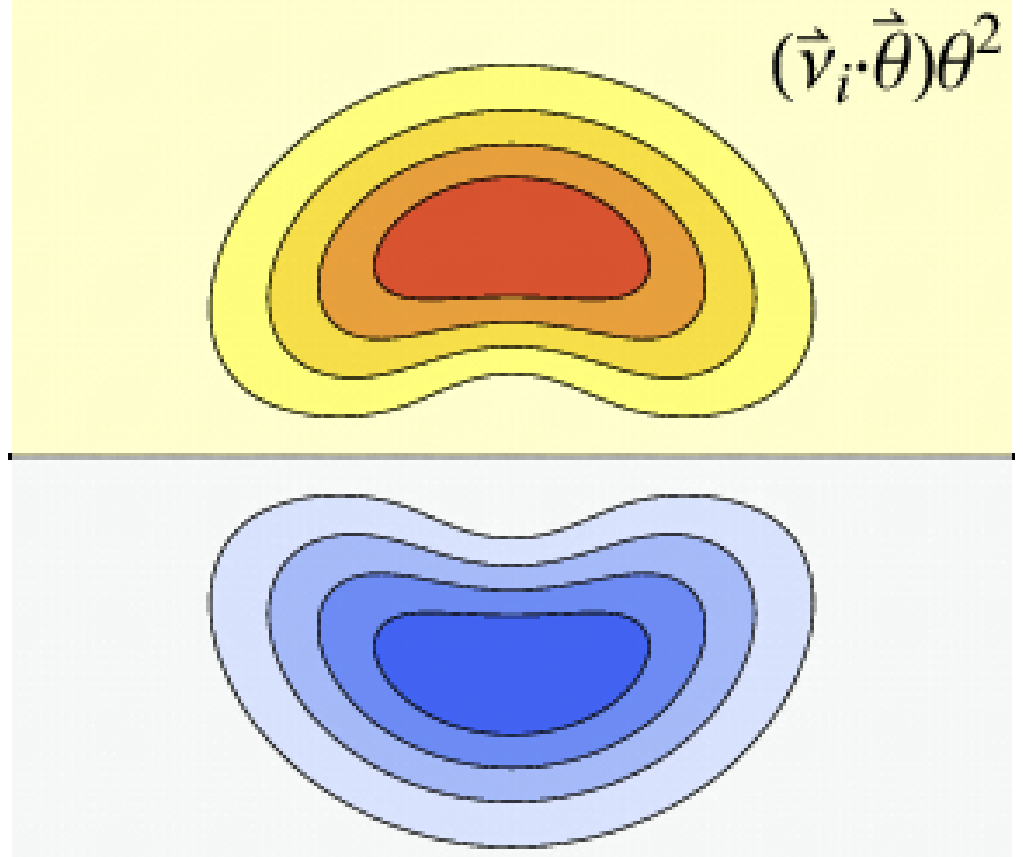}
	\includegraphics{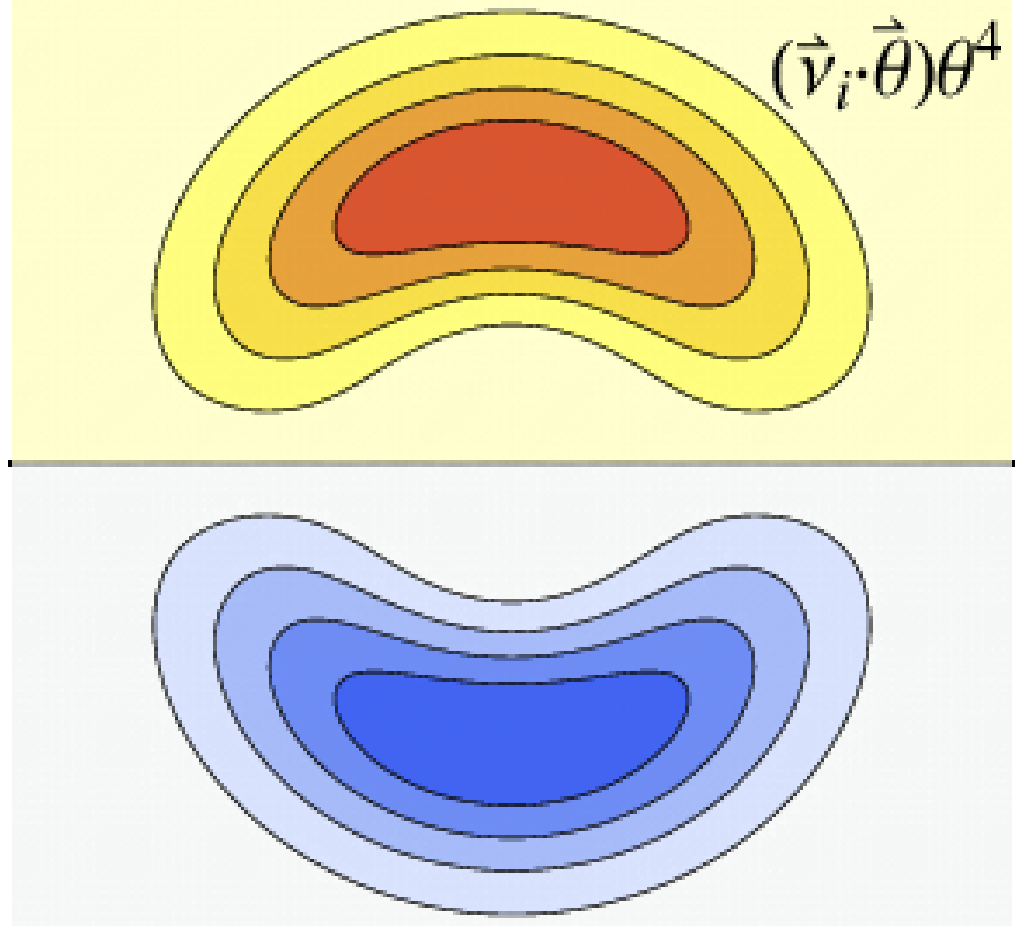}
\vspace{1.41 truein}
	\includegraphics{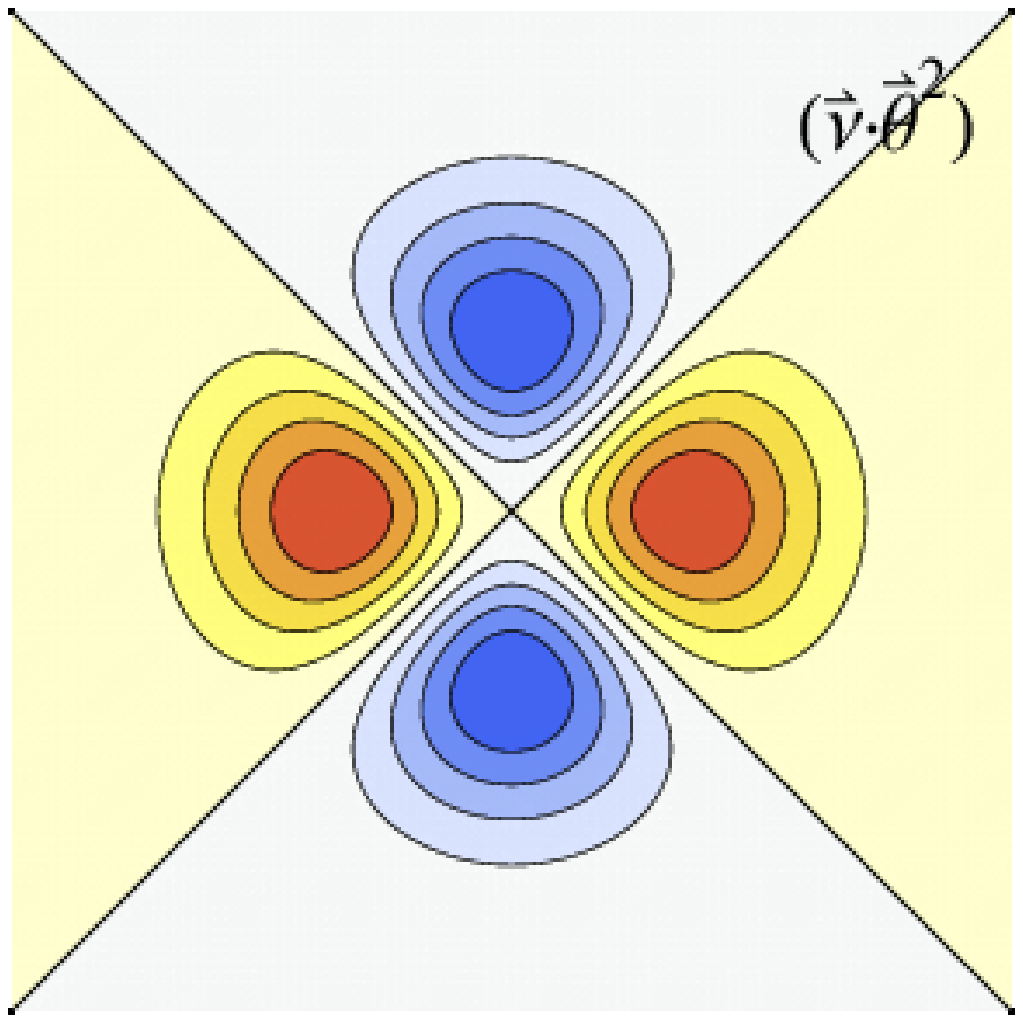}
	\includegraphics{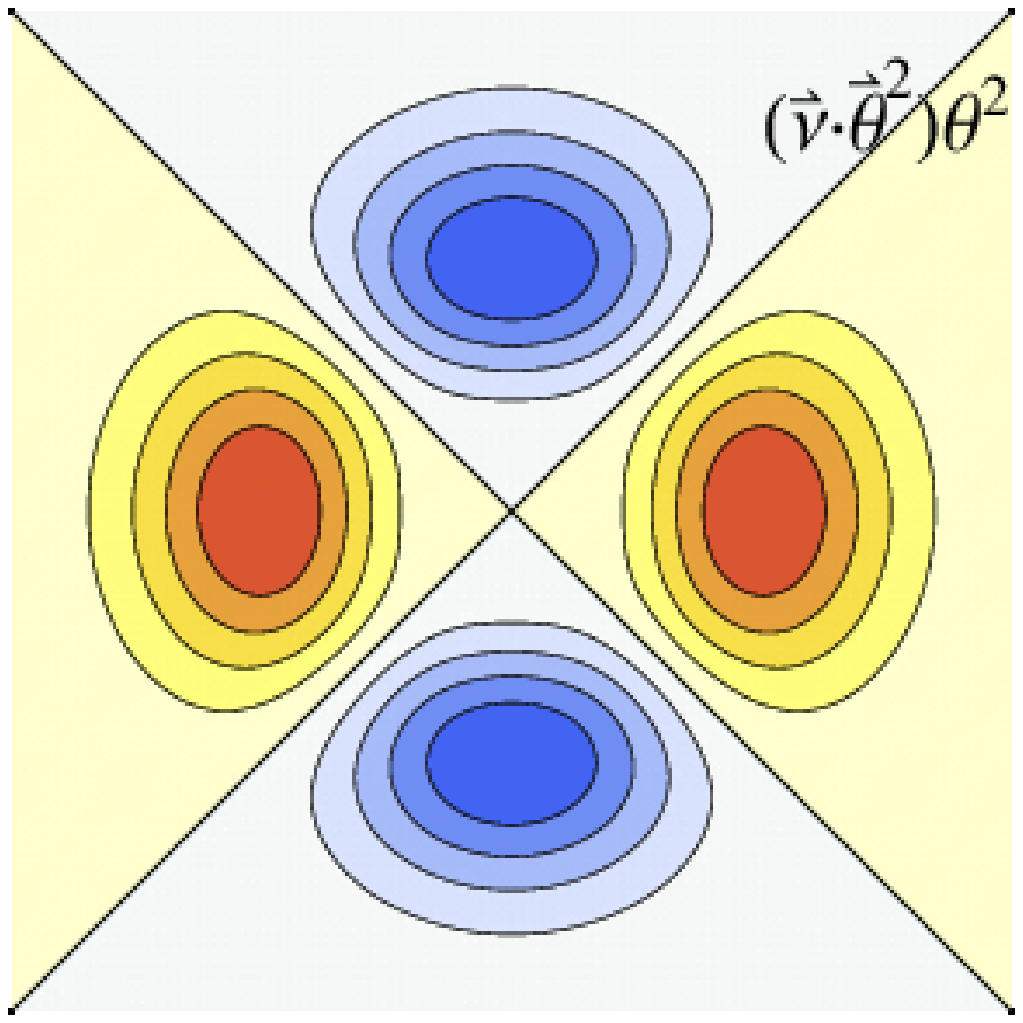}
	\includegraphics{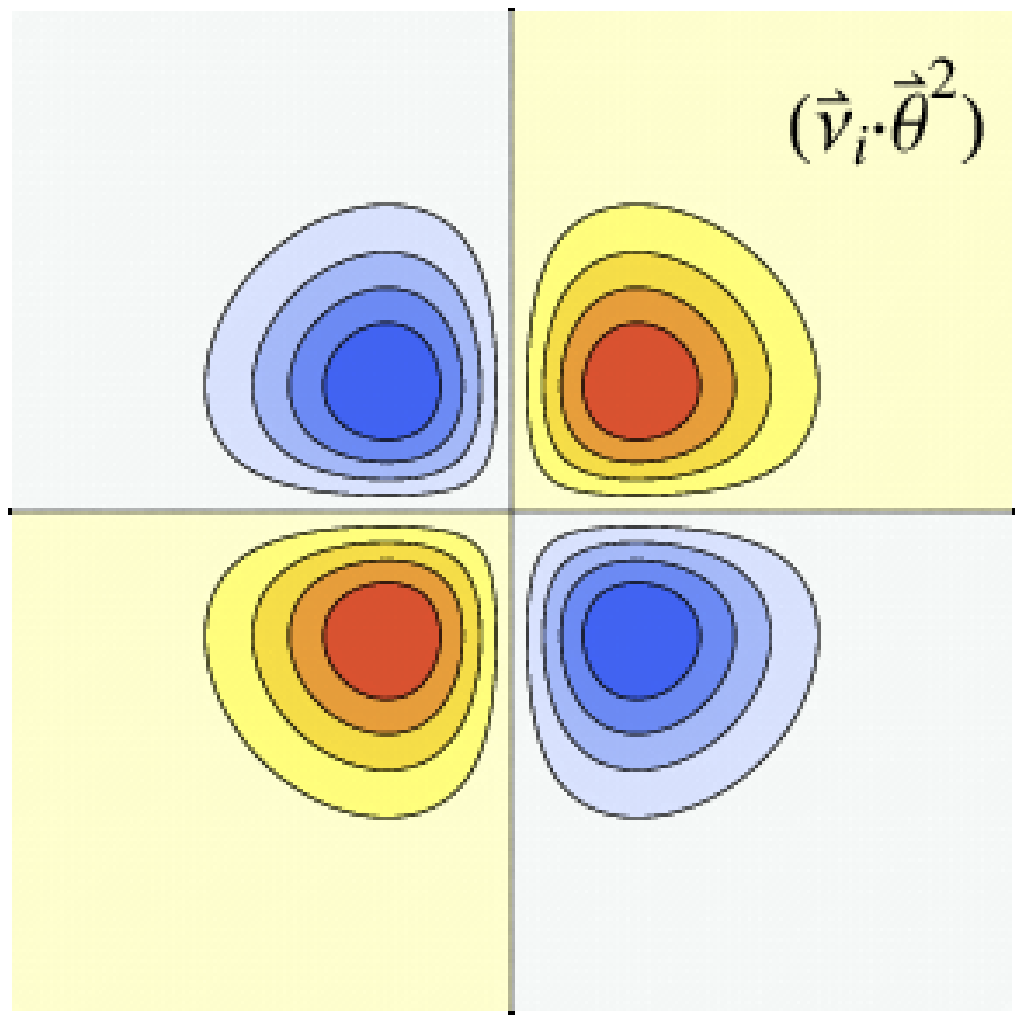}
	\includegraphics{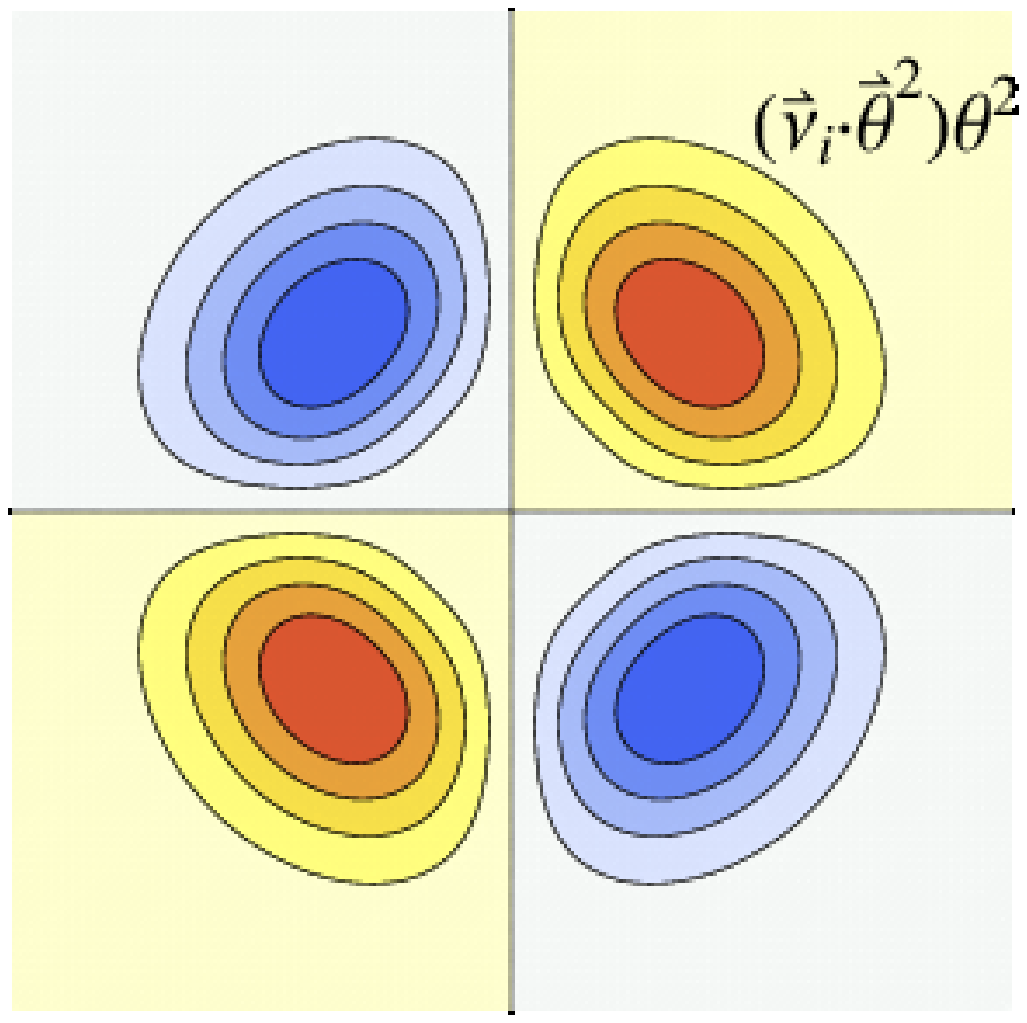}
\vspace{1.41 truein}
	\includegraphics{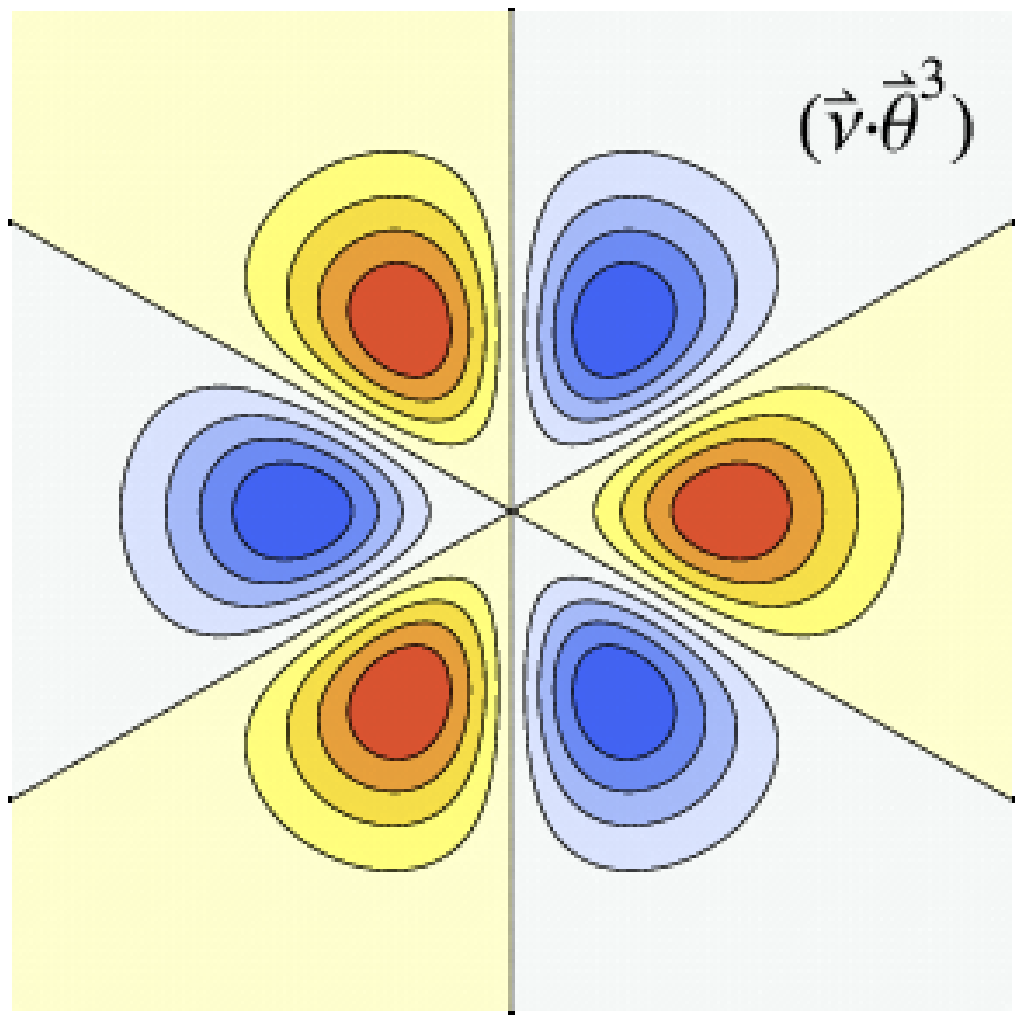}
	\includegraphics{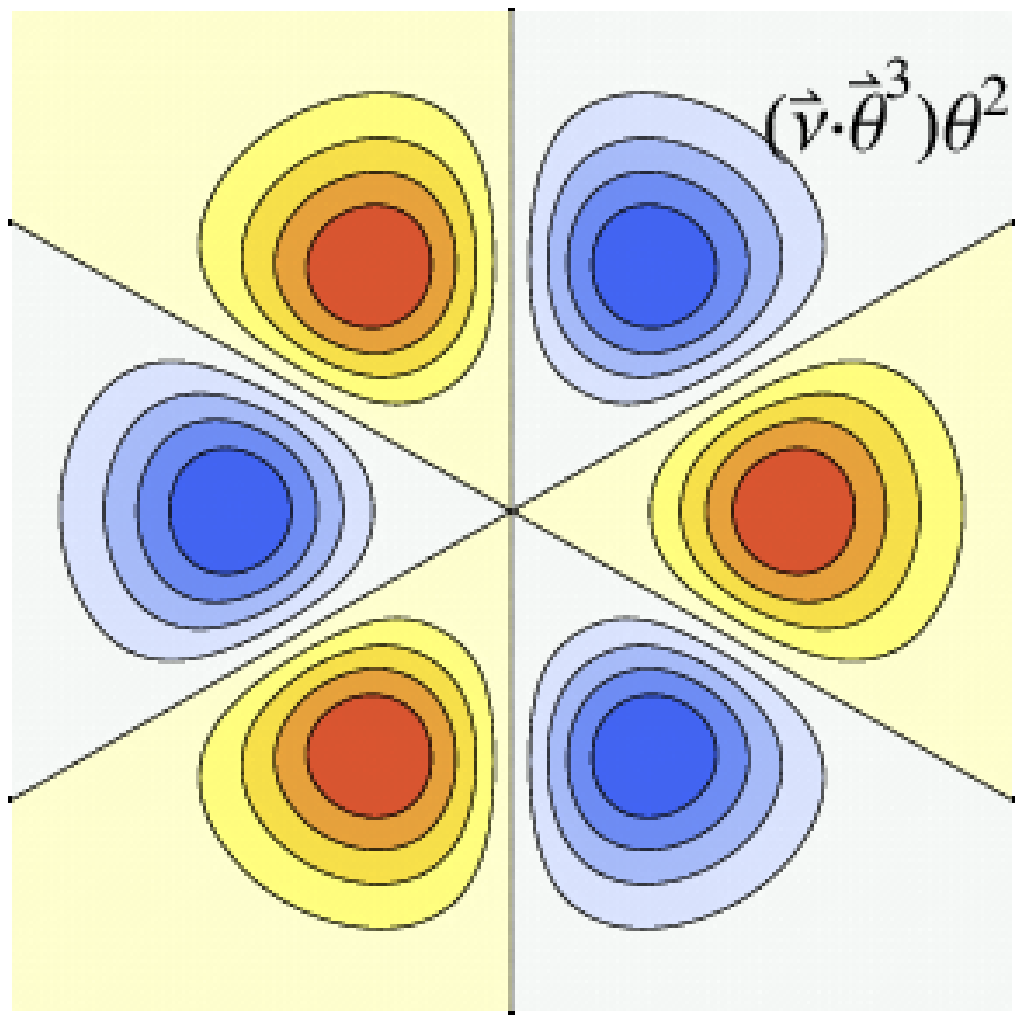}
	\includegraphics{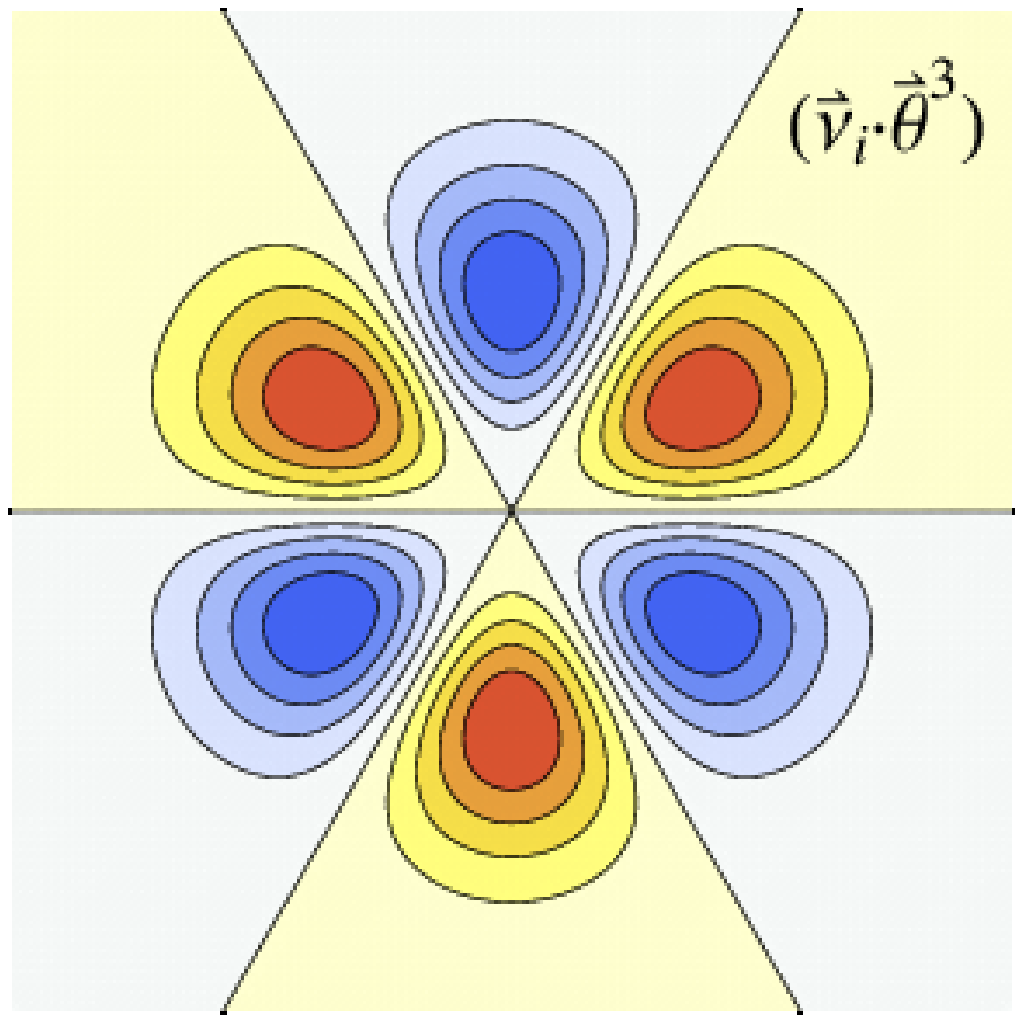}
	\includegraphics{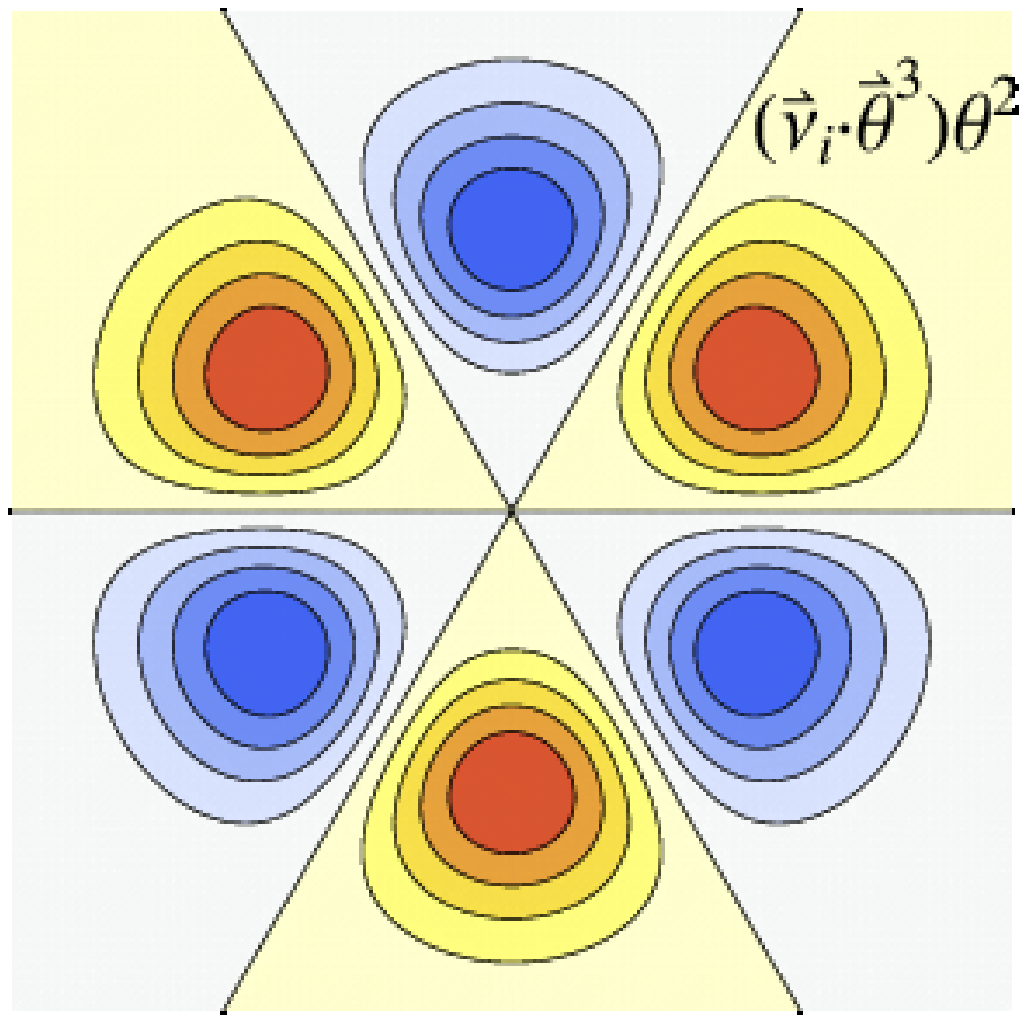}
\end{figure}
\end{landscape}

The relation between the lowest order shapelets and perturbations to the circular Gaussian function are given in table \ref{tab:shapelets_conversion}.  For reference, the Gaussian perturbations used in the model and given in the table are plotted in figure \ref{fig:gaussian_perturbations}.  The polar shapelets, which are essentially the orthonormal versions of the shown functions, are depicted in \citet{MasseyRefregier2005, MasseyRowe2007}.

Following \citet{MasseyRefregier2005}, but normalizing the flux to $I_0$, the function representing a galaxy of scale length $\alpha$ is

\begin{equation}
\label{eq:lensed_galaxy_model_shapelets} 
I(\vec{\theta}) = \frac{I_0}{2\sqrt{\pi}\alpha} \sum_{n=0}^{\infty}  \sum_{m=-n}^{n}  f_{nm} \chi_{nm}(\vec{\theta};\alpha)
\end{equation}

\noindent where $\chi_{nm}$ are the Polar shapelets and $f_{nm}$ are unitless coefficients.  The first subscript of the coefficient, $n$, denotes the radial dependence of the term, and the second, $m$, denotes the spin symmetry.

We note again here that the aberrations used in this work are all unitless quantities.  This differs from shapelets convention, specifically that in \citet{MasseyRefregier2005} wherein flexions are unitful and always appear in multiple with the galaxy scale length when appearing in a shapelet coefficient.

The non-zero shapelets coefficients needed to represent the linear expansion of the lensed galaxy of equation \eqref{eq:Lensed_galaxy_model_simple} are 

\begin{subequations}
\label{eq:shapelets_coefficients}
\begin{align}
\label{eq:shapelets_coefficients_simple}
f_{00} &= 1 \\ \nonumber
f_{22} &= \frac{1}{\sqrt{2}} \vec{\tilde{g}} \\ \nonumber
f_{31} &= \frac{1}{4\sqrt{2\text{ln}(2)}} \frac{1}{\sqrt{2}} \left(3 \vec{\tilde{F}}\right), \ f_{11} = \frac{1}{4\sqrt{2\text{ln}(2)}} \left(3 \vec{\tilde{F}}\right) \\ \nonumber
f_{33} &= \frac{1}{4\sqrt{2\text{ln}(2)}} \sqrt{\frac{3}{2}} \vec{\tilde{G}} 
\\ \nonumber
\\ \label{eq:shapelets_coefficients_perturbation}
f_{51} &=  \sqrt{3} \vec{p}, \ f_{31} = 3 \sqrt{2} \vec{p}, \ f_{11} = 3 \vec{p} \\ \nonumber
f_{53} &=  \sqrt{6} \vec{q}, \ f_{33} = 2 \sqrt{6} \vec{p} \\ \nonumber
&\text{where} \quad 
	 \vec{p} = \frac{1}{4\sqrt{2\text{ln}(2)}} \frac{1}{2} 
		\left( 3 \vec{\tilde{g}}\vec{F}^{*} + \vec{\tilde{g}}^{*}\vec{G} \right) 
\\ \nonumber
&\qquad \quad \ \  \vec{q} = \frac{1}{4\sqrt{2\text{ln}(2)}} \frac{1}{2}
	 	\left( 3\vec{\tilde{g}}\vec{F} \right).
\end{align}
\end{subequations}

\noindent Only quantities with positive spin coefficients are listed, as negative spin coefficients are simply the complex conjugates of their positive counterparts.    All other unlisted coefficients are zero, or will otherwise not affect the final measured spin one, two, or three image shape characteristics.

For clarity, we have broken up the shapelets coefficients into parts \eqref{eq:shapelets_coefficients_simple} and \eqref{eq:shapelets_coefficients_perturbation} which derive from the linear galaxy expansion terms in the model given by \eqref{eq:Lensed_galaxy_model_simple}, and the quadratic galaxy expansion terms in \eqref{eq:Lensed_galaxy_model_perturbation}, respectively.  The total coefficients are sums of all components.

Terms varying linearly on the galaxy, $\chi_{11}$, appear with non-zero coefficient, despite having no obvious counterpart in equation \eqref{eq:Lensed_galaxy_model_final}.  These terms appear because the shapelets basis is orthonormal, and thus $\chi_{31}$ and $\chi_{51}$ have components which vary linearly with radius.  As the lensing distortions are not orthonormal, the inclusion of the $\chi_{11}$ term is necessary in order to counter the part of the linearly varying $\chi_{31}$ and $\chi_{51}$ terms absent from the lensing model, see table \ref{tab:shapelets_conversion}.  However, measurements of flexion are made by probing terms which vary as radius cubed on the galaxy, $\chi_{31}$ and $\chi_{33}$ in this basis.  Thus $\chi_{11}$ is merely a remainder of sorts.  

\subsection{Model of a PSF}

As a Gaussian is a convenient model for an unlensed galaxy, it is also a convenient, simple approximation for a symmetric point spread function.  Additionally, we argued in \S \ref{sec:aberrations} that convergence and defocus are the same, shear and astigmatism are the same, F\--flexion and coma are the same, and G\--flexion and trefoil are the same.  However, the lensing aberrations are applied to the galaxy image, which is non-uniformly illuminated, but the telescope aberrations are applied to a uniformly illuminated wavefront.  We therefore cannot account for the telescope aberrations by applying deflections to the symmetric, atmospherically aberrated PSF image in the same way that we could treat the gravitational lens as applying deflections to the Gaussian, unlensed galaxy.

In order to account for the effects of the telescope aberrations on the PSF, we follow the work of \citet{JarvisSchechter2008}, and compute the moments imparted onto images by wavefront gradients in the pupil plane.  If we were to compute an infinite number of moments, we could reconstruct the exact form of the image.  But, in the spirit of informed approximation, we shall compute exact moments through third, and then approximate the form of the PSF as Gaussian plus perturbations, similar to that of the Galaxy model.  We choose to truncate our model at third moments, as the second and third image moments will affect shear and flexion measurements most significantly.

The PSF is the stellar image.  Ignoring atmospheric effects, which add a series of random delays to the wavefront, the light reaching the pupil from a star is parallel.  Summarizing from \citet{JarvisSchechter2008}, in the limit of geometric optics wavefront aberrations will deflect light rays from a star hitting different areas on the pupil into beams with slightly different directions.  If we assume no net tilt of the wavefront, an individual beam's deflection will be proportional to its final displacement from the star image's center in the image plane.\footnote{A net wavefront tilt will cause the entire star to be moved from its nominal position, shifting the central intensity; tilt imparts a net first moment.}  It thus follows that the moments of the stellar images are proportional to the deflections of light at the pupil, i.e. the gradients of the wavefront.

A wavefront delay across the pupil, $W^{\prime}$ and its gradient, $\vec{\nabla} W^{\prime}$, are given by the following functions of the normalized pupil coordinated $\vec{\rho}$ the radius of the pupil, $R$, and unitful aberrations defocus, $d^{\prime}$, astigmatism, $\vec{a}^{\prime}$, coma, $\vec{c}^{\prime}$, and trefoil, $\vec{t}^{\prime}$,

\begin{align}
\label{eq:delay_gradient}
W^{\prime} &= d^{\prime} \rho^2 + \vec{a}^{\prime} \cdot \vec{\rho}^2 + \vec{c}^{\prime} \cdot \rho^2\vec{\rho} + \vec{t}^{\prime} \cdot \vec{\rho}^3 
\\ \nonumber
\text{so } \vec{\nabla} W^{\prime} &= \left( 2\frac{d^{\prime}}{R}\vec{\rho} + 2\frac{\vec{a}^{\prime}}{R} \vec{\rho}^{*} + 2 \frac{\vec{c}^{\prime}}{R}  \rho^2 + \frac{\vec{c}^{\prime *}}{R} \vec{\rho}^2 + 3\frac{\vec{t}^{\prime}}{R} (\vec{\rho}^{*})^2  \right) 
\\ \nonumber
&= \zeta \left( 2d\vec{\rho} + 2\vec{a} \vec{\rho}^{*} + 2 \vec{c} \rho^2 + \vec{c}^{*}\vec{\rho}^2 + 3\vec{t}(\vec{\rho}^{*})^2  \right)
\\ \nonumber
&= \zeta \vec{\nabla} W.
\end{align}

\noindent We have introduced a unitless wavefront gradient $\vec{\nabla} W$ and unitless aberrations $d$, $a$, $c$, and $t$, which can be rendered unitful with the coefficient $\zeta$.  The scale $\zeta$ is arbitrary, but may be thought of as a `typical' ray displacement or image size and thus has units of the same-- angle of ray displacement, or equivalently angular image size.  In radians, $\zeta$ is given by the ratio of a typical magnitude for a wavefront delay on the pupil (possibly one wave, though there are many conventions) and the pupil radius.  The unprimed wavefront gradient $\vec{\nabla} W$, the defocus, $d$, astigmatism, $\vec{a}$, coma, $\vec{c}$, and trefoil, $\vec{t}$ are therefore unitless in all following discussion, however relating these unitless quantities to the physical delays in the wavefront at the telescope's pupil requires that one know the scale, $\zeta$, and pupil radius, $R$.  

Defining the zeroth moment to be one, and the first moments to be zero, i.e. normalized intensity and no tilt, there are four undetermined complex image moments up to and including third; a spin zero second moment, a spin two second moment, a spin one third moment, and a spin three third moment.  

\begin{align}
\label{eq:moments}
Q_0 &= \zeta^2 \int \int \left|\vec{\nabla} W \right|^2 \rho d\rho d\phi \\ \nonumber
\vec{Q_2} &= \zeta^2 \int \int \left(\vec{\nabla} W \right)^2 \rho d\rho d\phi \\ \nonumber
\vec{Q_1} &= \zeta^3 \int \int \left|\vec{\nabla} W \right|^2 \vec{\nabla} W \rho d\rho d\phi \\ \nonumber
\vec{Q_3} &= \zeta^3 \int \int \left(\vec{\nabla} W \right)^3 \rho d\rho d\phi, \\ \nonumber
\end{align}

Using the relations for the wavefront gradients presented in equations \eqref{eq:delay_gradient}, we find the resultant second and third moments for the aberrations are

\begin{align}
\label{eq:moments2}
Q_0 &= \zeta^2 \left( 2d^2 + 2a^2 + \frac{2}{3} c^2 + 3t^2 \right) \\ \nonumber
\vec{Q_2} &= \zeta^2 \left( 4\vec{a} d + \frac{1}{3} \vec{c}^2 + 2 \vec{c}^{*} \vec{t} \right) \\ \nonumber
\vec{Q_1} &= \zeta^3 \left( \frac{1}{3} \vec{c} \left( 8d^2 + 4a^2 + c^2 + 9t^2 \right)
				+ 4\vec{c}^{*}\vec{a}d + 8\vec{t}\vec{a}^{*}d 
				+ 4\vec{a}^2\vec{t}^{*} + (\vec{c}^{*})^2\vec{t} \right) \\ \nonumber
\vec{Q_3} &= \zeta^3 \left( 3\vec{t} \left(4d^2 + c^2 \right) 
				+ 4\vec{c}\vec{a}d + 4\vec{c}^{*}\vec{a}^2 \right).
\end{align}

We will work under the assumption that the atmospheric effect will simply add to the spin zero second moment.  Moreover, we shall assume that the atmosphere will only contribute to the spin zero moment.  $\vec{Q_1}$, $\vec{Q_2}$, and $\vec{Q_3}$ remain the same, but $Q_0$ becomes

\begin{equation}
\label{eq:moments_atm}
Q_0 = \zeta^2 \left( 2d^2 + 2a^2 + \frac{2}{3} c^2 + 3t^2 + 2S_{atm}^2 \right),
\end{equation}

\noindent where $S_{atm}^2$ is the unitless second moment caused by atmospherically induced semi-random wavefront delays at the pupil.  The factor of two is placed for convenience, so that the spin zero second moment of star aberrated by the atmosphere alone is $2\zeta^2 S_{atm}^2$.  As the atmospheric effects become dominant over the telescope effects, $S_{atm} \zeta$ approaches the measured Gaussian width of the PSF.

We define the Gaussian width of the PSF as $\alpha$ so that a measured spin zero second moment is $2\alpha^2$, if the PSF is truly Gaussian.  With this definition, the unitless Gaussian width is

\begin{equation}
\label{eq:alpha}
\frac{\alpha}{\zeta} =  \sqrt{ \left( d^2 + a^2 + \frac{1}{3} c^2 + \frac{3}{2}t^2 + S_{atm}^2 \right) }.
\end{equation}

\noindent Like the second moment from which it is derived, the unitless Gaussian width is a measurable quantity.

In the absence of other asymmetric aberrations, each spin $n$ wavefront delay produces a spin $n$ moment whose magnitude is proportional to the product of itself and the defocus to some power.  Therefore, analogous to the mappings from galaxy distortions $\vec{g}$, $\vec{F}$ and $\vec{G}$ to effective spin symmetric distortions $\vec{\tilde{g}}$, $\vec{\tilde{F}}$ and $\vec{\tilde{G}}$, we will define new, effective aberrations in tilde space which produce spin symmetric moments, namely

\begin{align}
\label{eq:mapping_aberrations} 
\vec{\tilde{a}} &= \frac{1}{4 \frac{\alpha^2}{\zeta^2}} 
		\left( 4\vec{a} d + \frac{1}{3} \vec{c}^2 + 2 \vec{c} \vec{t}^{*} \right) \\ \nonumber
\vec{\tilde{c}} &= \frac{1}{24 \frac{\alpha^3}{\zeta^3}} \frac{1}{3} \left( \frac{1}{3} \vec{c} \left( 8d^2 + 4a^2 + c^2 + 9t^2 \right)
				+ 4\vec{c}^{*}\vec{a}d + 8\vec{t}\vec{a}^{*}d 
				+ 4\vec{a}^2\vec{t}^{*} + (\vec{c}^{*})^2\vec{t} \right) \\ \nonumber 
\vec{\tilde{t}} &= \frac{1}{24 \frac{\alpha^3}{\zeta^3}} \left( 3\vec{t} \left(4d^2 + c^2 \right) 
				+ 4\vec{c}\vec{a}d + 4\vec{c}^{*}\vec{a}^2 \right).				
\end{align}

\begin{figure}[htb]
\epsscale{1.00}
\plotone{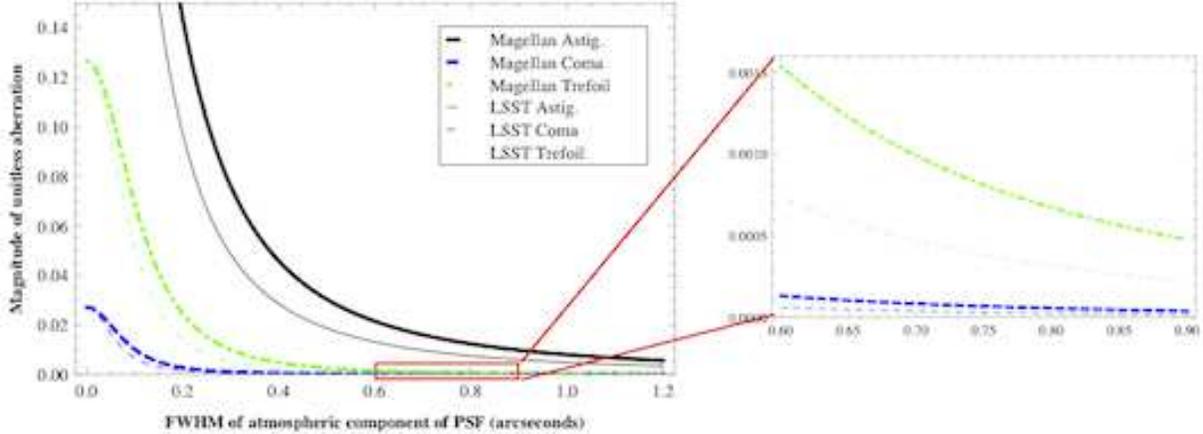}
\caption{Magnitude of the effective, unitless astigmatism $\tilde{a}$ (black, solid), coma $\tilde{c}$ (blue, dashed), and trefoil $\tilde{t}$ (green, dot-dashed) aberrations for one wave (600nm delay at pupil edge) of the unitful aberration plus equal defocus.  As a scale for the image size, $\zeta$, we have used $0\farcs1$, which also corresponds to 100nm of wavefront delay for 1m of mirror radius.  For a fixed wavefront delay, the magnitude of the {\it effective} aberration scales (inversely) with the width of the atmospheric smearing.  We have zoomed in on the region between atmospherically induced PSF widths of FWHM = $0\farcs6$ and $0\farcs9$, to show detail in the effective coma and trefoil aberrations in the regime of traditional ground based observations.  \label{fig:magnitudes_of_aberrations}}
\end{figure}

\noindent The magnitude of these effective aberrations for one wave (600nm) of astigmatic, comatic, or trefoil aberration mixed with one wave of defocus is shown in figure \ref{fig:magnitudes_of_aberrations} for the Magellan 6.5m telescopes and the proposed LSST 8.4m telescope.  For a constant wavefront delay due to telescope aberration, the magnitudes of the effective asymmetric aberrations decrease with increased atmospheric smearing.

Using the effective aberrations from equation \eqref{eq:mapping_aberrations}, the spin zero, one, two and three moments simplify to,

\begin{align}
\label{eq:moments3}
Q_0           = & \zeta^2  \left( 1+ \frac{d^2 + a^2 + \frac{1}{3} c^2 + \frac{3}{2} t^2}{S_{atm}^2} \right) 
		=  2 \alpha^2  \\ \nonumber
\vec{Q_2} = & \zeta^2  4 \frac{\alpha^2}{\zeta^2}  \left(\vec{\tilde{a}} \right) 
		=  4 \alpha^2  \left(\vec{\tilde{a}} \right) \\ \nonumber
\vec{Q_1} = & \zeta^3  24 \frac{\alpha^3}{\zeta^3} \left( 3\vec{\tilde{c}} \right) 
		=  24 \alpha^3  \left( 3\vec{\tilde{c}} \right) \\ \nonumber
\vec{Q_3} = & \zeta^3 24  \frac{\alpha^3}{\zeta^3} \left(  \vec{\tilde{t}} \right)  
		=  24 \alpha^3  \left( \vec{\tilde{t}} \right)
\end{align}

There are many possible functions which will produce these first few moments.  However, as stated in the beginning of this subsection, we seek only a first order approximation to the PSF, preferably one which can be easily represented as low order shapelets for convolution.  As a Gaussian is a sufficient model for a symmetric PSF caused by atmospheric smearing, and spin one through three distortions will most severely affect the lensing measurement, we choose a functional form consisting of a Gaussian plus spin one through three Gaussian perturbations which satisfies the above listed second and third moment constraints.  

We argue that the spin asymmetric Gaussian perturbations with lowest order radial dependence are the most physically relevant and thus terms with redundant spin symmetries but higher order radial dependences should be dropped in the model.  In the limit of geometric optics and in the absence of atmospheric effects, the telescope produces aberrations of finite extent.  Defocus images stars into finite circular rings, perfect images of the pupil.  Astigmatism does the same, unless in conjunction with defocus, which will convert the finite rings into finite ellipses.  Coma, in its extreme, produces `comet' images with a pointlike head and a dimmer, but still defined circular tail, an image of the outer ring of the pupil.  Any infinite extent of the aberrations is caused by diffraction within the telescope and smearing due to atmospheric seeing.  Therefore, the spin asymmetries introduced by telescope aberrations would logically affect terms with lower order radial dependence more strongly than terms with higher order radial dependence.  

Therefore we will retain the astigmatic, spin two term varying as $\vec{\theta}^2$ and the trefoil, spin three term varying as $\vec{\theta}^3$, dropping all spin two and three terms with higher radial dependence.  The comatic, spin one term has the same radial dependence on the pupil as trefoil (cubic, one order higher than shear's quadratic pupillary dependence), so it ought to have the same radial variation on the image.  We therefore retain the $\vec{\theta}\theta^2$ spin one term, discarding the others.  While this is an admittedly ad hoc approximation, it does allow us to create a completely constrained model using only second and third moments.

The complete model for the asymmetric point spread function as it will most significantly affect a lensed galaxy is therefore

\begin{align}
\label{eq:PSF_model} 
I(\vec{\theta}) = \frac{I_0}{2 \pi \alpha^2} 
	\text{exp}&\left(- \frac{\theta^2}{2 \alpha^2} \right) \times 
	\Big[ 1 +   \frac{1}{ \alpha^2 } 
		\left(\vec{\tilde{a}} \cdot \vec{\theta}^2 \right)
	\\ \nonumber
	& + \frac{1}{4 \sqrt{2 \text{ln}(2)} \alpha^3 } \left(
		3 \left( \vec{\tilde{c}} \cdot \vec{\theta} \right) \theta^2 
		+ \left( \vec{\tilde{t}} \cdot (\vec{\theta})^3 \right)
	\right)
	\Big],
\end{align}

\noindent exactly the same model as a the linear lensed galaxy, but with the telescope aberrations producing spin one, two, and three moments replacing lensing terms.  The measurable Gaussian width of the PSF is $\alpha$, which is equal to $S_{atm} \zeta$ in equations \eqref{eq:moments2} in the limit of no telescope aberrations. 

Though the aberrations $\vec{\tilde{a}}$, $\vec{\tilde{c}}$, and $\vec{\tilde{t}}$ are combinations of all the telescope aberrations, they are proportional to astigmatism, coma, and trefoil in the absence of other asymmetric aberrations, and produce spin two, one and three moments respectively.  Therefore we shall loosely refer to them as astigmatism, coma, and trefoil respectively.  In the limit that defocus is much larger than the asymmetric aberrations, this loose approximation improves.  Optionally (and optimally), one could measure the second moment, measure the tilde space aberrations, and correct the telescope-- this step would be straightforward and require no iterations.

Using the same vector conversions as for the galaxy model, one can infer that the shapelets coefficients needed to represent the PSF model are

\begin{align}
\label{eq:shapelets_coefficients_PSF}
f_{00} &= 1 \\ \nonumber
f_{22} &= \frac{1}{\sqrt{2}} \vec{\tilde{a}} \\ \nonumber
f_{31} &= \frac{1}{4\sqrt{2\text{ln}(2)}} \frac{1}{\sqrt{2}} \left(3 \vec{\tilde{c}}\right), \ f_{11} = \frac{1}{4\sqrt{2\text{ln}(2)}} \left(3 \vec{\tilde{c}}\right) \\ \nonumber
f_{33} &= \frac{1}{4\sqrt{2\text{ln}(2)}} \sqrt{\frac{3}{2}} \vec{\tilde{t}}.
\end{align}

%% file: section_4_extracting_gravitational_lensing_parameters.tex
\section{Extracting gravitational lensing parameters from measured values}

We convolve a lensed galaxy model of scale length $\eta$ and shear and flexions $\vec{\tilde{g}}$, $\vec{\tilde{F}}$, and $\vec{\tilde{G}}$ with an aberrated PSF model of scale length $\sigma$ and astigmatism, coma and trefoil, $\vec{\tilde{a}}$, $\vec{\tilde{c}}$, and $\vec{\tilde{t}}$ to determine the effects on the resultant image's apparent lensing characteristics, i.e what an observer measuring the image shape would naively take the true shear and flexion values to be if he were not accounting for the PSF.  Given these apparent flexion and shear values, we then provide a analytic formulae to extract the true lensing terms, given the PSF at the position of the galaxy.

We perform the convolution in two parts.  First we convolve the lensed galaxy model with a symmetric PSF model, aberrated only by symmetric atmospheric smearing.  Then we convolve the galaxy model with the asymmetric, telescope aberrated PSF model.

\subsection{Convolution of the lensed galaxy model with the symmetric, atmospherically aberrated PSF model}
\label{sec:symmetric}
 
The symmetric PSF is a simple Gaussian, or in the shapelets basis $f_{00} = 1$, and all other coefficients are zero.  

Using the shapelets representation of the galaxy model given in equation \eqref{eq:lensed_galaxy_model_shapelets} with coefficients in \eqref{eq:shapelets_coefficients}, it is straightforward to analytically compute the convolved galaxy form.  One can perform this convolution either by computing a minimal number of relatively simple integrals, or by switching to cartesian shapelets and performing the matrix manipulations detailed in \citet{RefregierBacon2003}.  As spin m shapelets will only map onto other spin m shapelets under convolution with a Gaussian, the mathematics are tractable using either method.

The scale length $\xi$ of the convolved galaxy is predictably equal to the quadratic sum of the PSF and galaxy Gaussian widths, $\sqrt{\sigma^2 + \eta^2}$.  This scale length dictates the scale length of the shapelets basis in which to optimally decompose the convolved image.  Using this basis, we find the shapelets coefficients for the convolved galaxy are

\begin{align}
\label{eq:shapelets_coefficients_symmetric} 
f_{00}^{\prime} &= 1\\ \nonumber
f_{22}^{\prime} &= \frac{1}{\sqrt{2}} \frac{1}{\xi^2} 
	\left(\vec{\tilde{g}} \eta^2 \right) \\ \nonumber
f_{31}^{\prime} &= \frac{1}{4\sqrt{2\text{ln}(2)}} \frac{1}{\sqrt{2}} \frac{1}{\xi^3}
	\left(3\vec{\tilde{F}} \eta^3 \right),
 \ f_{11}^{\prime} = \frac{1}{4\sqrt{2\text{ln}(2)} } \frac{1}{\xi} 
	\left(3 \vec{\tilde{F}} \eta \right) \left( 1 + \frac{\sigma^2}{\xi^2} \right)\\ \nonumber
f_{33}^{\prime} &= \frac{1}{4\sqrt{2\text{ln}(2)}} \sqrt{\frac{3}{2}} \frac{1}{\xi^3}
	\left(\vec{\tilde{G}} \eta^3 \right) \\ \nonumber
\\ \label{eq:convolved_shapelets_coefficients_perturbation}
f_{51}^{\prime} &=  \sqrt{3} \frac{1}{\xi^5} \left(\vec{p} \eta^5 \right),
 \ f_{31}^{\prime} = 3 \sqrt{2} \frac{1}{\xi^3} \left(\vec{p} \eta^3 \right) \left( 1 + \frac{\sigma^2}{\xi^2} \right),
 \ f_{11}^{\prime} = 3 \frac{1}{\xi} \left(\vec{p} \eta \right) \left( 1 + \frac{\sigma^2}{\xi^2} \right)^2 \\ \nonumber
f_{53}^{\prime} &=  \sqrt{6} \frac{1}{\xi^5} \left(\vec{q} \eta^5 \right) ,
 \ f_{33}^{\prime} = 2 \sqrt{6} \frac{1}{\xi^3} \left(\vec{q} \eta^3 \right) \left( 1 + \frac{\sigma^2}{\xi^2} \right),
\end{align}

\noindent where $\vec{p}$ and $\vec{q}$ are the functions of $\vec{F}$ and $\vec{G}$ given in equation \eqref{eq:shapelets_coefficients_perturbation}.

We wish to extract terms which vary as $\vec{\theta}^2$, $\vec{\theta}\theta^2$, and $\vec{\theta}^3$ as these are the shear and flexion like terms in the model, $\vec{g}^{\prime}$, $\vec{F}^{\prime}$, and $\vec{G}^{\prime}$.  Naively, one obtains these terms by simply gathering the $f_{22}^{\prime}$, $f_{31}^{\prime}$, and $f_{33}^{\prime}$ coefficients.  However, one must first remove the effects of the higher order variances which would be detected as a separate signal from the flexions, but `trickle down' into these terms by virtue of the orthogonality of the shapelets basis.  

We remove the $\vec{\theta}\theta^4$ and $\vec{\theta}^3\theta^2$ dependences, which would be detected as a separate signal with $\theta^5$ radial dependence, and compare the remaining $f_{22}^{\prime}$, $f_{31}^{\prime}$ and $f_{33}^{\prime}$ terms with the unconvolved galaxy model's coefficients.  We find the following mapping from intrinsic ellipticity and gravitational shear and flexion to the apparent signal in a seeing degraded image to be\footnote{The flexions $\vec{F}$ and $\vec{G}$ multiplying $\vec{\tilde{g}}$ in the non linear terms of \eqref{eq:convolution_mapping2} and \eqref{eq:convolution_mapping3} are intentionally gravitational flexions and not effective flexions $\vec{\tilde{F}}$ and $\vec{\tilde{G}}$, consistent with equation \eqref{eq:Lensed_galaxy_model_final}.}

\begin{subequations}
\label{eq:convolution_mapping}
\begin{align}
\label{eq:convolution_mapping1} 
\vec{g}^{\prime} \xi^2 &=
	\left(\vec{\tilde{g}} \eta^2 \right) 
\\ \label{eq:convolution_mapping2} 
\vec{F}^{\prime} \xi^3 &= 
	\left(\vec{\tilde{F}} \eta^3 \right) 
	+ 2 (\frac{\sigma}{\xi})^2 \left( 3 \vec{\tilde{g}}\vec{F}^{*} + \vec{\tilde{g}}^{*}\vec{G} \right) \eta^3 
\\ \label{eq:convolution_mapping3} 
\vec{G}^{\prime} \xi^3 &= 
	\left(\vec{\tilde{G}} \eta^3 \right) 
	+ 4 (\frac{\sigma}{\xi})^2 \left( 3\vec{\tilde{g}}\vec{F} \right) \eta^3.
\end{align}
\end{subequations}

To linear order in asymmetric terms, each apparent gravitational aberration is diluted by the ratio of the uncorrupted galaxy size to the measured galaxy size to a power equal to the radial dependence of the aberration on the galaxy.  

The quadratic expansion to the galaxy model, which had formerly only contributed to the shear and flexion-like signals by altering the origin and magnitude of the pre-convolved flexion signals, $\vec{\tilde{F}}$ and $\vec{\tilde{G}}$, contribute to the apparent signals again after convolution.  This new contribution is caused by a `mixing down' of the quintic radial signal to a cubic radial signal on the galaxy via convolution with the PSF.  These non-linear contributions to the apparent lensing signals are diluted by the PSF to the same power as the linear terms, but with an additional dilution factor of the ratio of the PSF size to the measured galaxy size, squared.

\begin{figure}[htbp]
\epsscale{0.60}
\plotone{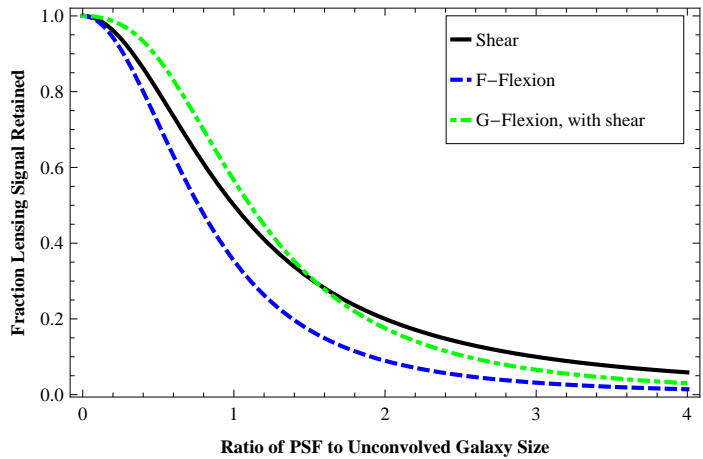}
\caption{The fraction of the shear/ellipticity (black, solid) and F\-- and G\--flexion (blue, dashed; green, dot\--dashed) signals retained after convolution with a symmetric PSF of increasing size.  The presence of intrinsic ellipticity or gravitational shear will generally alter the retained flexion signal, however for an SIS lens, the retained F-flexion is independent of shear.  The retained G\--flexion shown here is computed with shear and F\--flexion present in the lensed galaxy, boosting the observed ${G}^{\prime}$ signal.  The magnitude of the pre-degraded shear signal is $-0.3$, and it is aligned with $\tilde{F}$\--flexion and anti-aligned with $\tilde{G}$\--flexion. $\tilde{F}$\-- and $\tilde{G}$\--flexions are in a ratio of 1:-3, approximately consistent with a SIS of any ${r_{hl}}$:${\theta_{e}}$ ratio. \label{fig:fracsignalretained}}
\end{figure}

The ratio of the effective signal after convolution with a symmetric PSF (i.e. the measured signal) and the effective lensing signal before convolution is shown in figure \ref{fig:fracsignalretained}.  The horizontal axis contains the ratio of the PSF size to the unconvolved galaxy size.  For both shear and flexion, the larger the PSF, the less lensing signal is retained post convolution, however the variation with PSF size differs for shear and the flexions.  The fraction of $\tilde{F}$ and $\tilde{G}$\--flexions retained post-convolution will generally depend on how much shear is present and the ratio of F\-- and G\--flexions.  As seen in equation \eqref{eq:convolution_mapping2}, for an SIS the 1:-3 ratio of F\-- to G\--flexions will render the effect of convolution on $\tilde{F}$\--flexion independent of shear.  Therefore the accelerated dilution of $\tilde{F}$\--flexion relative to the shear is caused by the PSF alone.  The PSF causes the same dilution of $\tilde{G}$\--flexion, however, the shear enhances the measured $\tilde{G}$\--flexion in most lenses, counteracting the effect of the PSF, as can be seen in equation \eqref{eq:convolution_mapping3}.  This is consistent with figure \ref{fig:fracsignalretained}.

\begin{figure}[htb]
\vspace{1.625 truein}
\includegraphics{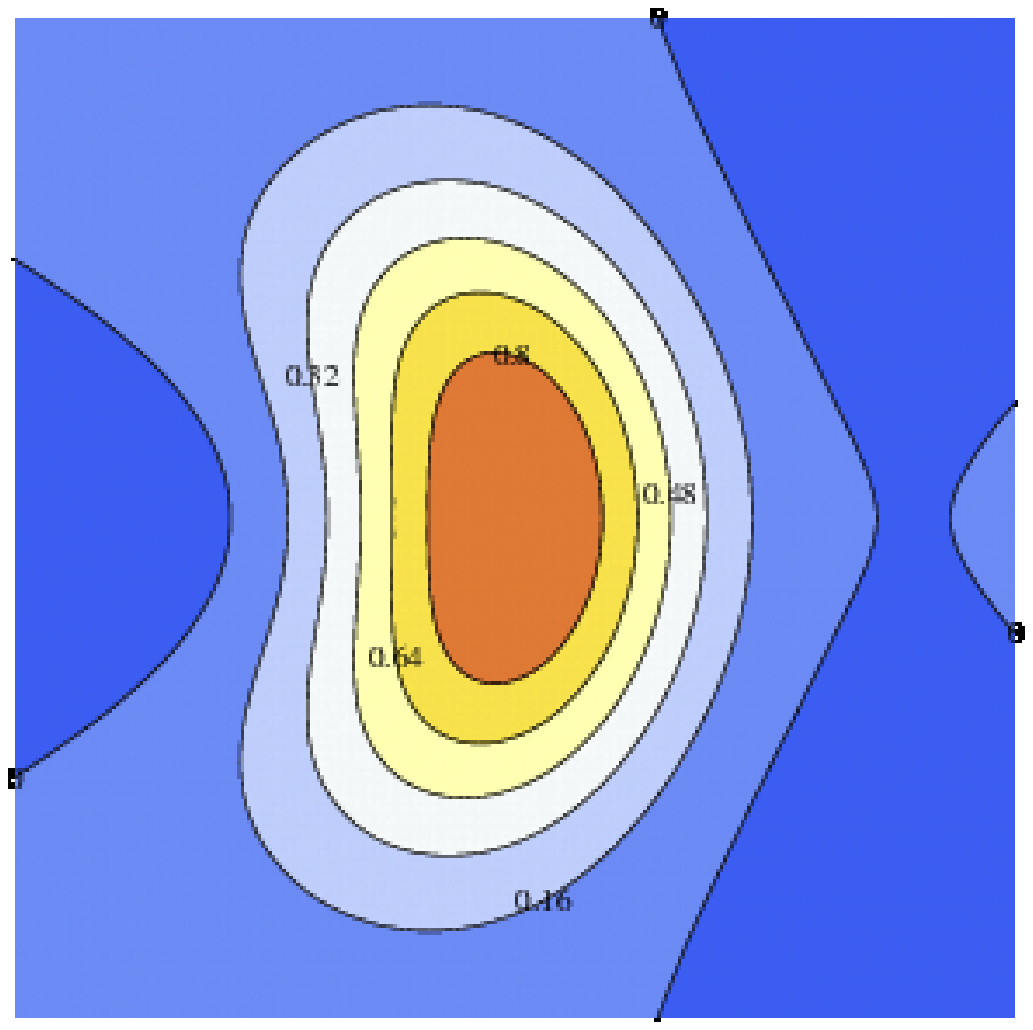}
\includegraphics{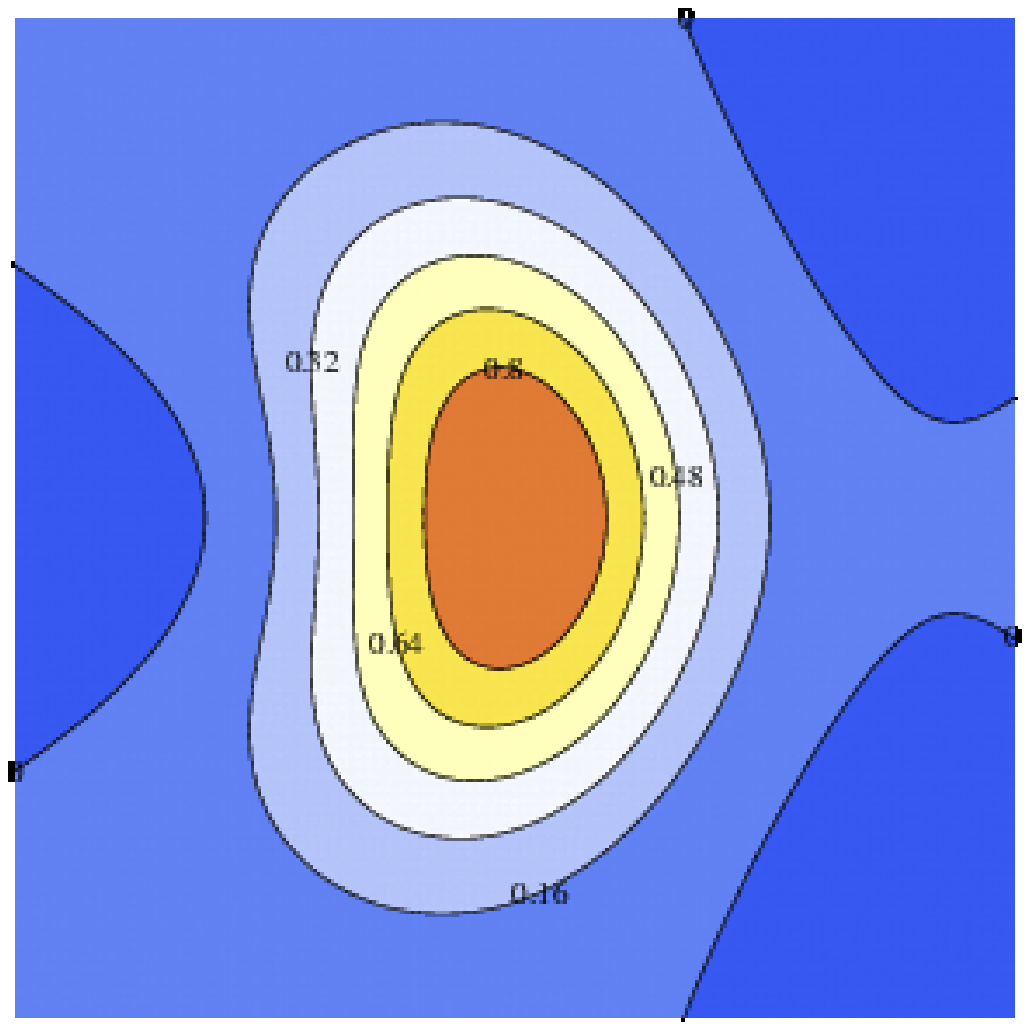}
\includegraphics{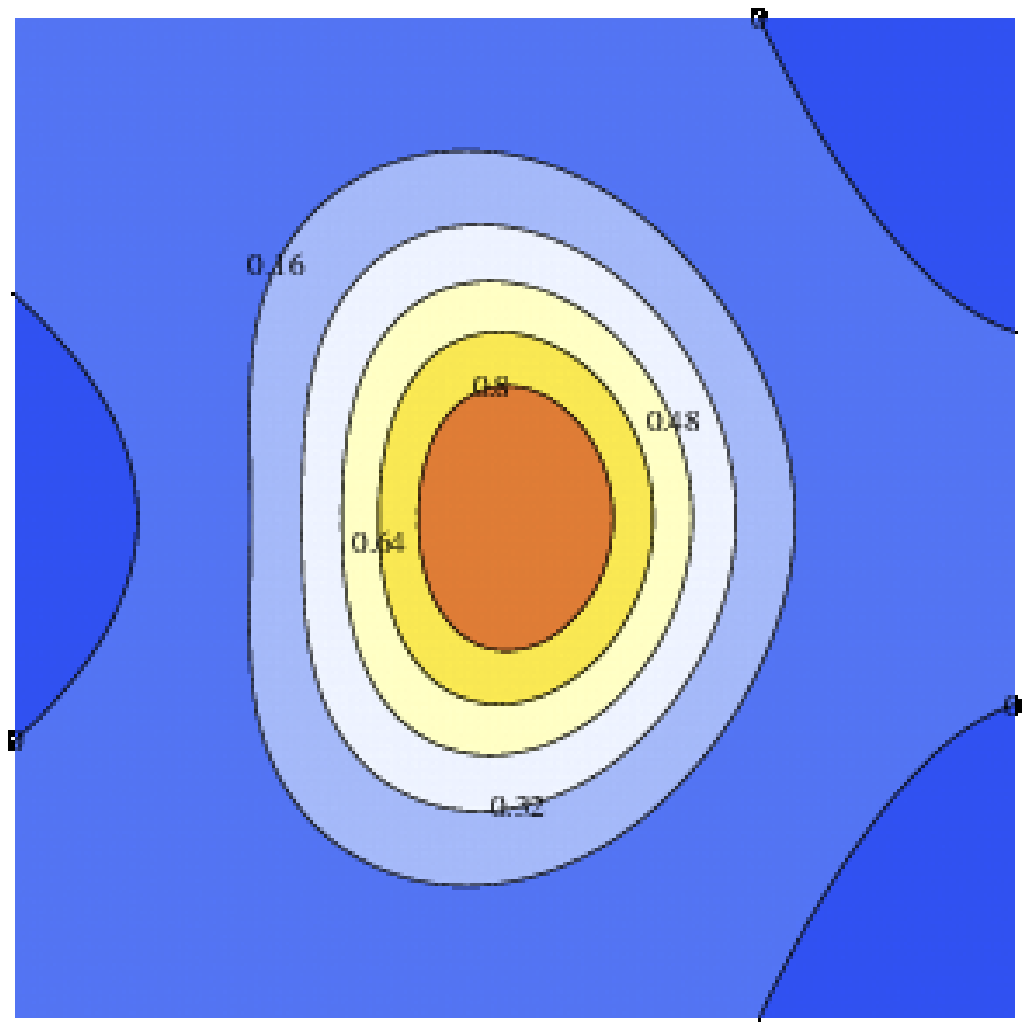}
\includegraphics{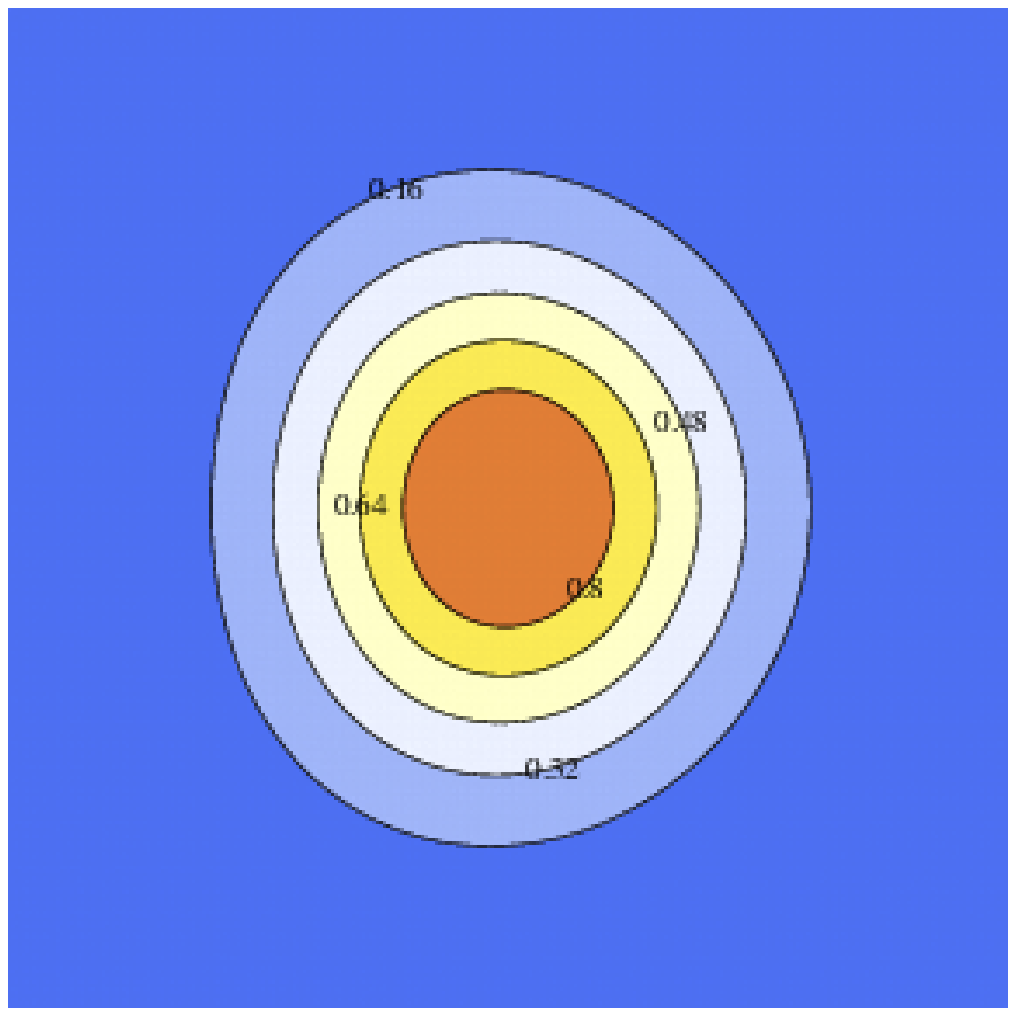}
\caption{A circular galaxy lensed by aligned shear $= -0.3$, F\--flexion $= -0.09$, and G\--flexion $= 0.27$.  (a) Unconvolved, and (b) convolved with PSFs with radii equal to half, (c) equal, and (d) twice the unlensed galaxy size. Ratios of shears and flexions simulate a SIS lens with $\frac{r_{hl}}{\theta_{e}} = \frac{1}{2}$.  Images have been rescaled to highlight differences in asymmetries. \label{fig:convolvedimages}}
\end{figure}

A contour plot of a lensed galaxy that is unconvolved, and convolved with PSFs of half, equal, and twice the unlensed galaxy size are shown in figure \ref{fig:convolvedimages}.  When the radius of the PSF is only half of the size of the unconvolved galaxy, most of the overall image shape is retained, consistent with figure \ref{fig:fracsignalretained}.  Once the radius of the PSF is equal to the size of the unconvolved galaxy, the shear and flexion signals become more noticeably diminished.  However, the spin three flexion signal is still apparent even when its spin one counterpart is nearly wiped out, because combinations of spin one flexion and shear can mix to create an additional apparent spin three flexion post-convolution.  Convolutions with yet larger PSFs circularize the final galaxy image.  These trends in signal dilution are in agreement with figure \ref{fig:fracsignalretained}.

\subsection{Convolution of the lensed galaxy with the asymmetrically aberrated PSF model}

The asymmetrically aberrated PSF is formally the same as the linear order galaxy model. By exploiting symmetries in the two models, and only considering asymmetric perturbations to second order, we can analytically compute the convolution of the lensed galaxy with the asymmetrically aberrated PSF induced by telescope aberrations.  As with the convolution of the galaxy and the symmetric PSF, this can be done either by computing a minimal number of relatively simple integrals, or by switching to cartesian shapelets and performing the matrix manipulations detailed in \citet{RefregierBacon2003}.  The shapelets coefficients for the convolved galaxy image are

\begin{subequations}
\label{eq:shapelets_coefficients_asymmetric} 
\begin{align}
\label{eq:shapelets_coefficients_asymmetric_simple} 
f_{00}^{\prime} &= 1
	+ \frac{\eta^2\sigma^2}{\xi^4}\left(\vec{\tilde{a}} \cdot \vec{\tilde{g}} \right)
\\ \nonumber
f_{22}^{\prime} &= \frac{1}{\sqrt{2}} \frac{1}{\xi^2} \Big[
	\left(\vec{\tilde{g}} \eta^2 + \vec{\tilde{a}} \sigma^2 \right) 
	+ \frac{9}{8} \eta \sigma 
		\left(\vec{\tilde{c}}\vec{\tilde{F}} 
			+ \frac{1}{\xi^4} \left( 3\vec{\tilde{c}}\vec{\tilde{F}} \eta^2 \sigma^2 
			- \vec{\tilde{c}}^{*}\vec{\tilde{G}} 
			- \vec{\tilde{F}}^{*}\vec{\tilde{t}} \right)
		\right)
	\Big]
\\ \nonumber
f_{31}^{\prime} &= \frac{1}{4\sqrt{2\text{ln}(2)}} \frac{1}{\sqrt{2}} \frac{1}{\xi^3} \Big[
	\left(3\vec{\tilde{F}} \eta^3 + \vec{\tilde{c}} \sigma^3 \right)
	\\ \nonumber
	&\qquad \qquad+ \frac{\eta \sigma}{\xi^2} \frac{1}{2} \Big(	
		\vec{\tilde{a}}\vec{\tilde{F}}^{*}\sigma(2\sigma^2 - \eta^2) 
		+ \vec{\tilde{g}}\vec{\tilde{c}}^{*}\eta  (2\eta^2 - \sigma^2)
		- \vec{\tilde{a}}^{*}\vec{\tilde{G}}(\sigma\eta^2)
		- \vec{\tilde{g}}^{*}\vec{\tilde{t}}  (\eta\sigma^2)
	\Big) 
	\Big],
\\ \nonumber
f_{11}^{\prime} &= \frac{1}{4\sqrt{2\text{ln}(2)} } \frac{1}{\xi} \Big[
	\left(3 \vec{\tilde{F}} \eta \left( 1 + \frac{\sigma^2}{\xi^2} \right) 
		+ 3 \vec{\tilde{c}} \sigma \left( 1 + \frac{\eta^2}{\xi^2} \right) \right) 
	\\ \nonumber
	& \qquad \qquad - \frac{\eta \sigma}{\xi^4} \frac{3}{2} \Big(	
		\vec{\tilde{a}}\vec{\tilde{F}}^{*}\sigma(4\sigma^2 + \eta^2) 
		+ \vec{\tilde{g}}\vec{\tilde{c}}^{*}\eta  (4\eta^2 + \sigma^2)
		- \vec{\tilde{a}}^{*}\vec{\tilde{G}}(\sigma\eta^2)
		- \vec{\tilde{g}}^{*}\vec{\tilde{t}}  (\eta\sigma^2)
	\Big)
	\Big]
\\ \nonumber
f_{33}^{\prime} &= \frac{1}{4\sqrt{2\text{ln}(2)}} \sqrt{\frac{3}{2}} \frac{1}{\xi^3} \Big[
	\left(\vec{\tilde{G}} \eta^3 + \vec{\tilde{t}} \sigma^3 \right) 
	+ \frac{\eta \sigma}{\xi^2} 6 \Big( 
		 \vec{\tilde{a}}\vec{\tilde{F}}\sigma^3 + \vec{\tilde{g}}\vec{\tilde{c}}\eta^3 
	\Big)
	\Big]
\\ \nonumber
\\ \label{eq:shapelets_coefficients_asymmetric_perturbation}
f_{51}^{\prime} &=  \sqrt{3} \frac{1}{\xi^5} \left(\vec{p} \eta^5 \right),
 \ f_{31}^{\prime} = 3 \sqrt{2} \frac{1}{\xi^3} \left(\vec{p} \eta^3 \right) \left( 1 + \frac{\sigma^2}{\xi^2} \right),
 \ f_{11}^{\prime} = 3 \frac{1}{\xi} \left(\vec{p} \eta \right) \left( 1 + \frac{\sigma^2}{\xi^2} \right)^2 \\ \nonumber
f_{53}^{\prime} &=  \sqrt{6} \frac{1}{\xi^5} \left(\vec{q} \eta^5 \right) ,
 \ f_{33}^{\prime} = 2 \sqrt{6} \frac{1}{\xi^3} \left(\vec{q} \eta^3 \right) \left( 1 + \frac{\sigma^2}{\xi^2} \right).
\end{align}
\end{subequations}

Again, we wish to extract the manifestations of shear and flexion-like variations on the galaxy, the $\vec{\theta}^2$, $\vec{\theta}\theta^2$, and $\vec{\theta}^3$ terms.  Following the steps in subsection \S\S \ref{sec:symmetric}, we first account for the higher order shape variations that will be detected as separate signals, and then compare the coefficients $f_{22}^{\prime}$, $f_{31}^{\prime}$, and $f_{33}^{\prime}$ to their unconvolved counterparts to obtain the terms which vary as shear and F\-- and G\--flexion respectively.  We find

\begin{subequations}
\label{eq:convolution_mapping_asym}
\begin{align}
\label{eq:convolution_mapping_asym1} 
\vec{g}^{\prime} \xi^2 &=
	\left\{\vec{\tilde{g}} \eta^2 + \vec{\tilde{a}} \sigma^2 \right\} 
	+ \Big\{
		\frac{9}{8} \eta \sigma 
		\left(\vec{\tilde{c}}\vec{\tilde{F}} 
			+ \frac{1}{\xi^4} \left( 3\vec{\tilde{c}}\vec{\tilde{F}} \eta^2 \sigma^2 
			- \vec{\tilde{c}}^{*}\vec{\tilde{G}}\eta^4 
			- \vec{\tilde{F}}^{*}\vec{\tilde{t}}\sigma^4 \right)
		\right)
	\Big\}
\\ \label{eq:convolution_mapping_asym2} 
\vec{F}^{\prime} \xi^3 &= 
	\left\{\vec{\tilde{F}} \eta^3 + \vec{\tilde{c}} \sigma^3 \right\} 
	+ \Big\{
		2 (\frac{\sigma}{\xi})^2 
			\left( 3 \vec{\tilde{g}}\vec{F}^{*} + \vec{\tilde{g}}^{*}\vec{G} \right) \eta^3
		\\ \nonumber 
		& \qquad + \frac{\eta \sigma}{\xi^2} \frac{1}{2} \Big(	
		   \vec{\tilde{a}}\vec{\tilde{F}}^{*}\sigma(2\sigma^2 - \eta^2) 
		+ \vec{\tilde{g}}\vec{\tilde{c}}^{*}\eta  (2\eta^2 - \sigma^2)
		- \vec{\tilde{a}}^{*}\vec{\tilde{G}}(\sigma\eta^2)
		- \vec{\tilde{g}}^{*}\vec{\tilde{t}}  (\eta\sigma^2)
		\Big) 
	\Big\}
\\ \label{eq:convolution_mapping_asym3} 
\vec{G}^{\prime} \xi^3 &= 
	\left\{\vec{\tilde{G}} \eta^3 + \vec{\tilde{t}} \sigma^3 \right\} 
	+ \Big\{
		4 (\frac{\sigma}{\xi})^2 \left( 3\vec{\tilde{g}}\vec{F} \right) \eta^3 
		+ \frac{\eta \sigma}{\xi^2} 6 \Big( 
		 \vec{\tilde{a}}\vec{\tilde{F}}\sigma^3 + \vec{\tilde{g}}\vec{\tilde{c}}\eta^3 
		 \Big)
	\Big\}.
\end{align}
\end{subequations}

The effects of telescope aberrations on the final convolved image can be broken down into terms which vary linearly with asymmetries due to lensing and telescope aberrations (first braced), and those which vary quadratically with lensing and telescope aberrations (second braced).  For flexion, the terms quadratic in asymmetries have two distinct origins:  \begin{itemize}
\item  The first group of non-linear terms vary as the product of shear and flexion (only with lensing aberrations, not with telescope aberrations).  These are carry-overs from the convolution of the quadratic expansion of the galaxy model and the symmetric part of the PSF, and as such are also present in equation \eqref{eq:convolution_mapping}, the mapping for the lensing terms under convolution with a symmetric PSF.  
\item The second group of non-linear terms are cross terms between the telescope aberrations $\vec{\tilde{a}}$, $\vec{\tilde{c}}$, and $\vec{\tilde{t}}$ and gravitational lensing aberrations $\vec{\tilde{g}}$, $\vec{\tilde{F}}$, and $\vec{\tilde{G}}$.  These terms are caused by mixing of spin signals under convolution to create signals with different spin symmetry.  The apparent shear signal is also influenced by such cross terms.
\end{itemize}

The linear contributions of the telescope aberrations in the first braced terms of equations \eqref{eq:convolution_mapping_asym1}, \eqref{eq:convolution_mapping_asym2}, and \eqref{eq:convolution_mapping_asym3}, highlight the similar forms of the PSF and lensing models- each lensing distortion adds to its telescope aberration counterpart (shear/astigmatism, F\--flexion/coma, G\--flexion/trefoil), but the lensing aberration is scaled by the size of the galaxy and the PSF aberration is scaled by the size of the PSF, each to the radial dependence of the aberration.  
The cross terms between the telescope and lensing aberrations in the second braced part of the equations also show the intrinsic symmetries between the two types of aberrations.  However, the parallelism between the lensing and telescope aberrations is broken by the fact that the galaxy model is expanded to quadratic order in lensing terms, while the PSF is truncated at linear order. 

\subsection{Interpretation of the apparent shear and flexions}

\begin{figure}[ht]
\vspace{2.05 truein}
\includegraphics{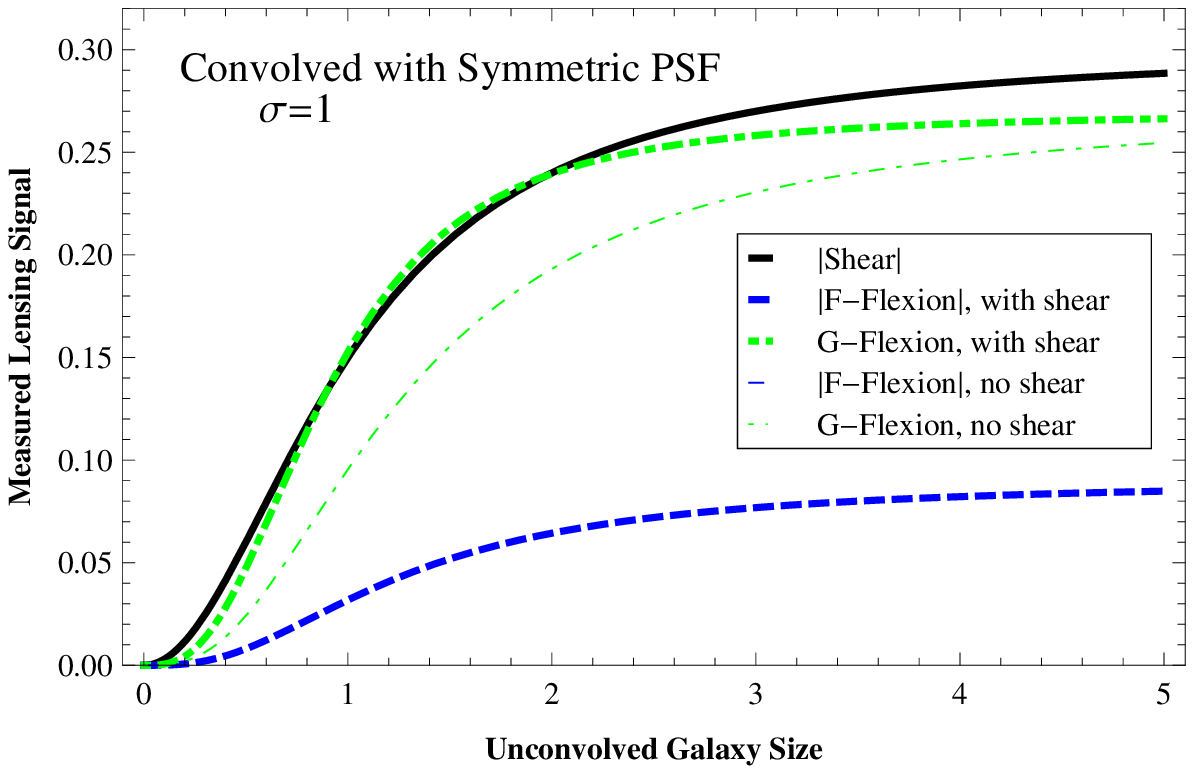}
\includegraphics{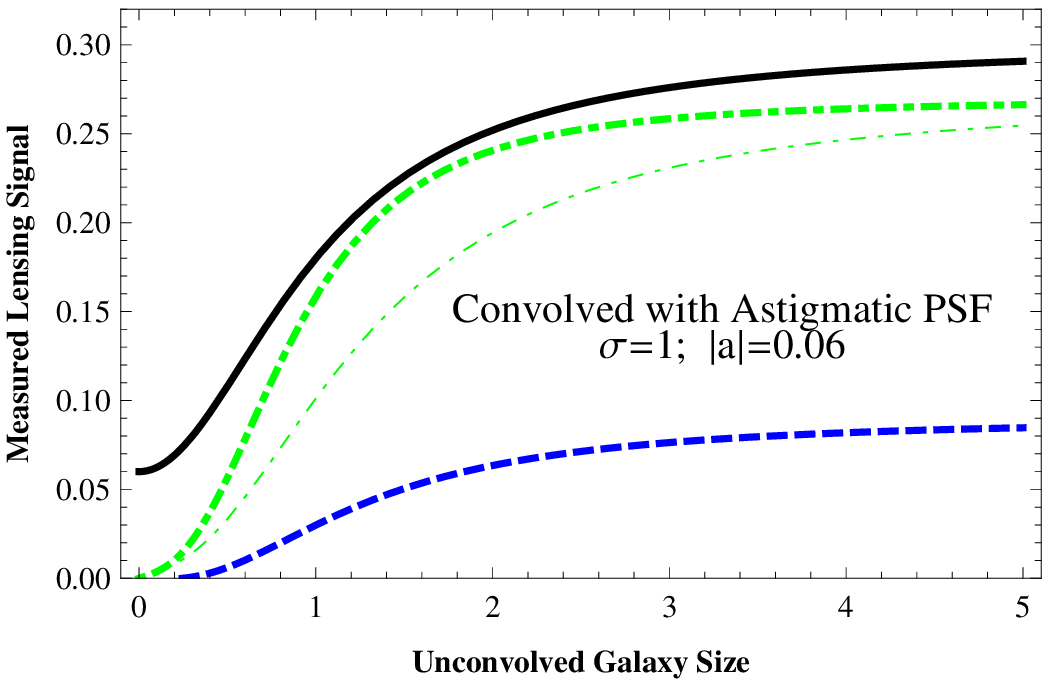}
\vspace{2.05 truein}
\includegraphics{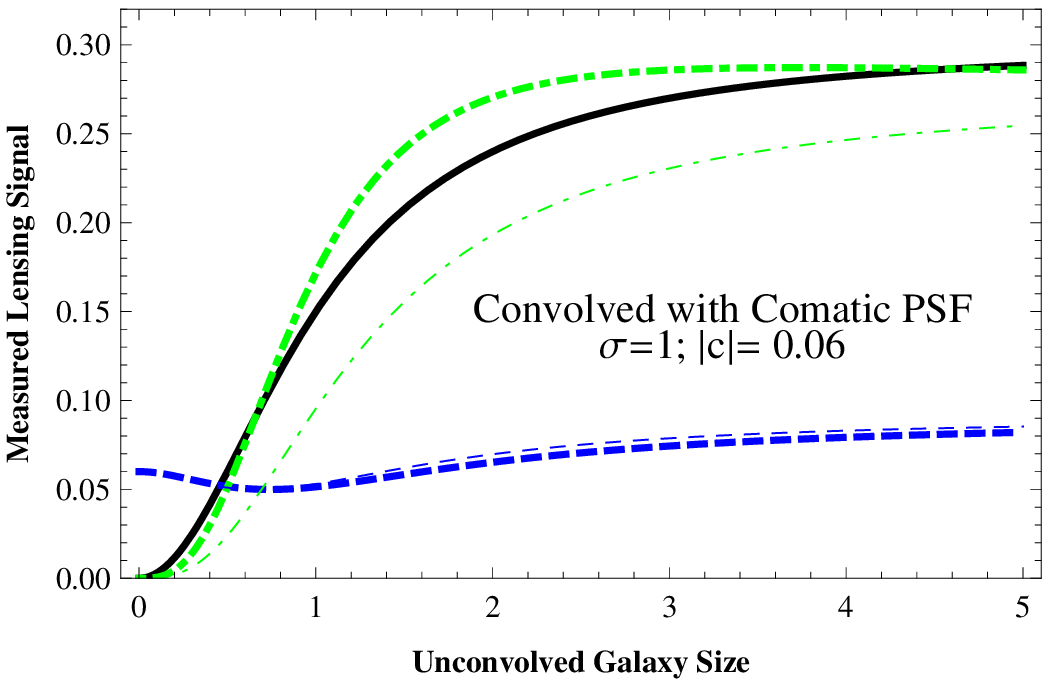}
\includegraphics{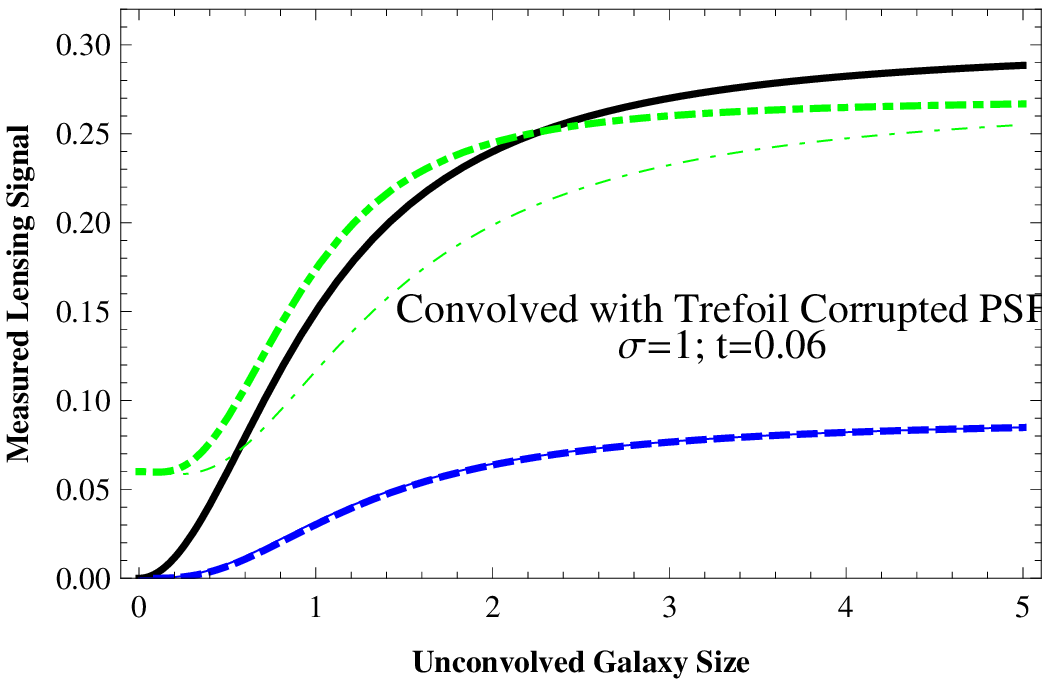}
\vspace{2.05 truein}
\includegraphics{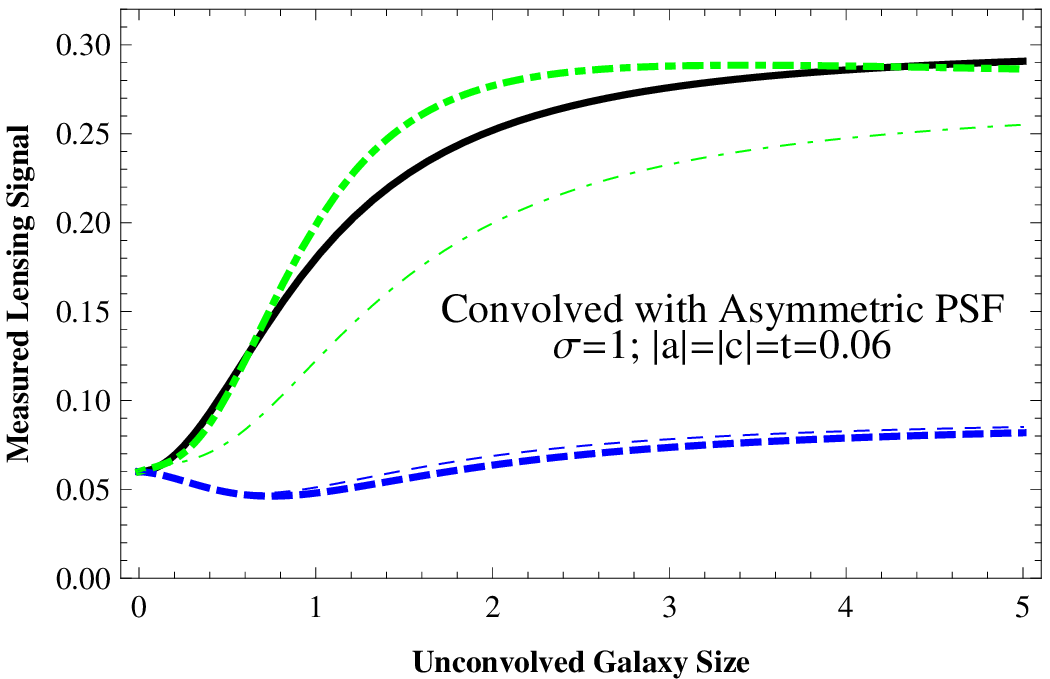}
\caption{Shown within each plot are the apparent shear (black, solid) and F\ and G\--flexion (blue, dashed; green, dot\--dashed) signals for a circular galaxy model lensed by aligned shear $= -0.3$, $\tilde{F}$\--flexion $= -0.09$, and $\tilde{G}$\--flexion $= 0.27$.  Also shown are the apparent F\-- and G\--flexion signals for the same galaxy model if shear were not present (thin blue, dashed; thin green, dot\--dashed), though these signals may be identical to their counterparts where shear is present.   The unit of the galaxy radius on the horizontal axis is the PSF half light radius, $\sigma$.  Top left to bottom right: the apparent signal after convolution with a PSF that is (a) symmetric, (b) astigmatic, ($\tilde{a}= -0.06$), (c) comatic, ($\tilde{c} = -0.06$), (d) corrupted by trefoil, ($\tilde{t} = 0.06$), and (e) corrupted by all of the listed aberrations.   Pre convolved values for the lensing parameters approximately simulate a SIS lens with $\frac{r_{hl}}{\theta_{e}} = \frac{1}{2}$.  Aberrations are large for the sake of illustration. \label{fig:convolvedasymmetricimages}}
\end{figure}
\afterpage{\clearpage}

Figure \ref{fig:convolvedasymmetricimages} shows the apparent shear and flexions on galaxies of varying intrinsic size for a PSF with fixed size and various asymmetries.  In the limit of zero galaxy size, the apparent shear and F\-- and G\--flexion signals are equal to the values of their corresponding telescope aberrations, astigmatism, coma, and trefoil.  This result is somewhat intuitive, as a galaxy of zero width is a star, which produces the PSF when imaged by the telescope.  In the other extreme, of very large galaxy size, the apparent lensing values approach the pre-convolved, effective lensing values and are virtually unaffected by the PSF and its aberrations.  This is also expected.  

To describe the behavior of the variations in the lensing terms between the zero and infinite galaxy size extremes for a fixed PSF size, we break down apparent shears and flexions into their contributions from linear and non-linear terms.

Referring back to figure \ref{fig:fracsignalretained}, in absence of any telescope aberration, atmospheric smearing causes the lensing signal to be diluted by the ratio of the unconvolved to the convolved galaxy size, to the radial power of the lensing aberration.  In absence of any asymmetric telescope aberration, this same relation holds here; shear approaches its true value as $\frac{\eta^2}{\xi^2}$ and flexion approaches its true value as $\frac{\eta^3}{\xi^3}$.  However, note that figure \ref{fig:convolvedasymmetricimages} depicts a fixed PSF and increasing galaxy size on the horizontal axis whereas its predecessor had the inverse ratio on the horizontal axis.

Now including telescope aberrations, but only allowing terms linear in asymmetries (first braced terms of equation \eqref{eq:convolution_mapping_asym}), we find that the effect of asymmetric telescope aberrations is to add an offset to the post-convolved lensing terms which varies as the ratio of the PSF size to the convolved galaxy size, again, to the radial power of the aberration.  For shear this is $\frac{\sigma^2}{\xi^2}$, and for flexion this is $\frac{\sigma^3}{\xi^3}$.  In the limit of a star, where $\xi=\sigma$, these constants are one, fully weighting the aberration; in the limit of a large galaxy, these constants approach zero, fully nulling the aberration.   

For the sake of comparing the relative magnitudes of the terms linear and non-linear in asymmetries, it is mathematically advantageous to separate out the effect of atmospheric dilution from each term in equation \eqref{eq:convolution_mapping_asym}.  Once the effect of the atmosphere has been uncoupled, these terms can then be thought of as perturbations to the pre-convolved shear and flexions, which are then diluted by the atmosphere with them.  This treatment of the effect of the atmosphere and telescope aberrations on the final apparent shears and flexions is perhaps less physically intuitive than the view where the effects of the telescope and atmosphere are coupled, but it does assist with the interpretation of the relative {\it sizes} of the biases introduced by different aberrations.  Therefore, we decouple the effect of the atmosphere from the other terms here.

\begin{subequations}
\label{eq:convolution_mapping_asym_bad}
\begin{align}
\label{eq:convolution_mapping_asym_bad1} 
\vec{g}^{\prime}  &= \frac{\eta^2}{\xi^2}\Big[
	\left\{\vec{\tilde{g}}  + \vec{\tilde{a}} \frac{\sigma^2}{\eta^2} \right\} 
	+ \Bigg\{
		\frac{9}{8}  
		\left(\vec{\tilde{c}}\vec{\tilde{F}} \frac{\sigma}{\eta}
			+\frac{\eta^4}{\xi^4} \left( 3\vec{\tilde{c}}\vec{\tilde{F}} \frac{\sigma^3}{\eta^3} 
			- \vec{\tilde{c}}^{*}\vec{\tilde{G}} \frac{\sigma}{\eta}
			- \vec{\tilde{F}}^{*}\vec{\tilde{t}} \frac{\sigma^5}{\eta^5} \right)
		\right)
	\Bigg\}
	\Big]
\\ \label{eq:convolution_mapping_asym_bad2} 
\vec{F}^{\prime}  &= \frac{ \eta^3}{\xi^3} \Big[
	\left\{\vec{\tilde{F}} + \vec{\tilde{c}}\frac{\sigma^3}{\eta^3} \right\} 
	+\frac{\eta^2}{\xi^2} \Bigg\{
		2 \left( 3 \vec{\tilde{g}}\vec{F}^{*} + \vec{\tilde{g}}^{*}\vec{G} \right)  
		\frac{\sigma^2}{\eta^2}
		\\ \nonumber 
		& \qquad \quad + \frac{1}{2} \Big(	
		   \vec{\tilde{a}}\vec{\tilde{F}}^{*}(2\frac{\sigma^4}{\eta^4} - \frac{\sigma^2}{\eta^2}) 
		+ \vec{\tilde{g}}\vec{\tilde{c}}^{*} (2 \frac{\sigma}{\eta} - \frac{\sigma^3}{\eta^3})
		- \vec{\tilde{a}}^{*}\vec{\tilde{G}}\frac{\sigma^2}{\eta^2}
		- \vec{\tilde{g}}^{*}\vec{\tilde{t}}  \frac{\sigma^3}{\eta^3}
		\Big) 
	\Bigg\}
	\Big]
\\ \label{eq:convolution_mapping_asym_bad3} 
\vec{G}^{\prime} &= \frac{ \eta^3}{\xi^3} \Big[
	\left\{\vec{\tilde{G}} + \vec{\tilde{t}} \frac{\sigma^3}{\eta^3} \right\} 
	+  \frac{\eta^2}{\xi^2} \Bigg\{
		4 \left( 3\vec{\tilde{g}}\vec{F} \right) \frac{\sigma^2}{\eta^2} 
		+6 \Big( 
		 \vec{\tilde{a}}\vec{\tilde{F}}\frac{\sigma^4}{\eta^4} + \vec{\tilde{g}}\vec{\tilde{c}}\frac{\sigma}{\eta}
		 \Big)
	\Bigg\}.
	\Big]
\end{align}
\end{subequations}

As the galaxy size becomes large with resect to the PSF, the ratio of the true to measured galaxy size, $\frac{\eta}{\xi}$, approaches one while the ratio of PSF to true galaxy size, $\frac{\sigma}{\eta}$ approaches zero.  Assuming that most observers will either appropriately weight (or simply discard) galaxies for which PSF dilution will effectively wipe out any lensing signal, we take $\frac{\sigma}{\eta}$ to be small for most (priority) galaxies in a lensing survey.\footnote{For a given PSF, the smallest galaxies will have the noisiest measurements of PSF-corrected flexions, as the error in the measurement must propagate when removing the effects of PSF dilution.  Thus the galaxies with the least noisy measurements will be the largest ones for which minimal dilution correction is required.} Accordingly we take $\frac{\eta}{\xi}$ to be non-negligible, order of one. 

Examining equation \eqref{eq:convolution_mapping_asym_bad}, for shear, astigmatism perturbs the lensing distortion at a `rate' of $\frac{\sigma^2}{\eta^2}$.  Likewise, coma and trefoil each perturb their respective flexions at rates of $\frac{\sigma^3}{\eta^3}$.  These effects of these perturbations drop off quickly for the large galaxies of primary interest.  By contrast, the terms non-linear in asymmetries given in the second braced terms of equations \eqref{eq:convolution_mapping_asym} and \eqref{eq:convolution_mapping_asym_bad} perturb their lensing distortions at rates up to $\frac{\sigma}{\eta}$.  Thus, the effects of these cross terms can manifest in the measurements of shears and flexions in galaxies of much larger galaxy sizes than the linear terms can.

Anywhere in figure \ref{fig:convolvedasymmetricimages} where the apparent flexion signal is different when computed with and without the presence of shear in the galaxy is a demonstration of the effect of the non-linear asymmetric terms on the apparent signal.  The apparent signal enhancement of G\--flexion in figure \ref{fig:convolvedasymmetricimages}c is an extreme effect; coma has been added to this PSF, not trefoil.

\subsection{Extraction of the PSF from the lensing terms}

If one can solve equations \eqref{eq:convolution_mapping_asym} for the pre-convolved lensing values, $\vec{\tilde{g}}$, $\vec{\tilde{F}}$, and $\vec{\tilde{G}}$, one can create a completely analytic method for deconvolving the effects of an asymmetric PSF from a measured galaxy image.  We use the simplifying assumption that telescope coma, trefoil, and astigmatism are small enough that cross terms between them and flexion might be ignored.    We do {\it not} make the same assumption for shear.  

Using the variable $\mu$ for $\frac{\sigma}{\xi}$, and removing all references to $\eta$ which is only measurable indirectly, we find the deconvolution of shear and flexion in terms of the properties of the PSF and the directly measurable properties of the galaxy,

\begin{subequations}
\label{eq:deconvolution_mapping} 
\begin{align}
\label{eq:deconvolution_mapping1} 
&\vec{\tilde{g}} = \left( \frac{1}{1- \mu^2} \right)
	\left( \vec{g}^{\prime} - \vec{\tilde{a}} \mu^2 \right) 
	\\  \label{eq:deconvolution_mapping2}
&\vec{\tilde{F}} 
	+ 2 \mu^2 
	\left( 3 \vec{\tilde{g}}\vec{F}^{*} + \vec{\tilde{g}}^{*}\vec{G} \right)
= \\ \nonumber
& \qquad	\left( \frac{1}{1- \mu^2} \right)^{3/2}
	\left(\vec{F}^{\prime} - \vec{\tilde{c}} \mu^3 \right) 
	 -  \mu \left( 1- \mu^2 \right)^{1/2} 	
		\left( \vec{\tilde{g}}\vec{\tilde{c}}^{*} \right)
	+ \frac{1}{2} \mu^3 \left( \frac{1}{1- \mu^2} \right)^{1/2}  
		\left( \vec{\tilde{g}}\vec{\tilde{c}}^{*} + \vec{\tilde{g}}^{*}\vec{\tilde{t}} \right)
	\\ \label{eq:deconvolution_mapping3} 
&\vec{\tilde{G}} +
	4 \mu^2 
	\left( 3\vec{\tilde{g}}\vec{F} \right) 
= \\ \nonumber
& \qquad	\left( \frac{1}{1- \mu^2} \right)^{3/2}
	\left(\vec{G}^{\prime} - \vec{\tilde{t}} \mu^3 \right) 
	 -  \mu \left( 1- \mu^2 \right)^{1/2}
		\left( 6\vec{\tilde{g}}\vec{\tilde{c}} \right).
\end{align}
\end{subequations}

\noindent  This solution is, of course, recursive, but can be approximated to second order in aberrations, in uniformity with the rest of this work.  

We hesitate in the case of either type of flexion to assign or plot a correction `factor' as a function of PSF size.  Ideally, such a correction factor could be used as a short-cut to convert from the apparent to the pre-convolved flexion values, or from the apparent to the `true' flexion values specified by the derivatives of the lensing potential.  Even for a completely symmetric PSF, such a factor must either ignore the mixing between shear and the two types of flexions that occurs both prior to convolution and during convolution, or assume some relation between the two types of flexion based on a particular lens model.  Only the shear signal, when lensed by an atmospherically aberrated PSF in absence of asymmetric telescope aberrations can be corrected by a simple factor.  Correction of either flexion signal requires knowledge of the shear and the other flexion.

%% file: section_5_conclusion.tex
\section{Conclusion}
Multiple influences bear upon the final measured values of the terms we think of as shear and F\-- and G\--flexions in galaxy images.  We summarize our findings here.
\begin{enumerate}
\item Mixing between shear and flexion alters the magnitude and origin of the F\--flexion\--like spin one signal and G\--flexion\--like spin three signal on the galaxy even prior to convolution with a PSF.  Intrinsic galaxy ellipticity also changes the magnitude of the pre-convolved flexion signals.
\item Under convolution with a symmetric PSF, the shear signal will drop off as the ratio of the unlensed to the lensed galaxy size, squared.
\item A symmetric PSF will have two effects on the measured flexion signals.  (a) It will cause those spin one and three variations on the galaxy with cubed radial dependence to drop off as the ratio of the unlensed to the lensed galaxy size, cubed.  This effect is analogous to the dilution in shear, but higher order.  (b) It will cause those spin one and three variations on the galaxy with quartic radial dependence to mix down into the corresponding spin one and three flexion\--like signals, possibly enhancing the flexion\--like signal on the galaxy.
\item Shear and F\-- and G\--flexion have telescope aberration counterparts, astigmatism, coma, and trefoil, with matching deflection properties.  For asymmetric PSFs, each apparent lensing aberration will approach the value of its corresponding telescope aberration in the limit that the unconvolved galaxy size is small compared to the PSF size.  When the seeing is equal to the width of the unsmeared galaxy, the contributions of the telescope aberrations and the contributions of the gravitational distortions to the final measured image are equally weighted.
\item Under convolution with an asymmetric PSF, the pre-convolved shear and flexions can mix with the PSF asymmetries to corrupt to final convolved signals of the other lensing terms.  For example, the apparent spin three signal in a convolved galaxy image might have contributions from spin two shear mixed with spin one coma.  This effect may be relatively large in certain PSF regimes.
\end{enumerate}
One must account for all of the above when using measurements of `shear' and `flexion' to reconstruct the true lensing parameters.  This work corroborates, and more importantly quantifies, the well-known signal dilution of shear caused by atmospheric seeing, and bias to the same introduced by spin two fields in the PSF (i.e. astigmatism).   Moreover, we have analyzed the effect of the PSF on F\-- and G\--flexion, and have found that the atmospheric dilution is not the same for these lensing terms as it is for shear.  Importantly, one must carefully account for cross terms between shear and flexions when reconstructing the flexion signals, a detail not required for accurately reconstructing the shear signal.

As always, a small PSF, either due to a steady, absent, or controlled atmosphere (e.g. with wide-field adaptive optics) is of primary importance to retaining the gravitational lensing signal.  However, a well-maintained telescope focus will be especially critical to obtaining unbiased flexion measurements, as focus will dictate the severity of the effective astigmatic, comatic, and trefoil aberrations.  These aberrations, which can create corruptive signals on their own or interplay with shear and flexion to create non-linear distortions, will be much harder to measure with certainty and recover from in post processing than simple signal dilution from the PSF or shear-flexion mixing.  The telescope that can best control its aberrations will surely be most suited to measure flexion.  For the rest, we must do the best we can to measure aberrations and account for them.

\medskip
\noindent
\acknowledgments
{\it Acknowledgements:} I would like to thank Paul L. Schechter for his guidance and many reviews of this paper.  This work was supported by the National Science Foundation through a Graduate Research Fellowship to Rebecca Sobel Levinson and through AST\--0607601.

%% file: section_7_appendix.tex
\section{Using vectors and complex numbers to represent lensing}
\label{sec:notation}

In weak lensing, convergence, shear, and flexion are often expressed as components of matrices and tensors acting upon vector coordinates in the lens plane.  A complex number formalism for flexions as vectors and `pseudovectors' with various spin symmetries was introduced to weak lensing by \citet{BaconGoldberg2006} in order to simplify the discussion of flexion.\footnote{As tensors, ellipticities have been treated near identically to pseudovectors since at least \citet{Kaiser1992}, however the complex number formalism wasn't explicitly used there.}  However, this complex formalism for vectors is typically only used to describe the lensing terms themselves; matrices and tensors are still relied upon to describe the distortions to images imparted by shears and flexions.  \citet{CainSchechter2011} is a notable exception.

Here we shall avoid matrices and tensors, as the physical origins and spin symmetries of the lensing terms can easily be obscured within them.  We shall instead rely solely on vectors to capture the effects of the lensing terms on images.  For ease of notation, we will use the complex number formalism to express these vectors and pseudovectors.  For those unfamiliar with imaginary number notation as a tool to manipulate vectors, we review it here.  

\subsection{Spin n vectors}

Lensing distortions have magnitude, direction, and spin symmetry.  They are pseudovectors that may be expressed as

\begin{align}
\label{eq:many_forms_of_a_vector}
 \vec{v} =  v_1+ i v_2 &= v e^{i n\phi} \\ \nonumber
 \text{where } v_1 &= v \text{cos}(n\phi) \\ \nonumber
 \text{and }  v_2 &= v \text{sin}(n\phi)
\end{align}

\noindent where $v$, $\phi$, and $n$ are the vector magnitude, direction, and spin.  A pseudovector's spin reflects its rotational symmetry; a pseudovector with spin $n$ requires a rotation of $\frac{2\pi}{n}$ to be mapped back onto itself.  Therefore an ordinary vector is spin one, requiring a full circle rotation before pointing back onto it's initial direction.  However the vector describing the magnitude and orientation of an equilateral triangle is spin three, as any rotation of $\frac{2\pi}{3}$ will map the object back onto itself.  A spin zero object is a scalar. 

The complex conjugate of a vector is simply another vector given by

\begin{equation}
\label{eq:cc_vector}
\vec{v}^{*} =  v_1- i v_2 = v e^{-i n\phi} \\
\end{equation}

The vector multiplication of two vectors $\vec{u}$ and $\vec{v}$ is the multiplication of the complex numbers used to express them,

\begin{align}
\label{eq:vector_multiplication}
\vec{v}\vec{u} =  (v_1 u_1 - v_2 u_2) + i (v_2 u_1 + v_1 u_2) = u v e^{i (n_v \phi_v + n_u \phi_u)}.
\end{align}

The dot product of two vectors in this notation works exactly like the dot product of two ordinary vectors in any other notation.  Namely

\begin{equation}
\label{eq:dot_product}
\vec{v} \cdot \vec{u} = (v_1 u_1 + v_2 u_2),
\end{equation}

\noindent the result being a real number.  Equivalently, a dot product can be expressed as

\begin{equation}
\label{eq:dot_product_alt}
\vec{v} \cdot \vec{u} = \frac{1}{2}\left( \vec{v}\vec{u}^{*} + \vec{v}^{*} \vec{u} \right),
\end{equation}

\noindent an expansion that will be used often in this paper to simplify expressions.  Thus, any vector taken in dot product with itself, or in multiple with its complex conjugate, will be expressed as its magnitude squared, in agreement with equations \eqref{eq:dot_product} and \eqref{eq:dot_product_alt}:  $\vec{v} \cdot \vec{v} = \vec{v}\vec{v}^{*} = v^2$.  However, a vector squared and its magnitude squared are very different, see equation \eqref{eq:vector_multiplication}, and so vector quantities in this paper will {\it always} be denoted as such, and all quantities not denoted as vectors may be assumed to be scalars.

\subsection{Partial derivatives}

Partial derivatives with respect to a vector are also vectors and will be denoted by a vector sign.  

A vector first derivative is given by

\begin{equation}
\label{eq:first_derivative}
\frac{\vec{\partial}}{\partial \theta} = \frac{\partial}{\partial \theta_x} + i \frac{\partial}{\partial \theta_y}
\end{equation}

\noindent and operates on a scalar, converting it into a spin one vector.  The vector first derivative is the gradient operator $\vec{\nabla}$ and may be written and referred to as such.

There are two second derivatives, one which is spin zero and will map a scalar onto another scalar, and one which is spin two and will map a scalar onto a vector.  The spin zero second derivative is the product of the first derivative (spin one) and its complex conjugate (spin negative one).  The spin two second derivative is the vector product of the first derivative and itself.  They are respectively expressed as

\begin{align}
\label{eq:second_derivatives}
\frac{ \partial^2 }{ \partial \theta^2 } &=
	\left(\frac{\vec{\partial}}{\partial \theta} \right)^{*} \frac{\vec{\partial}}{\partial \theta} =
	\frac{\partial^2}{\partial \theta_{x}^2} + \frac{\partial^2}{\partial \theta_{y}^2} \\
\frac{ \vec{\partial}^2 }{ \partial \theta^2 } &=  
	\frac{\vec{\partial}}{\partial \theta}\frac{\vec{\partial}}{\partial \theta} = 
	\left( \frac{\partial^2}{\partial \theta_{x}^2} - \frac{\partial^2}{\partial \theta_{y}^2} \right) 
	+ i \left( 2 \frac{\partial}{\partial \theta_x} \frac{\partial}{\partial \theta_y} \right).
\end{align}

\noindent Extensions of the same principles can be made for third, fourth, and higher order derivatives.  The spin one and spin three vector third derivatives which are needed to derive the F\-- and G\--flexions from the lensing potential are

\begin{align}
\label{eq:third_derivatives}
\frac{ \partial^2 }{ \partial \theta^2 } \frac{ \vec{\partial} }{ \partial \theta }&=
	\frac{\vec{\partial}}{\partial \theta} 
	\left(\frac{\vec{\partial}}{\partial \theta} \right)^{*} 
	\frac{\vec{\partial}}{\partial \theta} 
	=
	\left( \frac{\partial^3}{\partial \theta_{x}^3} 
		+ \frac{\partial^3}{\partial \theta_{x}\partial \theta_{y}^2} \right) 
	+ i \left( \frac{\partial^3}{\partial \theta_{x}^2 \partial \theta_{y}} 
		+ \frac{\partial^3}{\partial \theta_{y}^3} \right) \\
\frac{ \vec{\partial}^3 }{ \partial \theta^3 } &=  
	\frac{\vec{\partial}}{\partial \theta} 
	\frac{\vec{\partial}}{\partial \theta}
	\frac{\vec{\partial}}{\partial \theta} 
	= 
	\left( \frac{\partial^3}{\partial \theta_{x}^3} 
		-3 \frac{\partial^3}{\partial \theta_{x}\partial \theta_{y}^2} \right) 
	+ i \left(3 \frac{\partial^3}{\partial \theta_{x}^2 \partial \theta_{y}} 
		- \frac{\partial^3}{\partial \theta_{y}^3} \right).
\end{align}